\documentclass[aps,preprintnumbers,amsmath,amssymb,nofootinbib,eqsecnum, preprintnumbers]{revtex4}
\usepackage{eurosym}
\usepackage{amsfonts}
\usepackage{amsmath}
\usepackage{amssymb,epsf}
\usepackage{color}
\usepackage{graphicx}
\usepackage{natbib}
\usepackage{float}
\usepackage{caption}
\usepackage{subfig}

\usepackage{epstopdf}
\usepackage{hyperref}
\hypersetup{
    colorlinks=true,
    linkcolor=blue,
    filecolor=magenta,      
   citecolor=blue
}

\begin{document}
\title{Kerr-Newman Black Holes in Weyl-Cartan Theory: Shadows and EHT constraints}

\author{
 Khadije Jafarzade$^{1,2,3}$\footnote{email address: khadije.jafarzade@gmail.com},
 Seyed Hossein Hendi$^{1,2,4*}$\footnote{email address: hendi@shirazu.ac.ir}
 Mubasher Jamil$^{5*}$\footnote{email address: mjamil@sns.nust.edu.pk}, and 
 Sebastian Bahamonde$^{6,7}$\footnote{email address: sbahamondebeltran@gmail.com} }

 \affiliation{
 $^{1}$Department of Physics, School of Science,Shiraz University, Shiraz 71454, Iran\\
 $^{2}$Biruni Observatory, School of Science, Shiraz University, Shiraz 71454, Iran\\
 $^{3}$ICRANet-Mazandaran, University of Mazandaran, P. O. Box 47415-416, Babolsar, Iran\\
 $^4$Canadian Quantum Research Center 204-3002 32 Ave Vernon, BC V1T 2L7 Canada\\ 
 $^5$School of Natural Sciences, National University of Sciences and Technology, H-12, Islamabad 44000, Pakistan\\
 $^6$Department of Physics, Tokyo Institute of Technology 1-12-1 Ookayama, Meguro-ku, Tokyo 152-8551, Japan\\
 $^7$Kavli Institute for the Physics and Mathematics of the Universe (WPI), The University of Tokyo Institutes
for Advanced Study (UTIAS), The University of Tokyo, Kashiwa, Chiba 277-8583, Japan}

 \begin{abstract}
With the recent release of the black hole image of Sgr A* alongside the earlier image of M87*, one can achieve an in-depth understanding of gravitational physics at the horizon scale. According to the Event Horizon Telescope (EHT) collaboration, the observed image is consistent with the expected appearance of a Kerr black hole. In the present work, we consider Kerr-Newman black holes in Weyl-Cartan theory as a supermassive black hole (BH) and evaluate the parameters of the model with shadow size estimates done by the observations of M87* and Sgr A* from EHT. Such a study can be a possible way to distinguish Weyl-Cartan theory from general relativity and ensure the validity of the idea. Besides, we calculate the energy emission rate for the corresponding BH and discuss how the model's parameters affect the emission of particles around the black hole. With this investigation, we are able to examine the time evolution and lifetime of the black hole in such a theory of gravity.
\end{abstract}

\maketitle
\section{Introduction}
 
Black holes are the most fascinating predictions of general relativity which hold the key to answer several unresolved problems in theoretical physics including the nature of quantum gravity and the information loss paradox. For several decades after their prediction, the existence of black holes in the physical universe remained a mystery. However, two of the most remarkable predictions of Einstein’s general theory of relativity namely, the black hole shadows and gravitational waves generated by black hole-black hole merger have been verified successfully in recent years. The LIGO-VIRGO collaboration detected gravitational wave signals produced by the merger of two black holes at high redshift, which has opened a new window of gravitational wave astronomy \cite{ligo1,ligo2}. Remarkably, the Event Horizon Telescope has witnessed the formation of photon rings, which constitute the shadow, in the vicinity of the supermassive black hole in M87* galaxy and also Sgr A* supermassive black hole at the heart of Milky Way galaxy, for the first time \cite{sgr1,sgr2}. For M87* supermassive black hole, the EHT team has detected large scale magnetic fields and plasma near the black hole using polarized synchrotron radiation \cite{m87a,m87b}. In addition, the EHT team inferred a mass accretion rate onto the black hole in M87* of $(3-20)\times10^{-4} $ solar mass per year. Thus, the M87* central black hole is an active black hole associated with an accretion disk and a jet.

The existence of photon rings around these supermassive black holes suggests the existence of accretion disks constituting high energy particles in motion and eventually falling into the black hole. Some of the photons emitted from the accretion disk with specific critical energy can move in circular orbits near the black hole. Due to chaotic motions and collisions between particles, these photons can escape towards an external observer. The paths followed by the photons in circular motion and during their escape can help us identify the spacetime curvature in strong gravity regime. Inside the bright rings, there lies a black region which is termed the black hole shadow (to have a more intuitive understanding of the nature of black hole shadow, see Refs. \cite{Falcke:528,Mizuno:585,Bronzwaer:920,Perlick:947})  whose geometry, radius and angular diameter size can be helpful to constrain the black hole candidate models from different gravitational theories.  The topic of shadow has been studied in the wide context of black hole physics, such as regular BHs \cite{Kumaran:812}, BHs in the presence of non-linear electrodynamics \cite{Allahyari:003,Okyay:009}, BHs with scalar hair \cite{Khodadi:026}, BHs with extra dimensions \cite{Vagnozzi:02420}, BHs under the effect of generalized uncertainty principle \cite{Pantig:2023,Lambiase:679}, BHs in Metric-Affine Bumblebee gravity \cite{Lambiase:023}, and BHs in symmergent gravity \cite{Cimdiker:100900,Cimdiker:184100,Cimdiker:195003}.

Eventually, we want to understand if the observed black holes possess any spin, charge, scalar hair or any deformations that are compatible with our existing gravitational theories. Further, the same observations would be instructive to differentiate the general relativity and the alternative extended models of gravity and eventually rule out some of the existing gravity theories. In literature, the black hole shadows are modelled and constrained not only using black hole candidates but wormhole and naked singularity spacetimes as well \cite{whns1,whns2,whns3}. Since a small size black hole (such as Sgr A*) produces a higher spacetime curvature as compared to large size black hole (such as M87*), therefore observations of shadow of Sgr A* can impose stringent constraints on the free parameters of black hole spacetime candidates, for a detailed review on available constraints on black holes in general relativity and modified gravity, see \cite{Wang:202181,Ghosh:2023174,Zakharov:2022141,Zajacek:2019012,Zajacek:2018,Mbonye:1999339,Nayak:2012709,Gonzalez:200478,sunny}.


It should be noted that in preliminary studies in the context of BH shadow, the black hole was assumed to be eternal  (in the sense that spacetime was time-independent). Therefore, a static or stationary observer could see a time-independent shadow. But from modern observational results, we know that our universe is expanding with acceleration, indicating that shadow depends on time. Although the effect of the cosmological expansion is negligible for the black hole candidates at the center of the Milky Way galaxy and the centers of nearby galaxies, its effect on the shadow size is significant for galaxies at very long distances \cite{Perlick}. One method to explain the amazing accelerating expansion is through introducing an unknown energy component so-called dark energy which makes up about 70 percent of the universe. One of the well-known candidates for dark energy scenarios is the cosmological constant. The role of the cosmological constant in gravitational lensing has been the subject of focused studies in recent years \cite{Sh1,Sh2,Sh3,Sh4,Sh5,EHT1}. It has been known that the shadow and photon ring are caused by the light deflection or gravitational lensing by BHs. Thus, exploration of the cosmological constant effect on the BH can be of particular interest \cite{Sh6,Sh7,Sh8}. It is worth pointing out that even though the expansion of the Universe was based on a positive cosmological constant, there is some evidence revealing the fact that it can be associated with a negative cosmological constant. The first evidence is based on supernova data. Although there is strong observational evidence from high-redshift supernovae that a positive cosmological constant is the main cause of the expansion of the Universe, the supernova data themselves
derive a negative mass density in the Universe which can be equivalent to a negative cosmological constant \cite{Riess,Perlmutter}. Several galaxy cluster observations appear to have inferred the presence of a negative mass in cluster environments. In Ref. \cite{Farnes}, it was shown that the introduction of negative masses can lead to an AdS space. The second reason to consider a negative cosmological constant is the concept of stability of the accelerating universe. In Ref. \cite{Maedaa}, authors analyzed the possibility of de Sitter expanding spacetime with a constant internal space and proved that the de Sitter solution would be stable only in the presence of the negative cosmological constant. The third one is through observational Hubble constant data. As
we know, an interesting approach to examine the accelerated cosmic expansion and study properties of dark energy is via the Hubble constant which is measured as a function of cosmological redshift. The investigation of $H(z)$ behavior at low redshift data showed that the dark energy density has a negative minimum for certain redshift ranges which can be simply modeled through a negative cosmological constant \cite{Dutta12}.

The most natural family of black hole candidates to test and analyze the black hole shadows is the Kerr black hole and other spinning black holes with free parameters derived by solving the governing field equations of modified gravity theories. Separate investigations have been reported in the literature over the past several years to constrain black hole candidates in extended gravity theories using shadows. Some of these theories include Horava Lifshitz, Einstein-Aether, Gauss-Bonnet, massive gravity, generalized teleparallel, Einstein-Maxwell-Dilaton, Chern-Simons, Einstein-Yang-Mills, string theory, and loop quantum gravity inspired black holes (see references \cite{grav1,grav2,grav3,grav4,grav5,grav6,grav7,grav8,Sh6,grav10}). In this article, we intend to investigate constraints on the newly proposed Weyl-Cartan theory (WCT) spinning black holes using the EHT data.

WCT, a theoretical framework that extends general relativity by incorporating nonmetricity and torsion, offers intriguing prospects in the study of black hole physics \cite{Utiyama:101,Hehl:393,Hehl:1995,Hehl:2013}.   In other words, all three geometric properties, namely, curvature, torsion, and nonmetricity are nonzero in this theory of gravity and they are fixed dynamically by their equations of motion.  Unlike other approaches, such as Einstein-Cartan theory \cite{Cartan:1922,Cartan:1923,Einstein:1925,Einstein:1928} or teleparallel gravity \cite{Chatzistavrakidis:1034,Cai:103520,Tong:104002,Hohmann:376}, WCT allows for a more comprehensive description of the gravitational field by considering both the metric tensor and additional geometric quantities. Then, torsion is sourced by intrinsic spin and nonmetricity by intrinsic dilations and shears. This inclusion of nonmetricity and torsion enables a richer understanding of the dynamics near black holes, particularly in extreme conditions where spacetime curvature is significant. By accounting for these additional geometric properties, WCT provides a more nuanced depiction of the gravitational interaction, potentially offering insights into phenomena such as the behavior of matter and energy within the event horizon, the formation and evolution of black holes, and the resolution of long-standing puzzles like the information paradox. While further research and empirical validation are necessary to fully assess the merits of WCT in black hole physics, its unique approach holds promise for advancing our comprehension of these enigmatic cosmic entities. Moreover, it is expected that the intrinsic properties of matter (spin, dilations, and shears) might be relevant in strong gravity regimes, and then, supermassive black holes could in principle be endowed with those intrinsic properties.

This paper is organized as follows: In Sec. \ref{Sec. II}, we briefly describe the geometrical features of charged rotating black holes in WCT. In Sec. \ref{SubsecA}, we study the effect of the parameters of the model on null geodesics, which in turn affect the photon sphere radius and shadow radii. In Sec. \ref{SubsecB}, we examine energy emission rate  and the influence of the WC parameters on the emission of particles. The black hole shadow observables
and their applications in estimating the WC parameters are presented in
Sec. \ref{IV}. Moreover, constraints on the parameters of the model are deduced using the M87* and Sgr A* black hole shadow observational data for the inclination angles of $ 17^{\circ} $ and $ 50^{\circ} $. The paper is finally concluded with some closing remarks.

 \section{Rotating Black holes in Weyl-Cartan theory} \label{Sec. II}

Metric-affine gravity theories are gauge theories of gravity formulated within a non-Riemannian geometry, allowing non-trivial contributions from the torsion tensor and nonmetricity, defined as~\cite{Hehl:1995}
\begin{eqnarray}
       T^{\lambda}\,_{\mu \nu}=2\tilde{\Gamma}^{\lambda}\,_{[\mu \nu]}\,,\quad Q_{\lambda \mu \nu}=\tilde{\nabla}_{\lambda}g_{\mu \nu}\,,
\end{eqnarray}
where now the affine connection $\tilde{\Gamma}^\lambda{}_{\mu\nu}$ does not depend on the metric $g_{\mu\nu}$, as in the case of Riemannian geometries such as in GR, where the Levi-Civita connection $\Gamma^\lambda{}_{\mu\nu}$ is expressed in terms of the metric. Moreover, the Levi-Civita connection is symmetric, meaning that it does not contain torsion and also satisfies the so-called metric compatibility condition that ensures no nonmetricity $Q_{\lambda \mu \nu}=\tilde{\nabla}_{\lambda}g_{\mu \nu}=0$. Both quantities have their own geometrical interpretation and they not only contribute to the dynamics of the theories constructed under these geometries but also modify the matter sector. Since the torsion tensor is antisymmetric in its last two indices, it carries up to 24 degrees of freedom, whereas nonmetricity is symmetric in its first two indices, meaning that it has 40 degrees of freedom in a general setting. Then, the affine connection can be decomposed as
\begin{eqnarray}
   \tilde{\Gamma}^\lambda{}_{\mu\nu}=\Gamma^\lambda{}_{\mu\nu} +\frac{1}{2}\left(T^{\lambda}\,_{\mu \nu}-T_{\mu}\,^{\lambda}\,_{\nu}-T_{\nu}\,^{\lambda}\,_{\mu}\right)+\frac{1}{2}\left(Q^{\lambda}\,_{\mu \nu}-Q_{\mu}\,^{\lambda}\,_{\nu}-Q_{\nu}\,^{\lambda}\,_{\mu}\right)=\Gamma^\lambda{}_{\mu\nu}+N^\lambda{}_{\mu\nu}\,,
\end{eqnarray}
where $N^\lambda{}_{\mu\nu}$ is a tensor known as the distortion tensor, which measures the difference between the general connection and the Riemannian one. Since the connection is more general, the definition of the covariant derivative now carries extra degrees of freedom coming from torsion and nonmetricity. Further, the general curvature can be decomposed in the post-Riemannian form as follows:
\begin{eqnarray}
\tilde{R}^{\lambda}\,_{\rho\mu\nu}=R^{\lambda}\,_{\rho\mu\nu}+\nabla_{\mu}N^{\lambda}\,_{\rho \nu}-\nabla_{\nu}N^{\lambda}\,_{\rho \mu}+N^{\lambda}\,_{\sigma \mu}N^{\sigma}\,_{\rho \nu}-N^{\lambda}\,_{\sigma \nu}N^{\sigma}\,_{\rho \mu}\,.
\end{eqnarray}
In our present study, we will concentrate on a particular geometry within the general Metric-Affine gravity case, known as Weyl-Cartan geometry, where the connection contains curvature, torsion, and nonmetricity is only given by its Weyl part. This reduces nonmetricity to be written as a vector:
\begin{eqnarray}
    Q_{\lambda\mu\nu}=g_{\mu\nu}W_{\lambda}\,,
\end{eqnarray}
and then, nonmetricity is reduced to up to 4 degrees of freedom and is fully expressed by the Weyl vector $W_{\mu}=\frac{1}{4}\,Q_{\mu\nu}{}^{\nu}$. Up to now, we have only defined geometrical quantities. One can then further construct gravitational theories using curvature (the general one), torsion, and the Weyl part of nonmetricity. In particular, we will concentrate on the model given by the action~\cite{Bahamonde:2021qjk}
\begin{eqnarray}\label{LagrangianIrreducible}
S &=& \frac{1}{64\pi}\int d^4x \sqrt{-\,g}
\left.\Bigl[2\Lambda
-\,4R+\frac{1}{4}F_{\mu\nu}F^{\mu\nu}-6d_{1}\tilde{R}_{\lambda\left[\rho\mu\nu\right]}\tilde{R}^{\lambda\left[\rho\mu\nu\right]}-9d_{1}\tilde{R}_{\lambda\left[\rho\mu\nu\right]}\tilde{R}^{\mu\left[\lambda\nu\rho\right]}+8\,d_{1}\tilde{R}_{\left[\mu\nu\right]}\tilde{R}^{\left[\mu\nu\right]}
\Bigr.
\right.
\nonumber\\
& &
\left.
\Bigl.
\;\;\;\;\;\;\;\;\;\;\;\;\;\;\;\;\;\;\;\;\;\;\;\;\;\;\;\;\;+\,\frac{1}{8}\left(32e_{1}+13d_{1}\right)\tilde{R}^{\lambda}\,_{\lambda\mu\nu}\tilde{R}^{\rho}\,_{\rho}\,^{\mu\nu}-7d_{1}\tilde{R}_{\left[\mu\nu\right]}\tilde{R}^{\lambda}\,_{\lambda}\,^{\mu\nu}\Bigr]\right.\,.
\end{eqnarray}
The first three terms of the above action describe the standard Einstein-Maxwell equations with a cosmological constant since $R$ is the Riemannian Ricci scalar, $\Lambda$ is the cosmological constant, and $F_{\mu\nu}=\partial_{[\mu}A_{\nu]}$ is the electromagnetic tensor. The other terms appearing in the action provide modifications of GR given by field strength tensors constructed from the modified Bianchi identities:
\begin{eqnarray}
   \tilde{R}^{\lambda}\,_{[\mu \nu \rho]}+\tilde{\nabla}_{[\mu}T^{\lambda}\,_{\nu \rho]}+T^{\sigma}\,_{[\mu \nu}\,T^{\lambda}\,_{\rho] \sigma}=0\,,\quad \tilde{R}^{\left(\lambda\rho\right)}\,_{\mu\nu}=g^{\lambda\rho}\nabla_{[\nu}W_{\mu]}=\frac{1}{4}\,g^{\lambda\rho}\tilde{R}^{\sigma}\,_{\sigma\mu\nu}\,.
\end{eqnarray}
Clearly, the above theory introduces kinetic terms for torsion and nonmetricity, giving dynamics to those fields. Moreover, the above theory is reduced to GR if both torsion and nonmetricity vanish. This means that the modification of gravity is based solely on those geometrical quantities and is mediated by the constants $e_1$, which only depend on nonmetricity, and $d_1$, which provides corrections when torsion is nonzero. Note that $d_1\leq0$ to avoid ghost instabilities.

In this work, we are interested in this theory since black hole solutions have been found in a series of papers~\cite{Bahamonde:2020fnq,Bahamonde:2021,Bahamonde:2021qjk,Bahamonde:2022meb,Bahamonde:2023piz}. The solution that we are interested in assumes the decoupling limit $|a k_{s}| \ll 1$ for the torsional part (valid only in that limit), where $a$ is the rotation parameter and $k_{s}$ is the spin charge related to the fact that torsion now provides an intrinsic spin generating a gravitational field. The solution was obtained without any approximation in the Weyl part of nonmetricity. The complete form of the torsion tensor and the Weyl part of nonmetricity were displayed in Eqs. (72)-(77) and (85) in~\cite{Bahamonde:2021qjk}, respectively. Finally, the metric describes a rotating black hole solution whose spacetime is~\cite{Bahamonde:2021qjk}.
\begin{eqnarray}
ds^2&=&-\Big(\Psi(r,\theta)-\frac{1}{3}\frac{a^4\Lambda\cos^2\theta\sin^2\theta}{r^2+a^2\cos^2\theta}\Big)\,dt^2
+\frac{r^2+a^2\cos^{2}\theta}{\left(r^2+a^2\cos^{2}\theta\right)\Psi(r,\theta)+a^2\sin^{2}\theta}\,dr^2+\frac{r^2+a^2\cos^2\theta}{1+\frac{1}{3}a^2\Lambda\cos^2\theta}\,d\theta^2\nonumber\\
&&+\,\sin^2\theta\left[r^2+a^2+a^2(1-\Psi(r,\theta))\sin ^2\theta+\frac{a^2 \Lambda\cos^2\theta\left(r^2+a^2\right)^2}{3 \left(r^2+a^2\cos^2\theta\right)}\right]d\varphi^2\nonumber\\
&&-\,2a\sin^2\theta\left[1-\Psi(r,\theta )+\frac{a^2\Lambda\left(r^2+a^2\right)\cos^2\theta}{3\left(r^2+a^2\cos^2\theta\right)}\right]dtd\varphi\,,
\label{metric}
\end{eqnarray}
with
\begin{eqnarray}
\Psi(r,\theta)=1-\big\{2Mr-\bigl[q_{e}^{2}+q_{m}^{2}+d_{1}k^{2}_{s}-4e_{1}(k^{2}_{e}+k^{2}_{m})\bigr]+\frac{1}{3}\Lambda r^2(r^2+a^2)\big\}/(r^{2}+a^{2}\cos^{2}\theta)\,,
\label{metric2}
\end{eqnarray}

where, the quantity $ M $ represents the mass parameter, while $q_{e}$ and $q_{m}$ are related to the electric and magnetic charges, respectively. $ k_{s} $ denotes the intrinsic spin charge. The integration constants $k_{e}$ and $k_{m}$ represent, respectively, the electric and magnetic intrinsic dilation charges\footnote{Note that in our paper, we use the notation $k_s$, $k_e$, and $k_m$ to denote the spin charge, electric, and magnetic dilation charges, whereas in the mentioned reference, $\kappa_s$, $\kappa_{e,m}$, $\kappa_{d,m}$ were used to denote those quantities.}. $\Lambda$ refers to the cosmological constant. For $\Lambda <0$ ($\Lambda >0$), these solutions are interpreted as asymptotically anti-de-Sitter (de-Sitter) black holes.  Metric (\ref{metric}) describes a Kerr-Newman Weyl-Cartan geometry connected to the electric and magnetic dilation charges related to the nonmetricity tensor, as well as by a spin charge related to the torsion tensor. Note that the theory producing the above solution is part of WCT with the traceless part of nonmetricity being zero. Therefore, the shear charge is absent. From a phenomenological perspective, such a solution can represent the asymptotic gravitational field of a rotating black hole in a Weyl-Cartan regime with scale invariance \cite{Bahamonde:2021}, whose spin charge is much smaller than the external rotation, or vice versa ($ \vert ak_{s}\vert \ll 1 $). In fact, this condition characterizes the decoupling limit between the orbital and the spin angular momentum of the solution.

\begin{figure}[!htb]
\centering
\subfloat[$ a=0.5$, $ q_{e}=k_{e}=0.2 $ and $ e_{1}=0.5 $]{
        \includegraphics[width=0.31\textwidth]{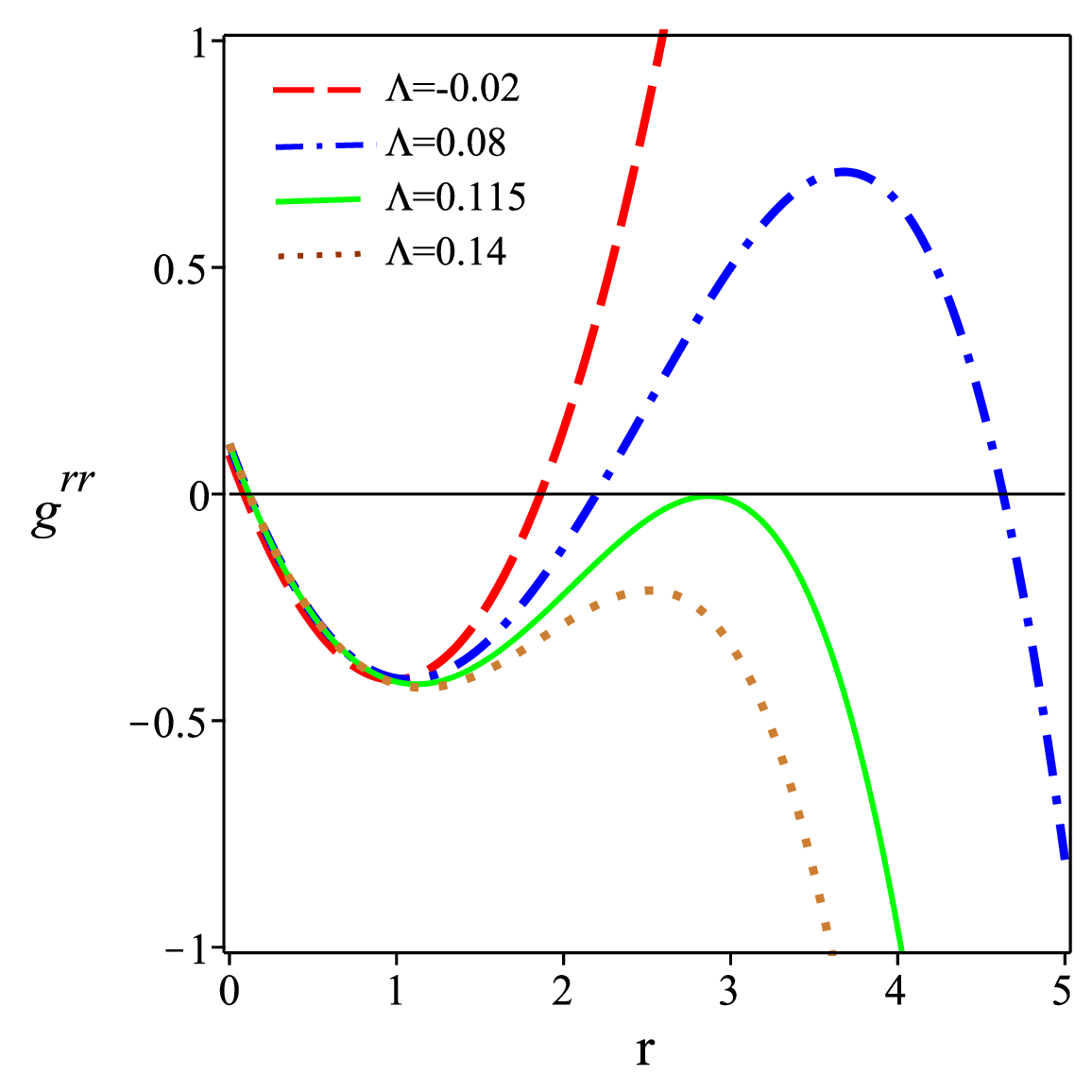}}
\subfloat[$e_{1}=0.5$, $ q_{e}=k_{e}=0.2 $ and $\Lambda=-0.05 $]{
        \includegraphics[width=0.31\textwidth]{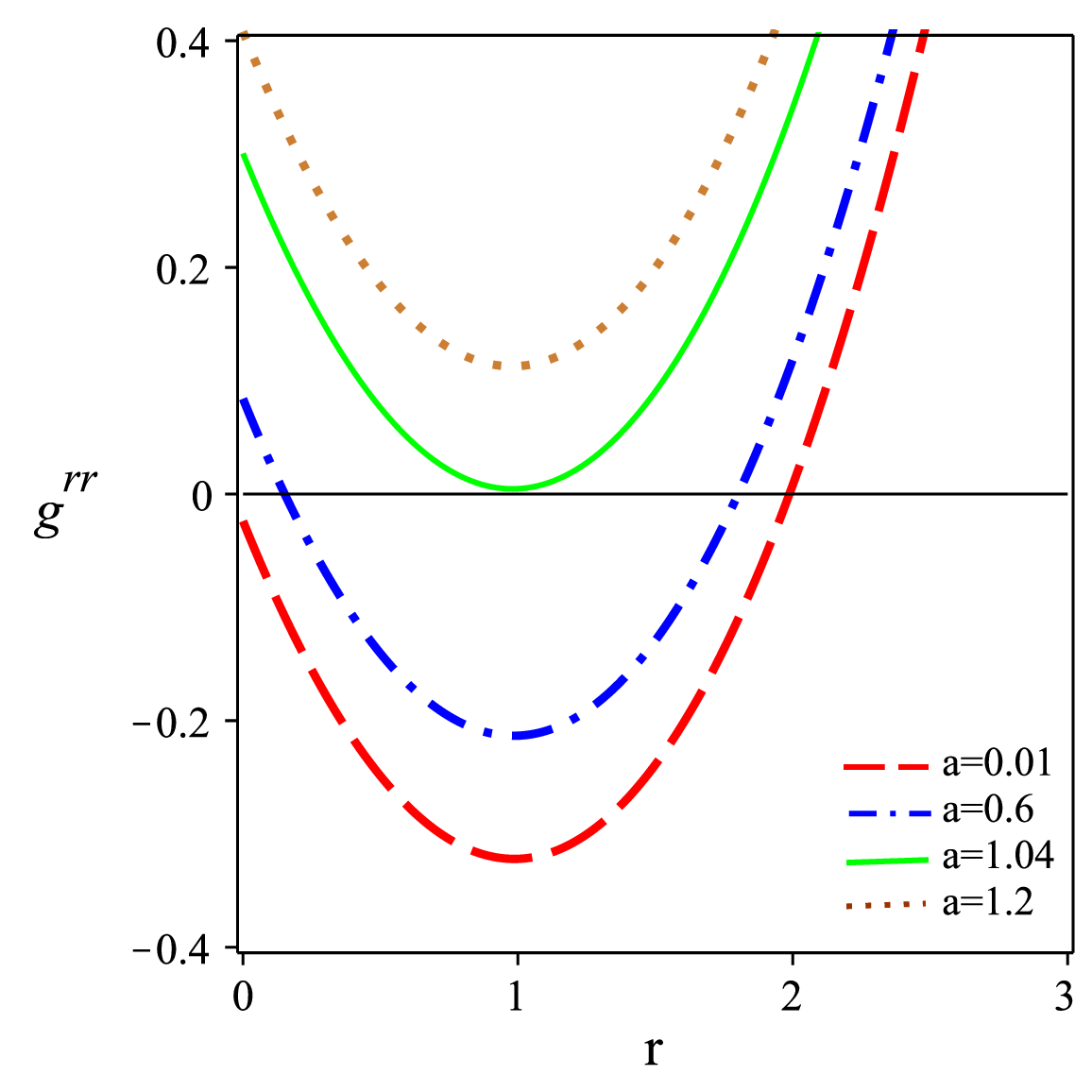}}
\subfloat[$a=e_{1}=0.5$, $k_{e}=0.2 $ and $\Lambda=-0.05 $]{
        \includegraphics[width=0.31\textwidth]{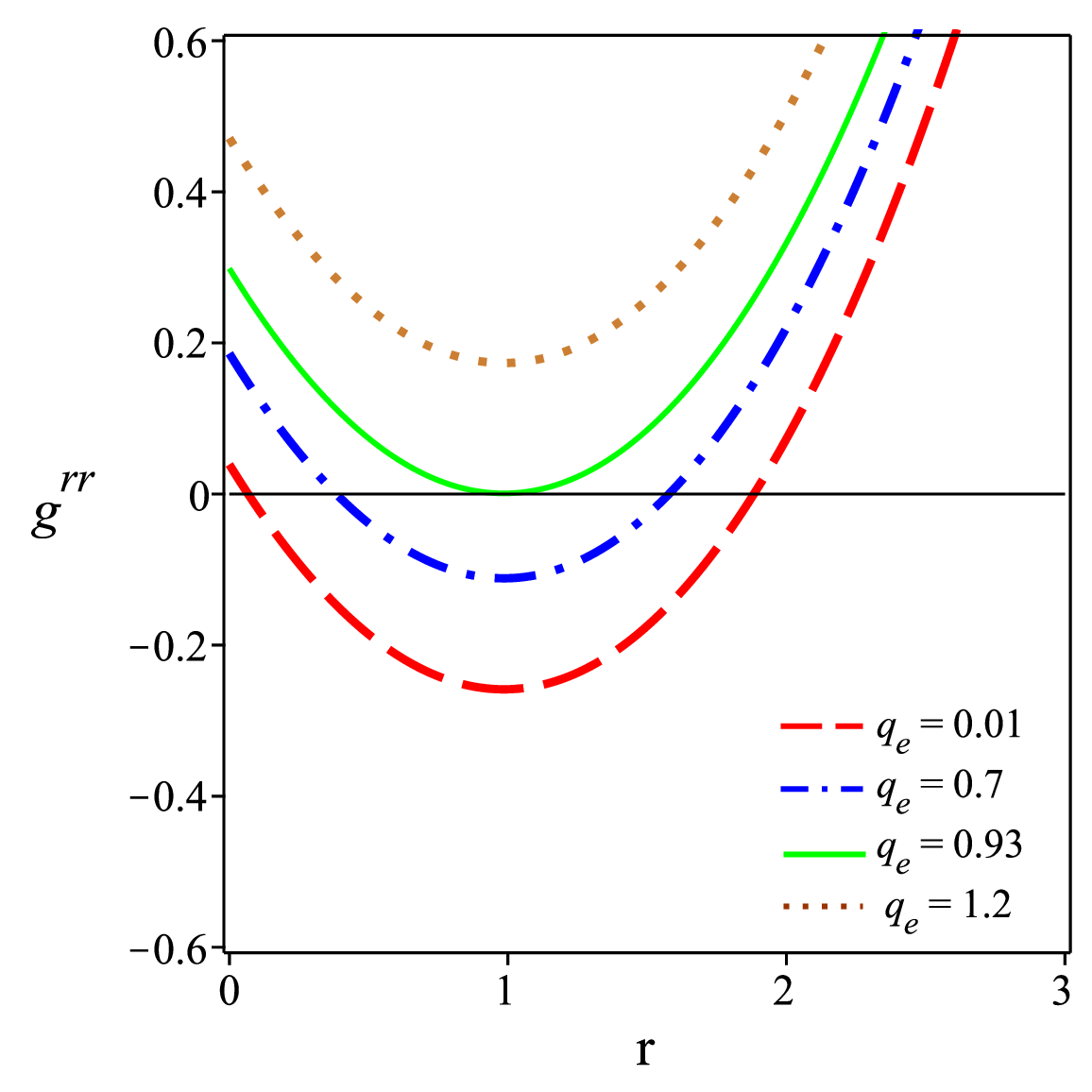}}\newline
\subfloat[$a=0.5$, $q_{e}=k_{e}=0.2 $ and $\Lambda=-0.05 $]{
        \includegraphics[width=0.31\textwidth]{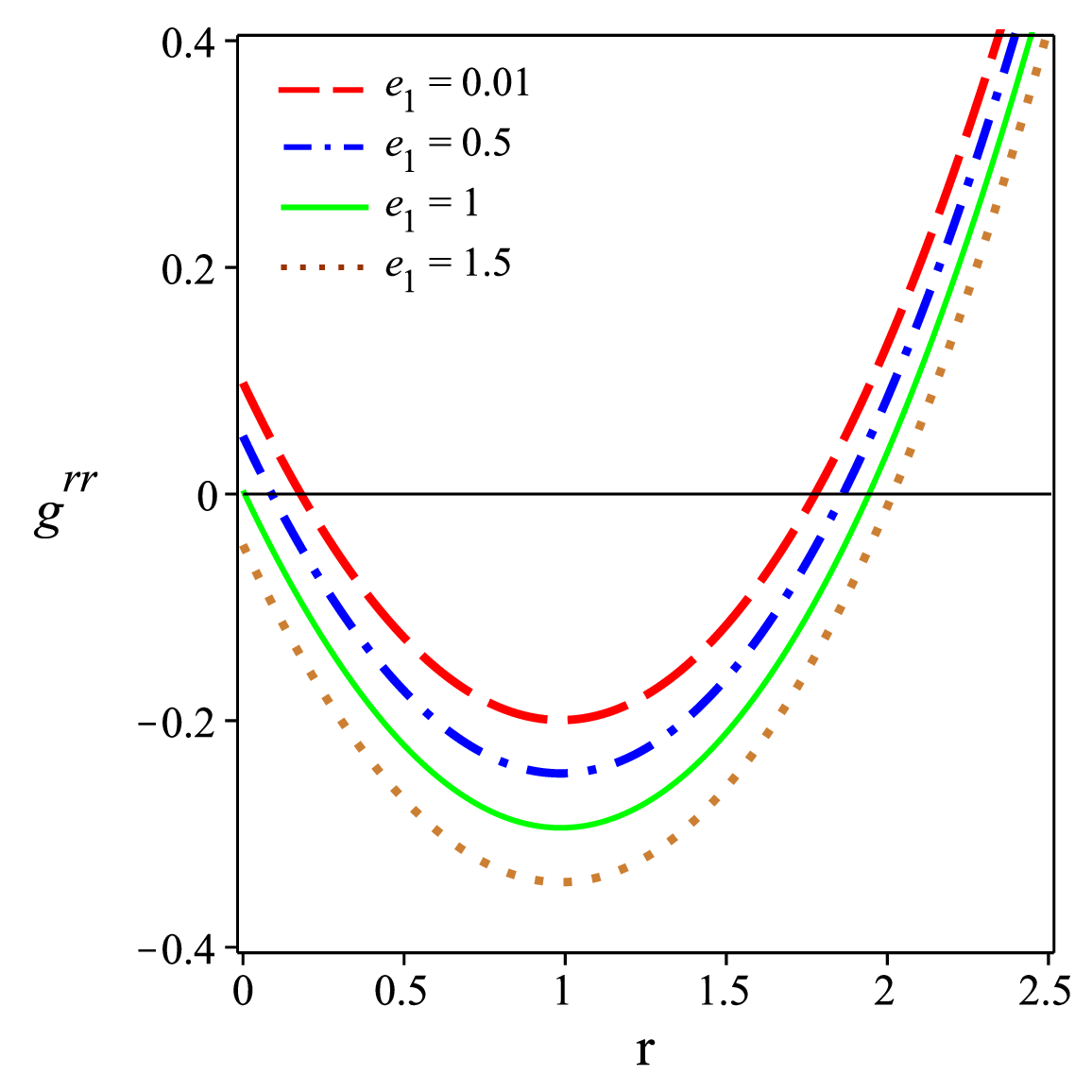}}
\subfloat[$a=e_{1}=0.5$, $q_{e}=0.2 $ and $\Lambda=-0.05 $]{
        \includegraphics[width=0.31\textwidth]{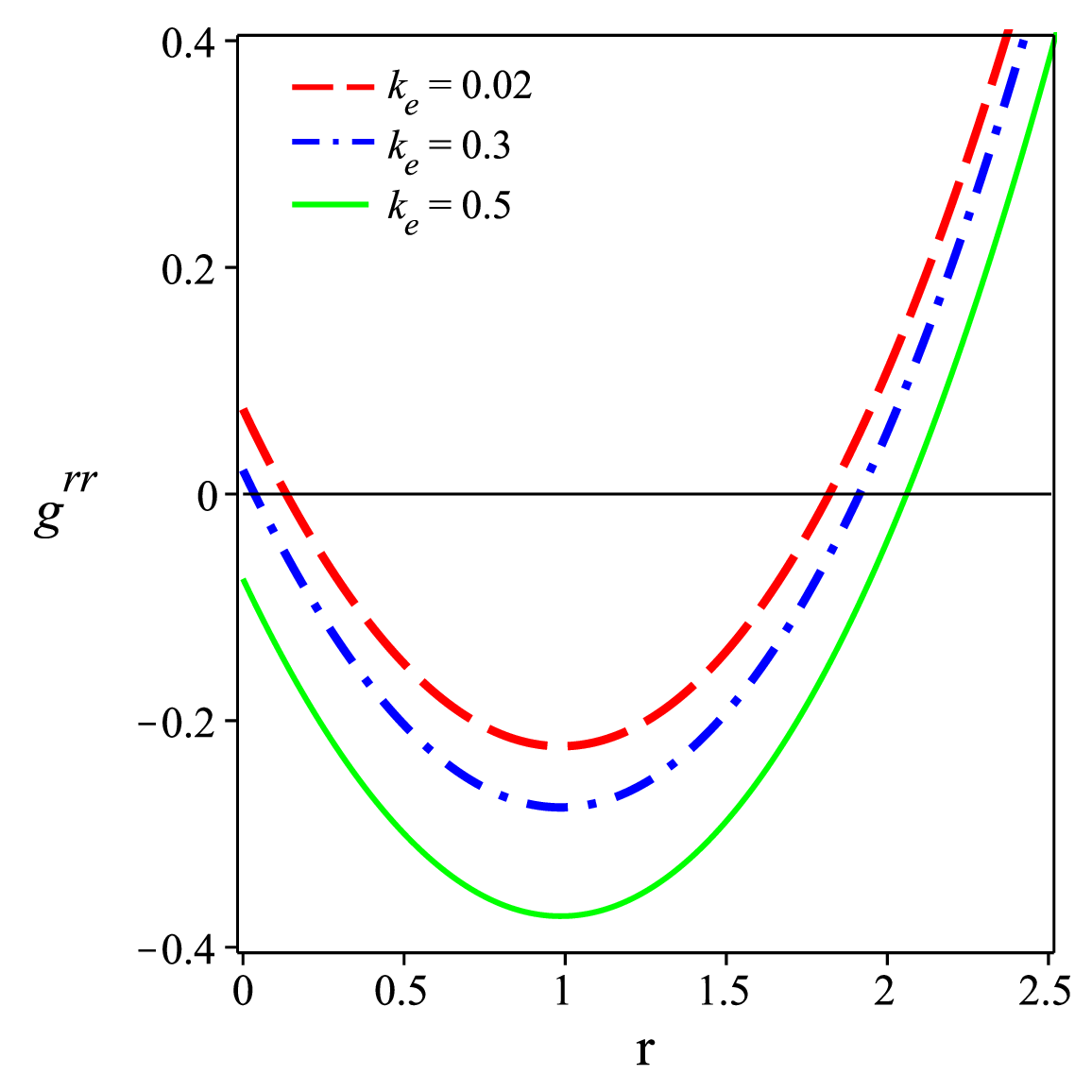}}\newline
\caption{The function $ g^{rr} $ versus $ r $ for $ k_{s}=10^{-6} $, $ q_{m}=k_{m}=0.2 $, $ d_{1}=0.5 $ and $M=1$.}
\label{Fig1a}
\end{figure}
\begin{figure}[!htb]
\centering
\subfloat[$ a=0.5$, $ q_{e}=k_{e}=0.2 $ and $ e_{1}=-0.5 $]{
        \includegraphics[width=0.31\textwidth]{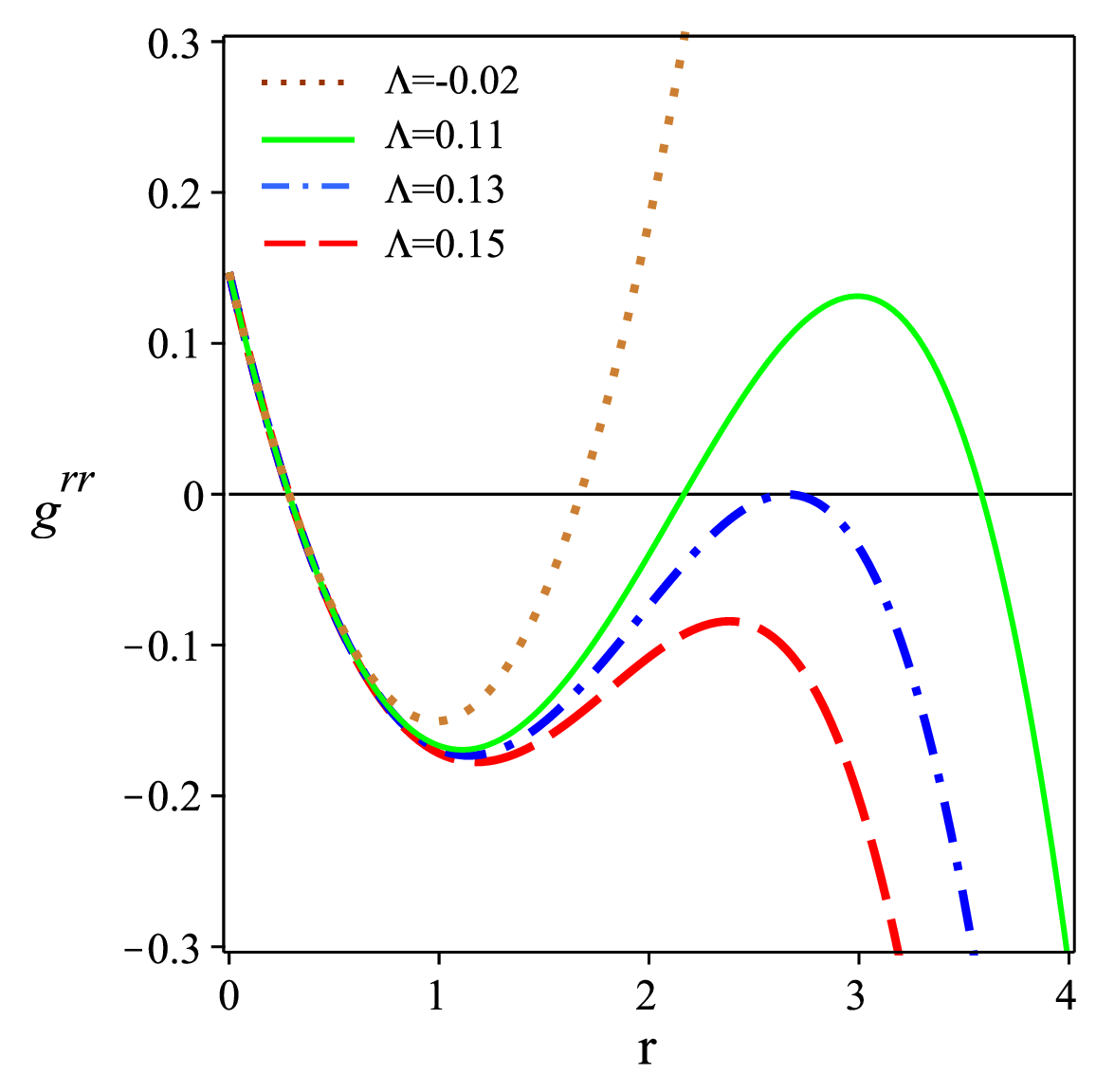}}
\subfloat[$e_{1}=-0.5$, $ q_{e}=k_{e}=0.2 $ and $\Lambda=-0.05 $]{
        \includegraphics[width=0.31\textwidth]{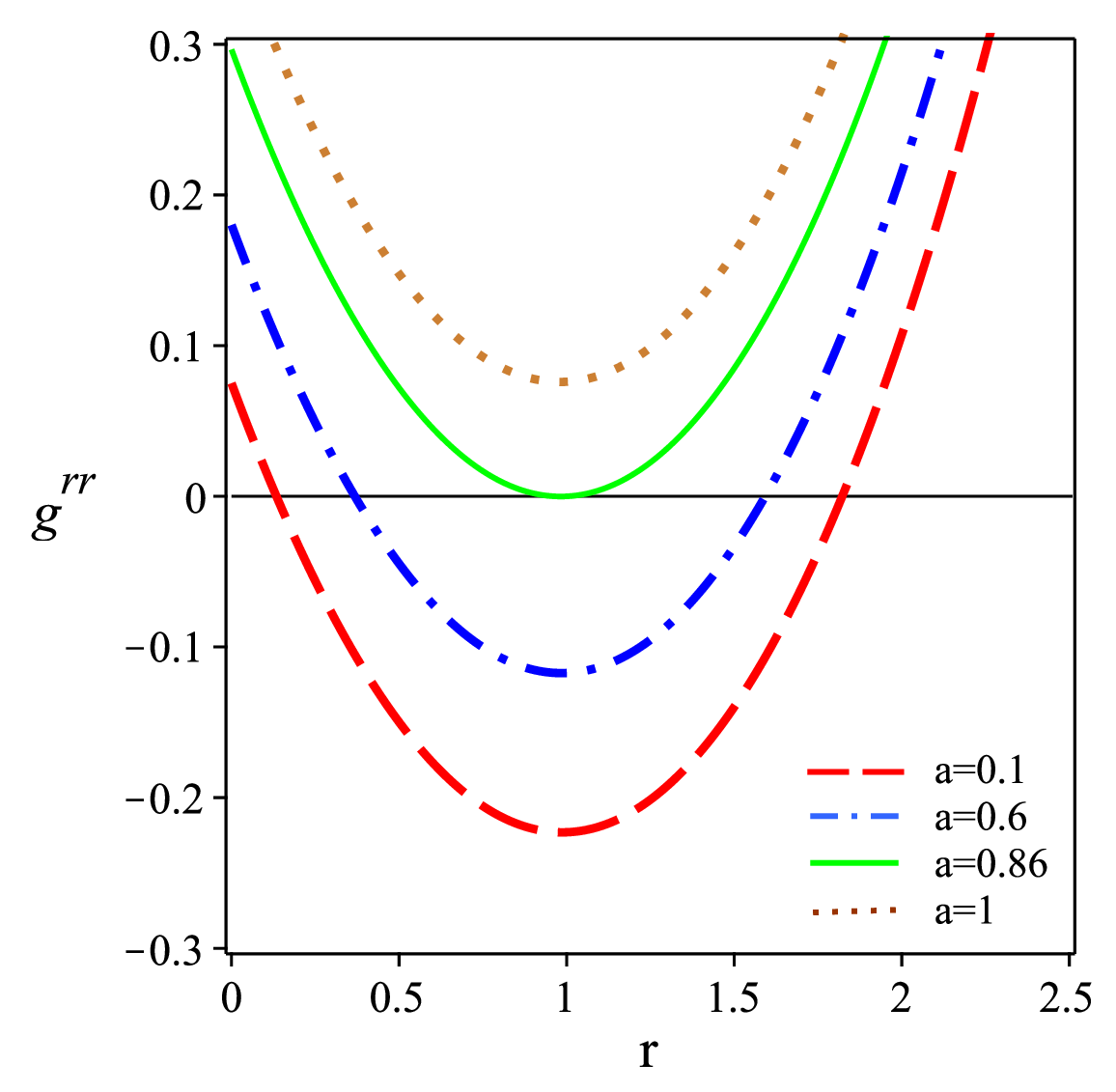}}
\subfloat[$a=0.5$, $e_{1}=-0.5$, $k_{e}=0.2 $ and $\Lambda=-0.02 $]{
        \includegraphics[width=0.31\textwidth]{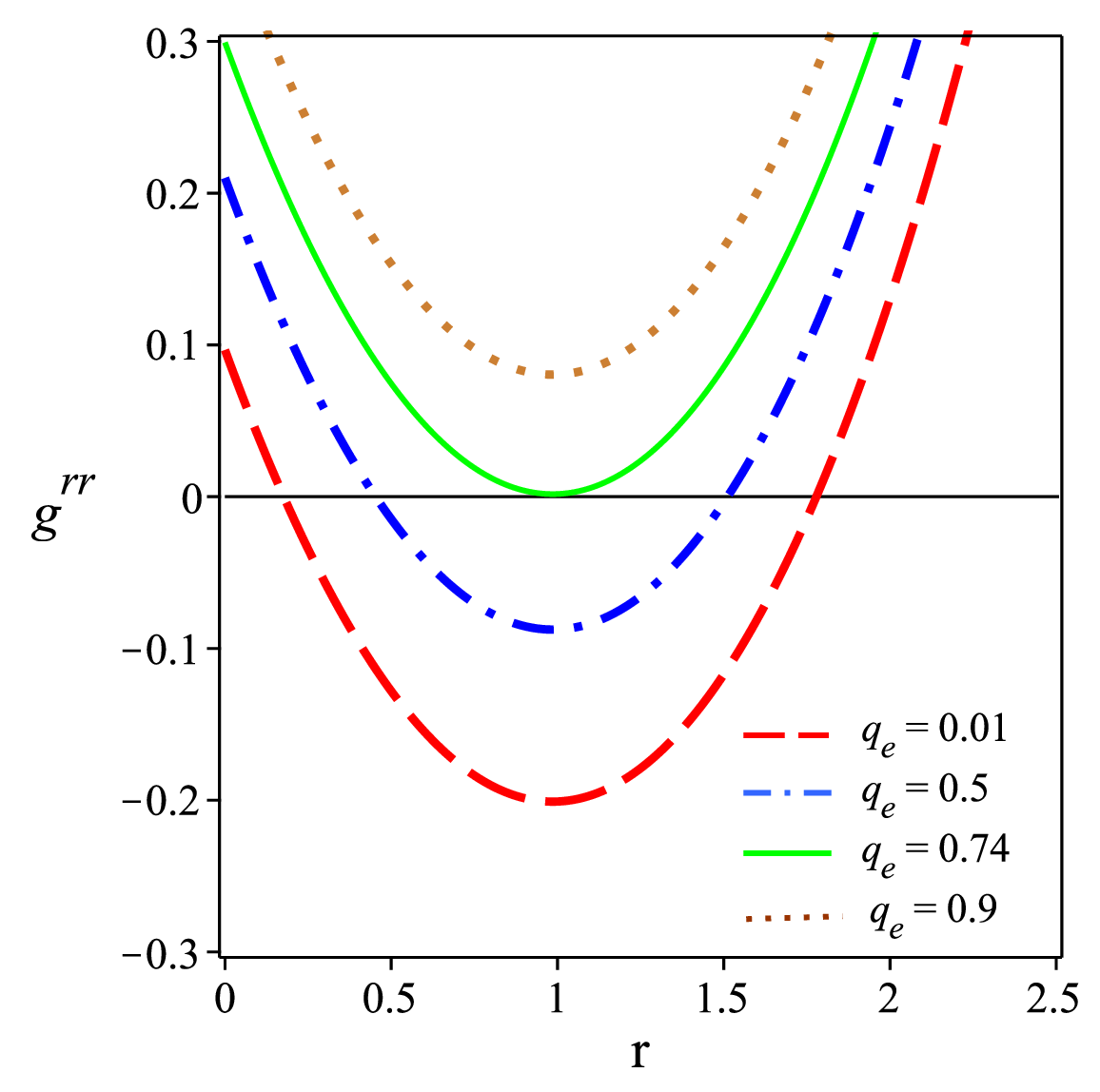}}\newline
\subfloat[$a=0.5$, $q_{e}=k_{e}=0.2 $ and $\Lambda=-0.02 $]{
        \includegraphics[width=0.32\textwidth]{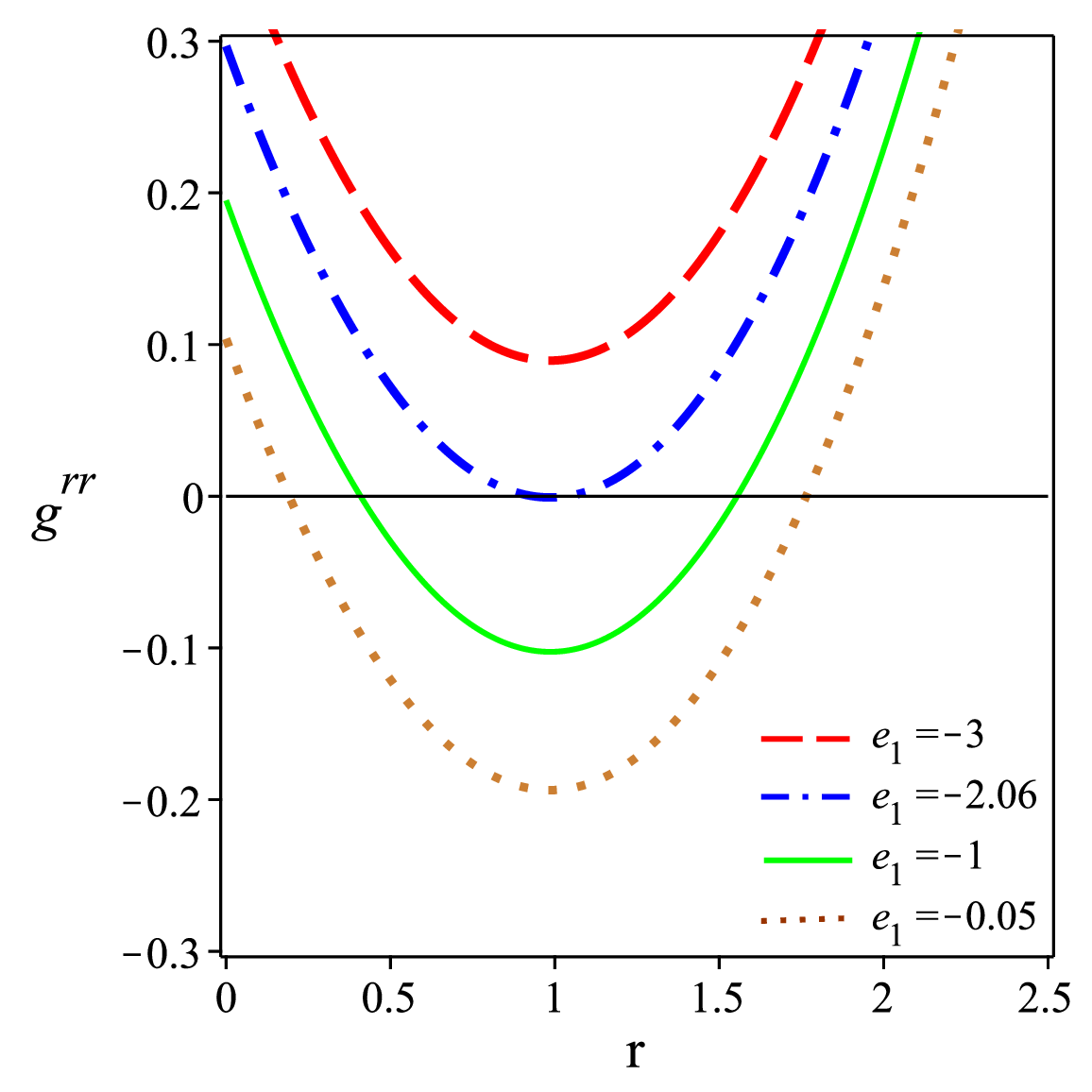}}
\subfloat[$a=0.5$, $e_{1}=-0.5$, $q_{e}=0.2 $ and $\Lambda=-0.02 $]{
        \includegraphics[width=0.31\textwidth]{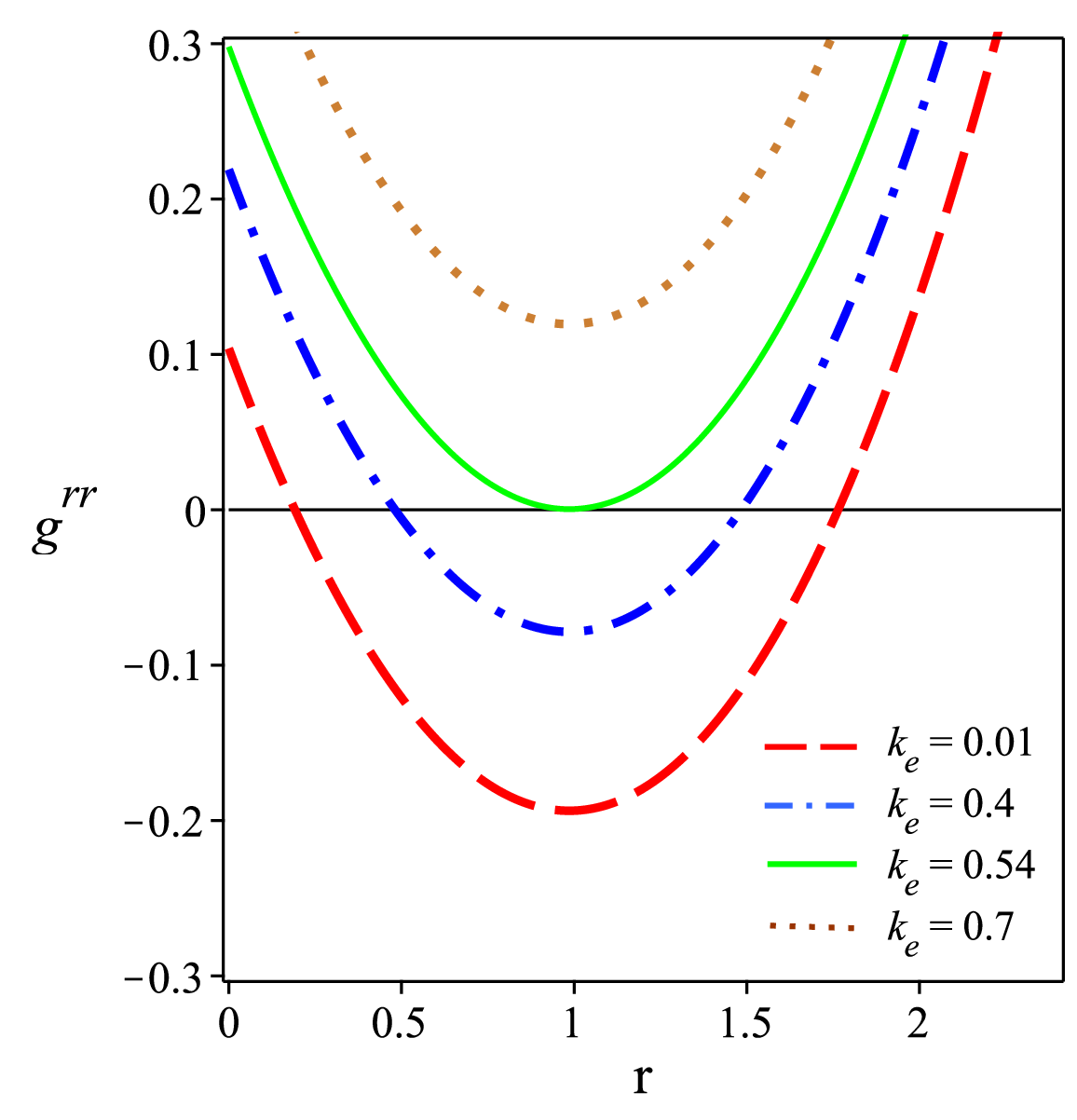}}\newline
\caption{The function $ g^{rr} $ versus $ r $ for $ k_{s}=10^{-6} $, $ q_{m}=k_{m}=0.2 $, $ d_{1}=-0.5 $ and $M=1$.}
\label{Fig1}
\end{figure}

It is worthwhile to investigate the geometrical structure and
physical properties of this solution. A singular solution
could be interpreted as a black hole solution if two conditions
are satisfied: I) The existence of a non-removable singularity.
II) Presence of at least one event horizon covering this
singularity. It should be noted that for regular black
holes, one only has to investigate the existence of the event
horizon. To address the first condition, we investigate the
Kretschmann scalar. Considering the metric (\ref{metric}), the
Kretschmann scalar is obtained as
\begin{eqnarray}
K=R_{\alpha\beta\gamma\delta}R^{\alpha\beta\gamma\delta}=\frac{H(a,\theta , q_{e}, q_{m}, k_{s}, k_{e}, k_{m}, \Lambda)}{(r^{2}+a^{2}\cos^{2}\theta)^{6}}
\end{eqnarray}
where $H$ is a polynomial function. We find that the Kretschmann
scalar diverges at $ r=0 $ and $ \theta =\pi/2 $ when $
r^{2}+a^{2}\cos^{2}\theta =0 $ and goes to $
\frac{8}{3}\Lambda^{2} $ as $ r\rightarrow \infty $. So  there is
a singularity at $ r=0 $. Now, we investigate the presence of
event horizon for the solution. Here we study two separate cases, the case of positive $d_{1}$ and $e_{1}$  (Case I), and the case of negative $d_{1}$ and $e_{1}$ (Case II).  As we know, the horizons are given
by the zeros of $ g^{-1}_{rr} $. To investigate the effects of
different parameters on number of roots, we have plotted different
diagrams in Figs. \ref{Fig1a} (case I) and \ref{Fig1} (case II). Fig. \ref{Fig1a} displays that the solution could admit
up to two roots if suitable values are considered for different
parameters. Since the event horizon is defined as outer root of $
g^{-1}_{rr} $ when its slope is positive, from Fig. \ref{Fig1a}(a),
one can find that the corresponding solution is covered by an
event horizon for $ \Lambda <0 $. Therefore, from now on, we will concentrate on anti-de-Sitter (AdS) black holes.  Figs. \ref{Fig1a}(b) and
\ref{Fig1a}(c) display that  the existence of root, hence black
hole solution is bounded by an upper limit of 
rotation parameter and electric charge. This
reveals the fact that for highly electrically charged solutions
and or fast-rotating ones, the roots of $ g^{rr} $ will disappear.
This results in solutions being a naked singularity. Therefore,
there is a certain value of electric charge and rotation parameter
that must not be crossed to remain the object as a black hole.
According to Figs. \ref{Fig1a}(d) and \ref{Fig1a}(e), it is clear that for all values of the parameter $ e_{1} $ and
electric dilation charge $k_{e}$, there is at least a root for the
function $ g^{rr}$. For very small values of these two parameters, two roots can be
observed, whereas, for intermediate or large values of this
parameter, there is only one root. From Figs. \ref{Fig1}(d) and \ref{Fig1}(e) can be seen that in contrast to case I, in case II,  the existence of the root is bounded by a lower (an upper) limit of the parameter $ e_{1} $ (electric dilation charge $k_{e}$) such that for $ e_{1}<-2.06 $ and $k_{e}>0.54$ no root is observed. Comparing Fig. \ref{Fig1a}(b) to Fig. \ref{Fig1}(b), one can find that more slowly rotating BH solutions in case I have only one root, whereas such BH solutions have two roots in case II (compare the red dashed curves of these two figures with each other). 

It is worth mentioning that according to our analysis, there is a difference between charged rotating BHs in Weyl-Cartan theory and the standard Kerr-Newman (KN) BH with electromagnetic fields of GR. For the standard KN case, there are three different cases for the metric function:  i) two real roots which are Cauchy and event horizons. i) One real root (extreme horizon). iii) Absence of real root (naked singularity). While for such BHs in Weyl-Cartan theory, four different cases can be observed for the metric function. i) Absence of an inner Cauchy horizon and the existence of a unique event horizon). ii) Cauchy and event horizons. iii) extreme horizon. iv) naked singularity. See Fig. \ref{Fig1a}(b) for better understanding. For more clarification, we explore the roots of a special case ($\Lambda=0$). In this case, the roots are obtained as
\begin{equation}\label{root}
r_{\pm}=M\pm\sqrt{M^{2}-A}
\end{equation}
where $ A= a^{2}+q_{e}^{2}+q_{m}^{2}+d_{1}k^{2}_{s}-4e_{1}(k^{2}_{e}+k^{2}_{m})$. For $ 0<A<M^{2} $, there are two real roots (Cauchy and event horizon). For the extreme case ($ A=M^{2} $), both horizons coincide. 
 For $ A>M^{2} $,the solution
does not contain any horizon (naked singularity). A special situation occurs for $A<0$ i.e. ($ 4e_{1}(k^{2}_{e}+k^{2}_{m})> a^{2}+q_{e}^{2}+q_{m}^{2}+d_{1}k^{2}_{s}$) such that the Cauchy horizon disappears and remains only an event horizon.

\begin{figure}[!htb]
\centering
\subfloat[ ]{
        \includegraphics[width=0.31\textwidth]{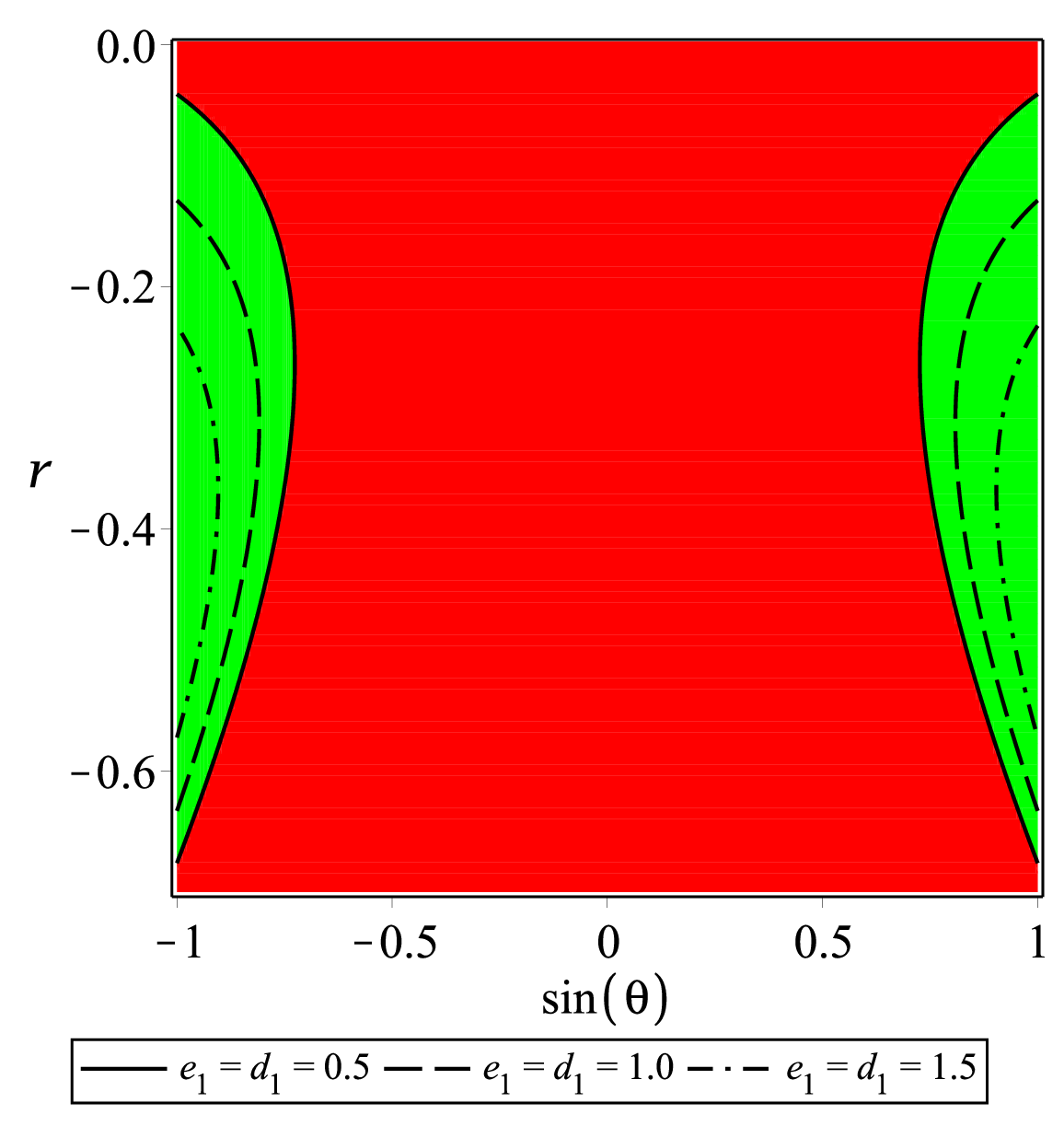}}
\subfloat[]{
        \includegraphics[width=0.31\textwidth]{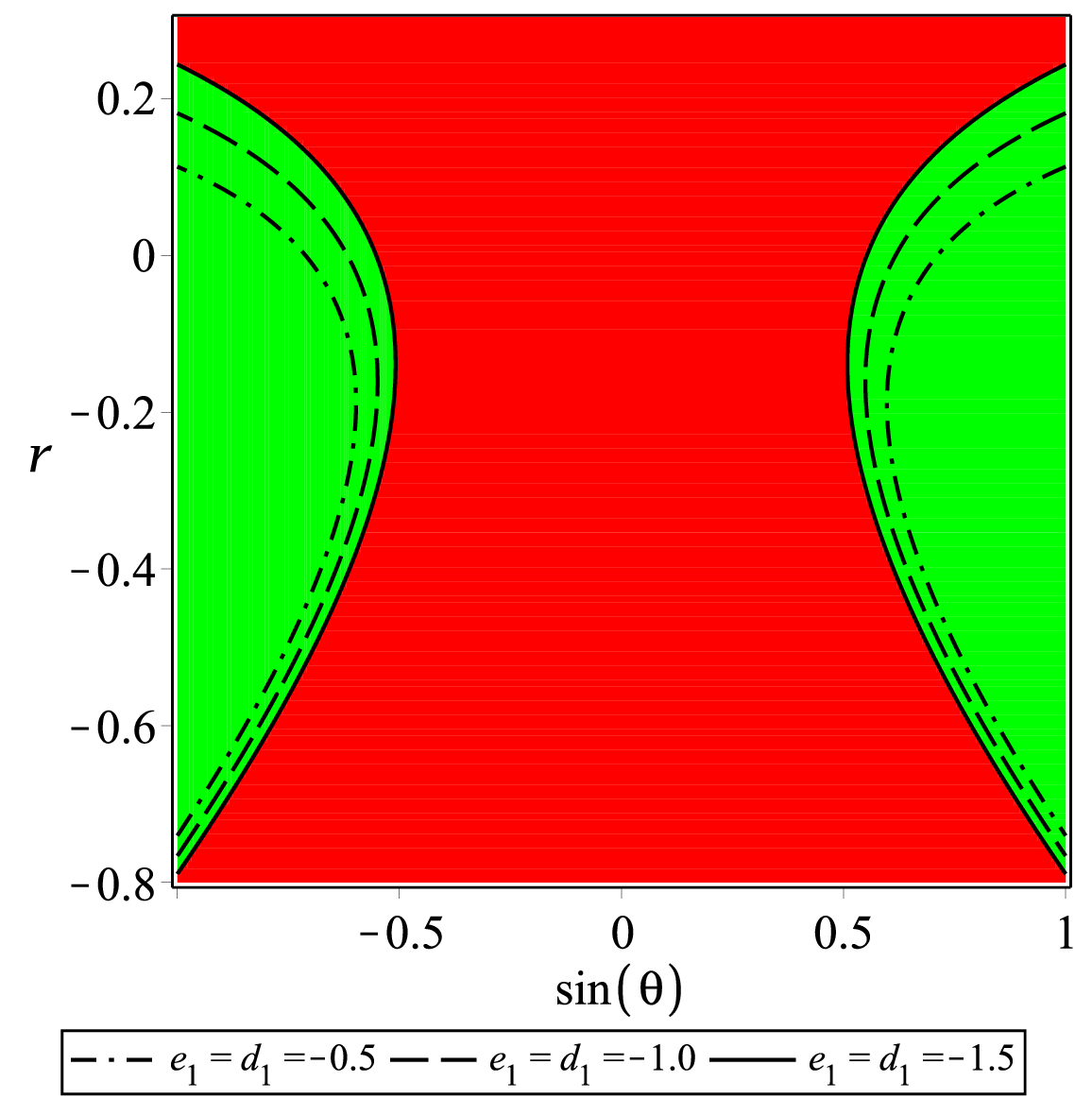}}
\subfloat[]{
        \includegraphics[width=0.325\textwidth]{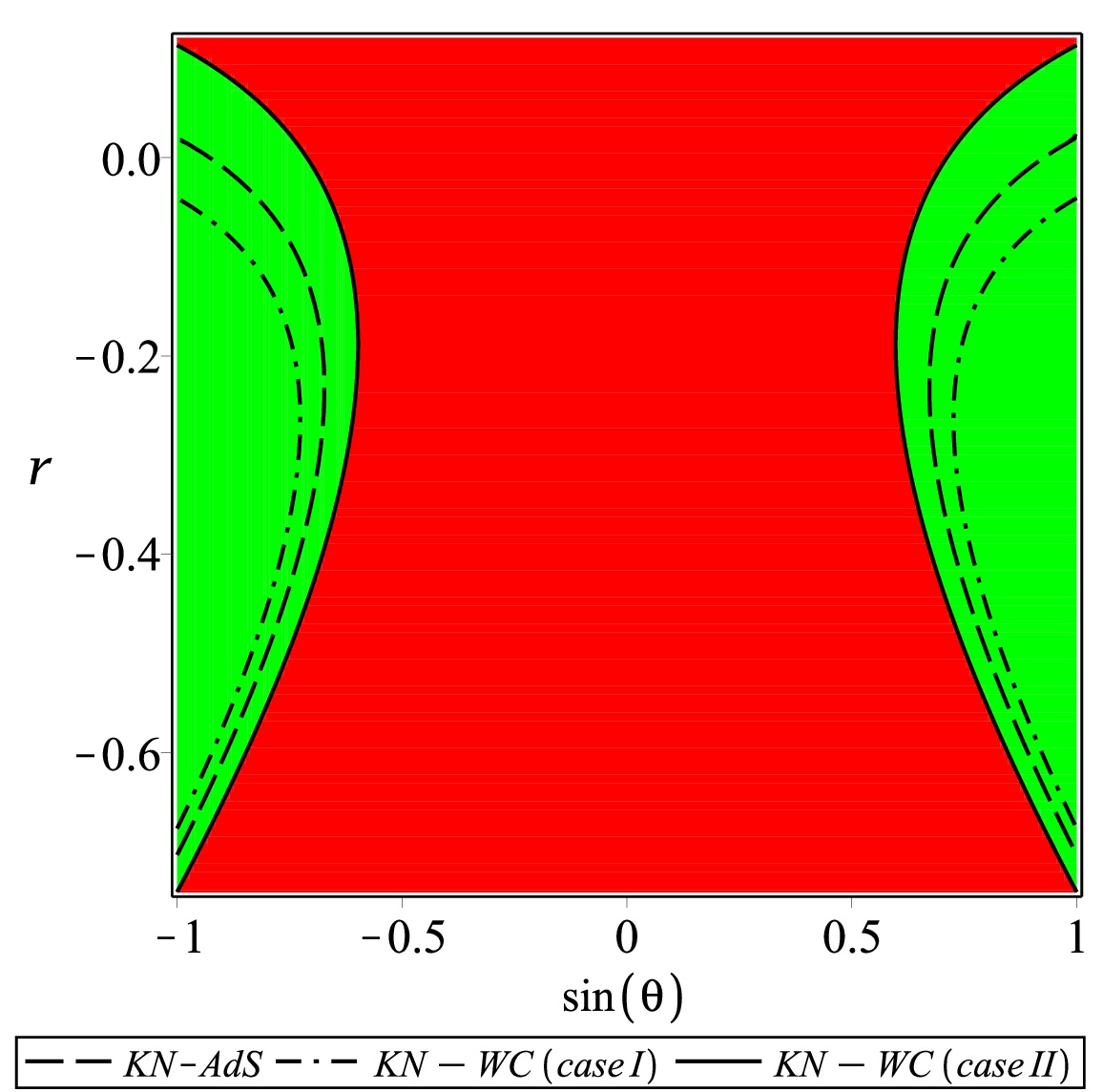}}\newline
\caption{The Causality violations/preservation regions of the black hole. Here we have fixed $a=0.5$, $ \Lambda=-0.2 $, $ k_{s}=10^{-6} $, $ q_{e}=q_{m}=k_{e}=k_{m}=0.2 $ and $M=1$.}
\label{ctc}
\end{figure}

Another physical property of the black hole is the Causality issue. The formulation of the fundamental laws of physics is based on the preservation of causality. A Closed Timelike Curve (CTC) is a closed curve whose tangent is everywhere timelike. Therefore, the existence of CTCs implies the violation of causality relations. In fact, CTC allows time travel, in the sense that an observer who travels on a trajectory in spacetime along this curve may return to an event before his departure. To counter the pathology, Hawking proposed the Chronology Protection Conjecture, which states: The laws of physics do not allow the appearance of closed timelike curves. It has been proved that causality violations occur in Kerr and Kerr-Newman black holes \cite{Gott:1126}. Causality violations and CTCs are possible if $ g_{\varphi \varphi}>0 $ \cite{Azreg:640,Contreras:802}. To more precise study of this issue, we have plotted Fig. (\ref{ctc}) which is a plot of $r$ versus $sin (\theta)$. According to Eq. (\ref{metric}), the sign of $ g_{\varphi \varphi} $ depends on the sign of parameters $ e_{1} $ and $ d_{1} $. Fig. \ref{ctc}(a) shows Causality violations/preservation regions for case I. The green region is where $ g_{\varphi \varphi}>0 $, representing that Causality violations and CTC occur in this region. While red region is where Causality relation preserves. As we see, CTCs exist only for $ r<0 $ which is similar to the behavior of the Kerr black hole (see Ref. \cite{Azreg:640} for more details). Regarding the case II, it can be seen from Fig. \ref{ctc}(b) that CTCs
are possible for $r>0$ as well. Such behavior is similar to that of Kerr-Newman black hole (see Ref. \cite{Azreg:640}). A comparison between the KN-AdS black hole and KN-AdS black holes in WC theory is depicted in Fig. \ref{ctc}(c). From this figure is clear that the smallest (biggest) region of CTC is related to case I (case II).

\section{Optical Features of the Black Hole Spacetime}
In this section, we present an in-depth study of the optical
features of the black hole in WCT, given by the
solution (\ref{metric2}), such as the shadow geometrical shape and
the energy emission rate. Taking into account these optical
quantities, we inspect the effects on the black hole solution by
varying the parameters of the theory.

\subsection{Particle motion and BH shadow}
\label{SubsecA}
In WCT, in general, the particle motion is different than the geodesic one. In the theory that we are interested, it was verified in \cite{Bahamonde:2021qjk} that for bosonic particles (such as light), the non-geodesic motion of the particles coincides with the geodesic equation. This means that for particles with spin, the geodesic would be modified. For our study, we will just concentrate on particle motion related to light, so that, the geodesic equation holds for our theory. 

When a black hole is situated between a bright object and an
observer it will cast a shadow. The apparent shape of a black hole
is defined by the boundary of the shadow. To study the black hole
shadow, we need to investigate the motion of a particle in the BH
background. To do so, we take into account the Hamilton-Jacobi
method in the black hole  spacetime as
\cite{Carter:1968,Decanini:2011}
\begin{equation}
\frac{\partial S}{\partial \sigma}=- \frac{1}{2}g^{\mu\nu}\frac{\partial S}{\partial x^{\mu}}\frac{\partial S}{%
\partial x^{\nu}},
\label{HJ1}
\end{equation}
where $\sigma $ is an affine parameter, $ g^{\mu \nu} $ is the metric tensor and $S $ is Jacobian action which takes the form
\begin{equation}
S=-\frac{1}{2}m^2\sigma+Et-L\varphi+S_r(r)+S_{\theta}(\theta),
\label{HJ2}
\end{equation}
in which $ m $ is the mass of the test particle and $S_r(r)$ and
$S_\theta(\theta)$ are the function of $r$ and $\theta$. Here, $
E=p_{t} $ and $ L= p_{\varphi}$ are the conserved energy and angular
momentum, respectively. Using the variable separable method and
inserting Eq. (\ref{HJ2}) into Eq. (\ref{HJ1}), we obtain the
equations of motion for photon ($m=0$) as
\begin{align}
\label{HJ3}
&\Sigma\frac{dt}{d\sigma}= \frac{r^2+a^2}{\Delta (r)}[E(r^2+a^2)-aL]-\frac{a}{\Delta (\theta)}\left[aE\sin^2\theta-L \right] ,\\
&\Sigma\frac{dr}{d\sigma}=\sqrt{\mathcal{R}(r)},\\
&\Sigma\frac{d\theta}{d\sigma}=\sqrt{\Theta(\theta)},\\
\label{HJ4}
&\Sigma\frac{d\varphi}{d\sigma}=\frac{a}{\Delta (r)}[E(r^2+a^2)-aL]-\frac{1}{\Delta (\theta)}\left(aE-\frac{L}{\sin^2\theta}\right),
\end{align}
where $ \Delta (r) $ and $ \Delta (\theta) $ are defined as
\begin{align}
\label{HJ6}
&\Delta (r)=(1-\frac{1}{3}\Lambda r^{2})(r^2+a^2)-2Mr+q_{e}^{2}+q_{m}^{2}+d_{1}k_{s}^{2}-4e_{1}(k_{e}^{2}+k_{m}^{2}) ,\\
&\Delta (\theta)=1+\frac{1}{3}\Lambda a^{2}\cos^2\theta,
\end{align}
also the expression $\mathcal{R}(r)$ and $\Theta(\theta)$ are given by
\begin{align}
\label{HJ5}
&\mathcal{R}(r)=[E(r^2+a^2)-aL]^2-\Delta (r)\left[ \frac{(aE-L)^2}{\Delta (\theta)}+\mathcal{K}\right] ,\\
&\Theta(\theta)=\mathcal{K}\Delta (\theta)-\left(  \dfrac{L^2}{\sin^2\theta}-a^2E^2  \right) \cos^2\theta,
\end{align}
in which $ \mathcal{K} $ is a constant of separation called Carter
constant. These equations define the propagation of photon around
the spacetime of the corresponding rotating black hole. For
$\mathcal{K}=0$, $\theta$-motion is suppressed, and all photon
orbits are restricted only to a plane ($ \theta=\pi/2 $), yielding
unstable circular orbits at the equatorial plane. Obtaining the
boundary of the black hole shadow demands the study of the radial
equation. We can rewrite the radial geodesic equation in terms of
the effective potential $V_{\text{eff}}$ as
\begin{equation*}
\Sigma^2\left(\frac{dr}{d\sigma}\right)^2+V_{\text{eff}}=0.
\end{equation*}

Introducing two impact parameters $\xi$ and $\eta$ \cite{Chandrasekhar1g} as
\begin{equation}
\xi=L/E,  \;\; \;\; \;\;   \eta=\mathcal{K}/E^2,
\end{equation}
the effective potential $ V_{\text{eff}} $ takes the following form
\begin{equation}\label{veff}
V_{\text{eff}}=\Delta (r)\left( \eta +\frac{(a-\xi)^2}{\Delta (\theta)}\right) -(r^2+a^2-a\;\xi)^2,
\end{equation}
where we have replaced $V_{\text{eff}}/E^2$ by $V_{\text{eff}}$.
Due to the constraint $ \frac{dr}{d\sigma}\geq 0 $, we expect that
the effective potential satisfies $ V_{\text{eff}}\leq 0 $. When
incoming photons, which are coming towards the black hole from a
light source, get near the black hole, they follow the three
possible trajectories. Either they have negative effective
potential and fall into the BH inevitably, or their effective
potential is positive and they are scattered away from the black
hole, or their effective potential is zero and they make a
circular orbit near the black hole. The boundary of the shadow is
mainly determined by the circular photon orbit, which satisfies
the following conditions
\begin{equation}\label{cond}
V_{\text{eff}}(r_{ph})=0,\quad~~~\frac{dV_{\text{eff}}(r_{ph})}{dr}=0,
\end{equation}
whereas instability of orbits obeys condition
\begin{equation}
\frac{d^{2}V_{\text{eff}}(r_{ph})}{dr^{2}}<0,
\end{equation}
\begin{table*}[htb!]
\caption{The event horizon ($r_{e}$), photon sphere radius ($r_{ph}$) and
shadow radius ($r_{sh}$) for the variation of $a$, $q_{e}$, $k_{e} $, $e_{1}$ and $\Lambda$ for $ q_{m}=k_{m}=0.2 $, $d_{1}=0.5 $, $ k_{s}=10^{-6} $, and  $M =1$.}
\label{table1a}\centering
\begin{tabular}{||c|c|c|c|c||}
\hline
{\footnotesize $a$ \hspace{0.3cm}} & \hspace{0.3cm}$0.1$ \hspace{0.3cm}
& \hspace{0.3cm} $0.4$\hspace{0.3cm} & \hspace{0.3cm} $0.7$\hspace{0.3cm} &
\hspace{0.3cm}$1.04$\hspace{0.3cm} \\ \hline
$r_{e}$ ($q_{e}=k_{e}=0.2$, $e_{1}=0.5$, $\Lambda =-0.02 $) & $1.9831$
& $1.9096$ & $1.7222$ & $0.98+0.12I$ \\ \hline
$r_{ph}$ ($q_{e}=k_{e}=0.2$, $e_{1}=0.5$, $\Lambda =-0.02 $) & $3.0483 $
& $2.9852$ & $2.8321$ & $2.4708$ \\ \hline
$r_{sh}$ ($q_{e}=k_{e}=0.2$, $e_{1}=0.5$, $\Lambda =-0.02 $) & $4.8347 $
& $4.8078$ & $4.7414$ & $4.5678$ \\ \hline
$r_{ph}>r_{e}$ & \checkmark & \checkmark & \checkmark & $\times$ \\ \hline
$r_{sh}>r_{ph}$ & \checkmark & \checkmark & \checkmark & \checkmark \\
\hline\hline
&  &  &  &  \\
{\footnotesize $q_{e}$ \hspace{0.3cm}} & \hspace{0.3cm} $0.1$ \hspace{0.3cm} &
\hspace{0.3cm} $0.4$\hspace{0.3cm} & \hspace{0.3cm} $0.7$\hspace{0.3cm} &
\hspace{0.3cm} $0.93$\hspace{0.3cm} \\ \hline
$r_{e}$ ($a=e_{1}=0.5$, $k_{e}=0.2$, $\Lambda =-0.02 $) & $1.8781$ & $1.7969$ & $1.5782$ & $0.98+0.05I$ \\ \hline
$r_{ph}$ ($a=e_{1}=0.5$, $k_{e}=0.2$, $\Lambda =-0.02 $) & $2.9665 $ & $2.8588$ & $2.5838$ & $2.1332$ \\ \hline
$r_{sh}$ ($a=e_{1}=0.5$, $k_{e}=0.2$, $\Lambda =-0.02 $) & $4.8113 $ & $4.7053$ & $4.4368$ & $4.0185$ \\ \hline
$r_{ph}>r_{e}$ & \checkmark & \checkmark & \checkmark & $\times$ \\ \hline
$r_{sh}>r_{ph}$ &  \checkmark & \checkmark & \checkmark & \checkmark \\
\hline\hline
&  &  &  &  \\
{\footnotesize $k_{e}$ \hspace{0.3cm}} & \hspace{0.3cm}$0.2$ \hspace{0.3cm} &
\hspace{0.3cm} $0.8$\hspace{0.3cm} & \hspace{0.3cm} $1.4$\hspace{0.3cm} &
\hspace{0.3cm}$2$\hspace{0.3cm} \\ \hline
$r_{e}$ ($a=e_{1}=0.5$, $q_{e}=0.2$, $\Lambda =-0.02 $) & $1.8625 $ & $2.3438$ & $3.0241$ & $3.7275$ \\ \hline
$r_{ph}$ ($a=e_{1}=0.5$, $q_{e}=0.2$, $\Lambda =-0.02 $) & $2.9456 $ & $3.6244$ & $4.6340$ & $5.7440$ \\ \hline
$r_{sh}$ ($a=e_{1}=0.5$, $q_{e}=0.2$, $\Lambda =-0.02 $) & $4.7907 $ & $5.4513$ & $6.3657$ & $7.2426$ \\ \hline
$r_{ph}>r_{e}$ & \checkmark & \checkmark & \checkmark & \checkmark \\ \hline
$r_{sh}>r_{ph}$ & \checkmark & \checkmark & \checkmark & \checkmark \\
\hline\hline
&  &  &  &  \\
{\footnotesize $e_{1}$ \hspace{0.3cm}} & \hspace{0.3cm}$0.01$ \hspace{0.3cm} &
\hspace{0.3cm} $0.5$\hspace{0.3cm} & \hspace{0.3cm} $1$\hspace{0.3cm} &
\hspace{0.3cm}$1.5$\hspace{0.3cm} \\ \hline
$r_{e}$ ($a=0.5$, $q_{e}=k_{e}=0.2$, $\Lambda =-0.02 $) & $1.7756 $ & $1.8625$ & $1.9427$ & $2.0163$ \\ \hline
$r_{ph}$ ($a=0.5$, $q_{e}=k_{e}=0.2$, $\Lambda =-0.02 $) & $2.8311 $ & $2.9456$ & $3.0537$ & $3.1546$ \\ \hline
$r_{sh}$ ($a=0.5$, $q_{e}=k_{e}=0.2$, $\Lambda =-0.02 $) & $4.6742 $ & $4.7868$ & $4.8930$ & $4.9920$ \\ \hline
$r_{ph}>r_{e}$ & \checkmark & \checkmark & \checkmark & \checkmark \\ \hline
$r_{sh}>r_{ph}$ & \checkmark & \checkmark & \checkmark & \checkmark \\
\hline\hline
&  &  & &  \\
{\footnotesize $\Lambda$ \hspace{0.3cm}} & \hspace{0.3cm}$0.01$ \hspace{%
0.3cm} & \hspace{0.3cm} $-0.05$\hspace{0.3cm} & \hspace{0.3cm} $-0.1$\hspace{%
0.3cm} & \hspace{0.3cm}$-0.22$\hspace{0.3cm} \\ \hline
$r_{e}$ ($ a=e_{1}=0.5 $, $q_{e}=k_{e}=0.2$) & $16.2154 $ & $1.8007$ & $1.7177$ & $1.5666$ \\ \hline
$r_{ph}$ ($ a=e_{1}=0.5 $, $q_{e}=k_{e}=0.2$) & $2.9378 $ & $2.9534$ & $2.9666$ & $3.0013$ \\ \hline
$r_{sh}$ ($ a=e_{1}=0.5 $, $q_{e}=k_{e}=0.2$) & $5.4347 $ & $4.3317$ & $3.7930$ & $2.9920$ \\ \hline
$r_{ph}>r_{e}$ & $\times$ & \checkmark & \checkmark & \checkmark \\ \hline
$r_{sh}>r_{ph}$ & \checkmark & \checkmark & \checkmark & $\times$ \\
\hline\hline
\end{tabular}%
\end{table*}
\begin{table*}[htb!]
\caption{The event horizon ($r_{e}$), photon sphere radius ($r_{ph}$) and
shadow radius ($r_{sh}$) for the variation of $a$, $q_{e}$, $k_{e} $, $e_{1}$ and $\Lambda$ for $ q_{m}=k_{m}=0.2 $, $d_{1}=-0.5 $, $ k_{s}=10^{-6} $, and  $M =1$.}
\label{table1}\centering
\begin{tabular}{||c|c|c|c|c||}
\hline
{\footnotesize $a$ \hspace{0.3cm}} & \hspace{0.3cm}$0.1$ \hspace{0.3cm}
& \hspace{0.3cm} $0.3$\hspace{0.3cm} & \hspace{0.3cm} $0.5$\hspace{0.3cm} &
\hspace{0.3cm}$0.87$\hspace{0.3cm} \\ \hline
$r_{e}$ ($q_{e}=k_{e}=0.2$, $e_{1}=-0.5$, $\Lambda =-0.02 $) & $1.8223$
& $1.7757$ & $1.6731$ & $0.98+0.08I$ \\ \hline
$r_{ph}$ ($q_{e}=k_{e}=0.2$, $e_{1}=-0.5$, $\Lambda =-0.02 $) & $2.8255 $
& $2.7854$ & $2.6999$ & $2.3501$ \\ \hline
$r_{sh}$ ($q_{e}=k_{e}=0.2$, $e_{1}=-0.5$, $\Lambda =-0.02 $) & $4.6088 $
& $4.5877$ & $4.5459$ & $4.3740$ \\ \hline
$r_{ph}>r_{e}$ & \checkmark & \checkmark & \checkmark & $\times$ \\ \hline
$r_{sh}>r_{ph}$ & \checkmark & \checkmark & \checkmark & \checkmark \\
\hline\hline
&  &  &  &  \\
{\footnotesize $q_{e}$ \hspace{0.3cm}} & \hspace{0.3cm} $0.1$ \hspace{0.3cm} &
\hspace{0.3cm} $0.3$\hspace{0.3cm} & \hspace{0.3cm} $0.5$\hspace{0.3cm} &
\hspace{0.3cm} $0.74$\hspace{0.3cm} \\ \hline
~$r_{e}$ ($a=0.5$, $e_{1}=-0.5$, $k_{e}=0.2$, $\Lambda =-0.02 $)~ & $1.6931$ & $1.6383$ & $1.5112$ & $0.98+0.07I$ \\ \hline
~$r_{ph}$ ($a=0.5$, $e_{1}=-0.5$, $k_{e}=0.2$, $\Lambda =-0.02 $)~ & $2.7251 $ & $2.6567$ & $2.5055$ & $2.1287$ \\ \hline
~$r_{sh}$ ($a=0.5$, $e_{1}=-0.5$, $k_{e}=0.2$, $\Lambda =-0.02 $)~ & $4.5705 $ & $4.5038$ & $4.3580$ & $4.0147$ \\ \hline
$r_{ph}>r_{e}$ & \checkmark & \checkmark & \checkmark & $\times$ \\ \hline
$r_{sh}>r_{ph}$ &  \checkmark & \checkmark & \checkmark & \checkmark \\
\hline\hline
&  &  &  &  \\
{\footnotesize $k_{e}$ \hspace{0.3cm}} & \hspace{0.3cm}$0.01$ \hspace{0.3cm} &
\hspace{0.3cm} $0.2$\hspace{0.3cm} & \hspace{0.3cm} $0.4$\hspace{0.3cm} &
\hspace{0.3cm}$0.54$\hspace{0.3cm} \\ \hline
~$r_{e}$ ($a=0.5$, $e_{1}=-0.5$, $q_{e}=0.2$, $\Lambda =-0.02 $)~ & $1.7251 $ & $1.6731$ & $1.4837$ & $0.98+0.03I$ \\ \hline
~$r_{ph}$ ($a=0.5$, $e_{1}=-0.5$, $q_{e}=0.2$, $\Lambda =-0.02 $)~ & $2.7658 $ & $2.6999$ & $2.4744$ & $2.1360$ \\ \hline
~$r_{sh}$ ($a=0.5$, $e_{1}=-0.5$, $q_{e}=0.2$, $\Lambda =-0.02 $)~ & $4.6103 $ & $4.5459$ & $4.3284$ & $4.0209$ \\ \hline
$r_{ph}>r_{e}$ & \checkmark & \checkmark & \checkmark & $\times$ \\ \hline
$r_{sh}>r_{ph}$ & \checkmark & \checkmark & \checkmark & \checkmark \\
\hline\hline
&  &  &  &  \\
{\footnotesize $e_{1}$ \hspace{0.3cm}} & \hspace{0.3cm}$-2.07$ \hspace{0.3cm} &
\hspace{0.3cm} $-1$\hspace{0.3cm} & \hspace{0.3cm} $-0.5$\hspace{0.3cm} &
\hspace{0.3cm}$-0.01$\hspace{0.3cm} \\ \hline
$r_{e}$ ($a=0.5$, $q_{e}=k_{e}=0.2$, $\Lambda =-0.02 $) & $0.98+0.02I $ & $1.5541$ & $1.6731$ & $1.7719$ \\ \hline
$r_{ph}$ ($a=0.5$, $q_{e}=k_{e}=0.2$, $\Lambda =-0.02 $) & $2.1373 $ & $2.5552$ & $2.6999$ & $2.8262$ \\ \hline
$r_{sh}$ ($a=0.5$, $q_{e}=k_{e}=0.2$, $\Lambda =-0.02 $) & $4.0220 $ & $4.4056$ & $4.5459$ & $4.6694$ \\ \hline
$r_{ph}>r_{e}$ & $\times$ & \checkmark & \checkmark & \checkmark \\ \hline
$r_{sh}>r_{ph}$ & \checkmark & \checkmark & \checkmark & \checkmark \\
\hline\hline
&  &  & &  \\
{\footnotesize $\Lambda$ \hspace{0.3cm}} & \hspace{0.3cm}$-0.05$ \hspace{%
0.3cm} & \hspace{0.3cm} $-0.1$\hspace{0.3cm} & \hspace{0.3cm} $-0.15$\hspace{%
0.3cm} & \hspace{0.3cm}$-0.27$\hspace{0.3cm} \\ \hline
$r_{e}$ ($ a=0.5 $, $ e_{1}=-0.5 $, $q_{e}=k_{e}=0.2$) & $1.61994 $ & $1.5469$ & $1.4874$ & $1.3779$ \\ \hline
$r_{ph}$ ($ a=0.5 $, $ e_{1}=-0.5 $, $q_{e}=k_{e}=0.2$) & $2.7080 $ & $2.7214$ & $2.7349$ & $2.7679$ \\ \hline
$r_{sh}$ ($ a=0.5 $, $ e_{1}=-0.5 $, $q_{e}=k_{e}=0.2$) & $4.1443 $ & $3.6582$ & $3.3083$ & $2.7533$ \\ \hline
$r_{ph}>r_{e}$ & \checkmark & \checkmark & \checkmark & \checkmark \\ \hline
$r_{sh}>r_{ph}$ & \checkmark & \checkmark & \checkmark & $\times$ \\
\hline\hline
\end{tabular}%
\end{table*}


Solving  Eq. (\ref{cond}), the critical impact parameters which
determine the contour of the shadow for the photon orbits around
the black hole are obtained as
\begin{eqnarray}
\xi_{c}&=& \frac{(r_{ph}^{2}+a^{2})\Delta'(r_{ph})-4r_{ph}\Delta (r_{ph})}{a\Delta'(r_{ph})},\\
\eta_{c}&=& \frac{r_{ph}^{2}\left[ 16\Delta (r_{ph})(a^{2}\Delta (\theta)-\Delta (r_{ph}))+8r_{ph}\Delta (r_{ph})\Delta'(r_{ph})-r_{ph}^{2}\Delta'(r_{ph})^{2}\right] }{a^{2}\Delta'(r_{ph})^{2}\Delta (\theta)}.
\end{eqnarray}

To obtain the apparent shape of the BH shadow, we introduce the
celestial coordinates $  x$ and $ y $ as follows \cite{Vazquez1g}
\begin{eqnarray}
\label{celestial:1a}
x&=&\lim_{r_0\rightarrow\infty}\left(-r_0^2\sin\theta_0\frac{d\varphi}{dr}\Big|_{(r_0,\theta_0)}\right),\\ \nonumber
y&=&\lim_{r_0\rightarrow\infty}\left(r_0^2\frac{d\theta}{dr}\Big|_{(r_0,\theta_0)}\right),
\end{eqnarray}
in which $ \theta_{0} $ is the inclination angle. We take the
limit $ r\rightarrow \infty $, because the distance between the
observer and the black hole is very large. For the corresponding
solution, celestial coordinates take the form
\begin{equation}
x=\frac{\Delta (\theta)^{-1}(a\sin\theta_0-\xi\csc\theta_0)}
{\sqrt{1-\frac{\Lambda}{3}\left[\eta +\frac{(a-\xi)^{2}}{\Delta (\theta)} \right] }}\,,\qquad
y=\pm \sqrt{\frac{\eta \Delta (\theta)+a^2\cos^2\theta_0-\xi^2\cot^2\theta_0}{1-\frac{\Lambda}{3}\left[\eta +\frac{(a-\xi)^{2}}{\Delta (\theta)} \right] }}\,.\label{xy}
\end{equation}

Here, we employ the definition adopted in Ref. \cite{Feng1g}. From
Eq. (\ref{xy}), it is clear that the shape of the shadow is
dependent on the inclination angle $\theta_0$. For north pole
$\theta_{0} =0$ (or equivalent south pole $\theta_{0}=\pi$), the
boundary of the shadow is a perfect circle. The photons that form
the shadow boundary satisfy
\begin{equation}
\xi(r^0_{\rm ph})=0\,,
\end{equation}
and the shadow radius is
\begin{equation}
r_{\rm sh}=\sqrt{\frac{a^2 + \Delta_{0}\eta(r^0_{\rm ph})}{1-\frac{\Lambda}{3}\left[\eta +\frac{a^{2}}{\Delta_{0}} \right] }}
,\label{Rsh1}
\end{equation}
where $ \Delta_{0}=1+\frac{1}{3}\Lambda a^{2} $. For
$\theta_{0}\ne 0$ or $\pi$, the shape of shadow is no longer
round, but distorted. The maximum distortion occurs for
$\theta_0=\pi/2$ in the equatorial plane. In such a situation, the
horizontal x-direction is squeezed, while the vertical y-direction
is elongated, but the shadow remains convex. For $y=0$, there are
two real solutions
\begin{equation}
\eta(r_{\rm ph}^\pm)=0\,,\qquad \hbox{with}\qquad r_{\rm ph}^+\ge  r_{\rm ph}^-
\end{equation}

The size of the shadow is obtained as
\begin{equation}
r_{\rm sh} = \frac{1}{2} \big(x(r_{\rm ph}^{+}) - x(r_{\rm ph}^{-})\big)\,.\label{Rsh2}
\end{equation}

For general inclination angle, $r_{\rm ph}^\pm$ can be determined
by requiring $ y(r, \theta_0)\Big|_{r=r_{\rm ph}^\pm}=0$, and the
shadow size is then again formally given by (\ref{Rsh2}). It is
important that the roots of the above equation $ r^{\pm}_{ph} $
must be chosen that both are outside of the horizon. When $
r^{-}_{ph}$ is smaller than the horizon, one should define $
r^{-}_{ph}=r_{e}$,  where $ r_{e} $ is radius related to the event
horizon. For $ r^{-}_{ph}=r^{+}_{ph}$, the shadow becomes round
sphere.

To observe an acceptable optical behavior, we need to investigate
the condition $r_{e} <r_{ph} < r_{sh}$. It helps us to find
admissible regions of parameters of the model for having an
acceptable physical result. Since it is not possible to solve the
equations analytically, we employ numerical methods to obtain the
photon sphere radius and shadow radii. Several values of the event
horizon, the photon sphere radius, and shadow radius are listed in
table \ref{table1a} (case I) and table \ref{table1} (case II). From table \ref{table1a}, one can find that the increase of the
rotation parameter and electric charge lead to an imaginary event
horizon, meaning that an acceptable optical result can be observed
only for limited regions of these two parameters. Regarding the
effect of these two parameters on the event horizon, photon sphere
radii and shadow size, we notice that both parameters have a
decreasing contribution to these quantities. A remarkable point
regarding the magnetic charge effect is that according to our
investigation, its effect on these three quantities is the same as
that of electric charge. The study of the electric dilation charge
($ k_{e} $) effect shows that the increase of this parameter leads
to increasing every three quantities and also an acceptable
optical behavior can be observed for all values of this parameter.
From the fourth row of table \ref{table1a}, it is evident that the
effect of parameter $e_{1}$ on these three quantities is just like
the $k_{e}$ effect. Investigating the influence of the
cosmological constant on the radius of the event horizon, photon
sphere radii, and shadow size, we find that increasing this
parameter makes decreasing (increasing) the event horizon and
shadow radii (photon sphere radius). According to table \ref{table1}, unlike case I, in case II, acceptable optical behavior cannot be observed for all values of $k_{e}$ and $e_{1}$. In other words, only for limited regions of these two parameters, an acceptable optical result can be found. Also, in case II, both parameters  $k_{e}$ and $ e_{1} $ have a decreasing contribution to the event
horizon, the photon sphere radius, and shadow radius which is opposite to the behavior observed in the case I.

\begin{figure}[!htb]
\centering
\subfloat[$ a=0.8$, $ q_{e}=k_{e}=0.2 $, $ e_{1}=-0.5$  and $ \Lambda=-0.02 $]{
        \includegraphics[width=0.31\textwidth]{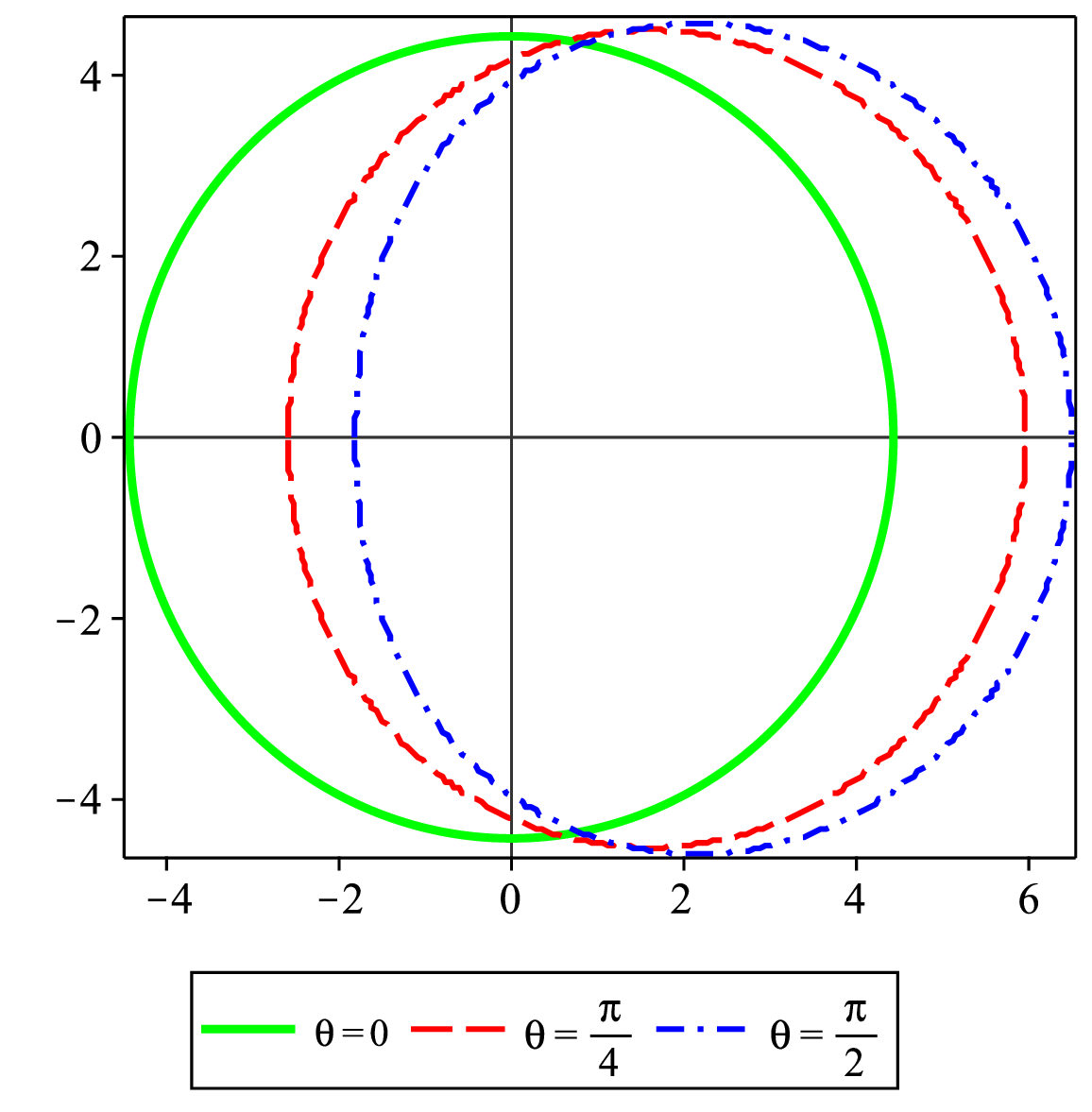}}
\subfloat[$\theta=\pi /2$, $q_{e}=k_{e}=0.2 $, $ e_{1}=-0.5$ and $\Lambda=-0.02 $]{
     \includegraphics[width=0.33\textwidth]{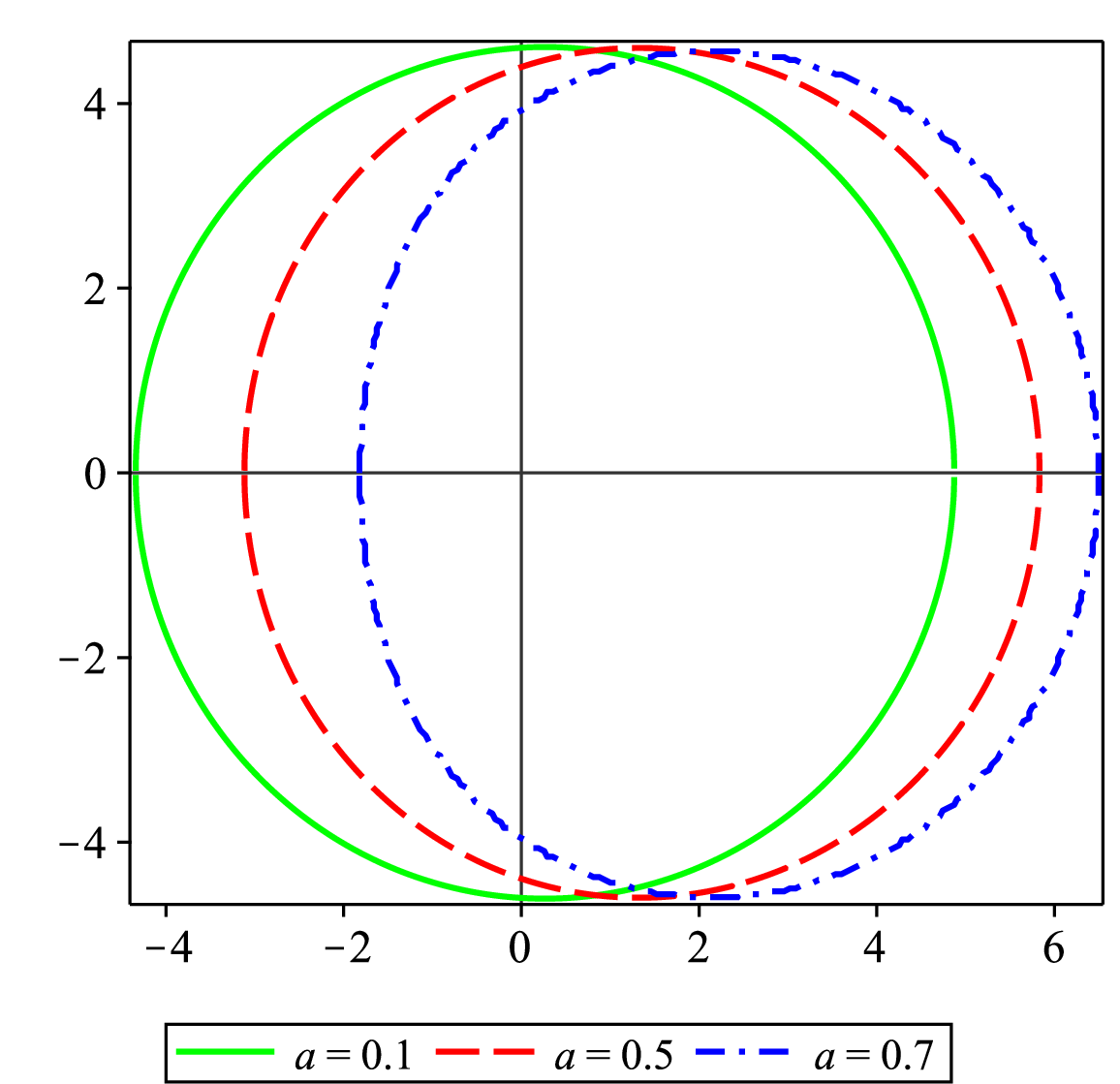}}
\subfloat[$\theta=\pi /2$, $ a=0.8$, $k_{e}=0.2 $, $ e_{1}=-0.5$ and $ \Lambda=-0.02 $]{
        \includegraphics[width=0.31\textwidth]{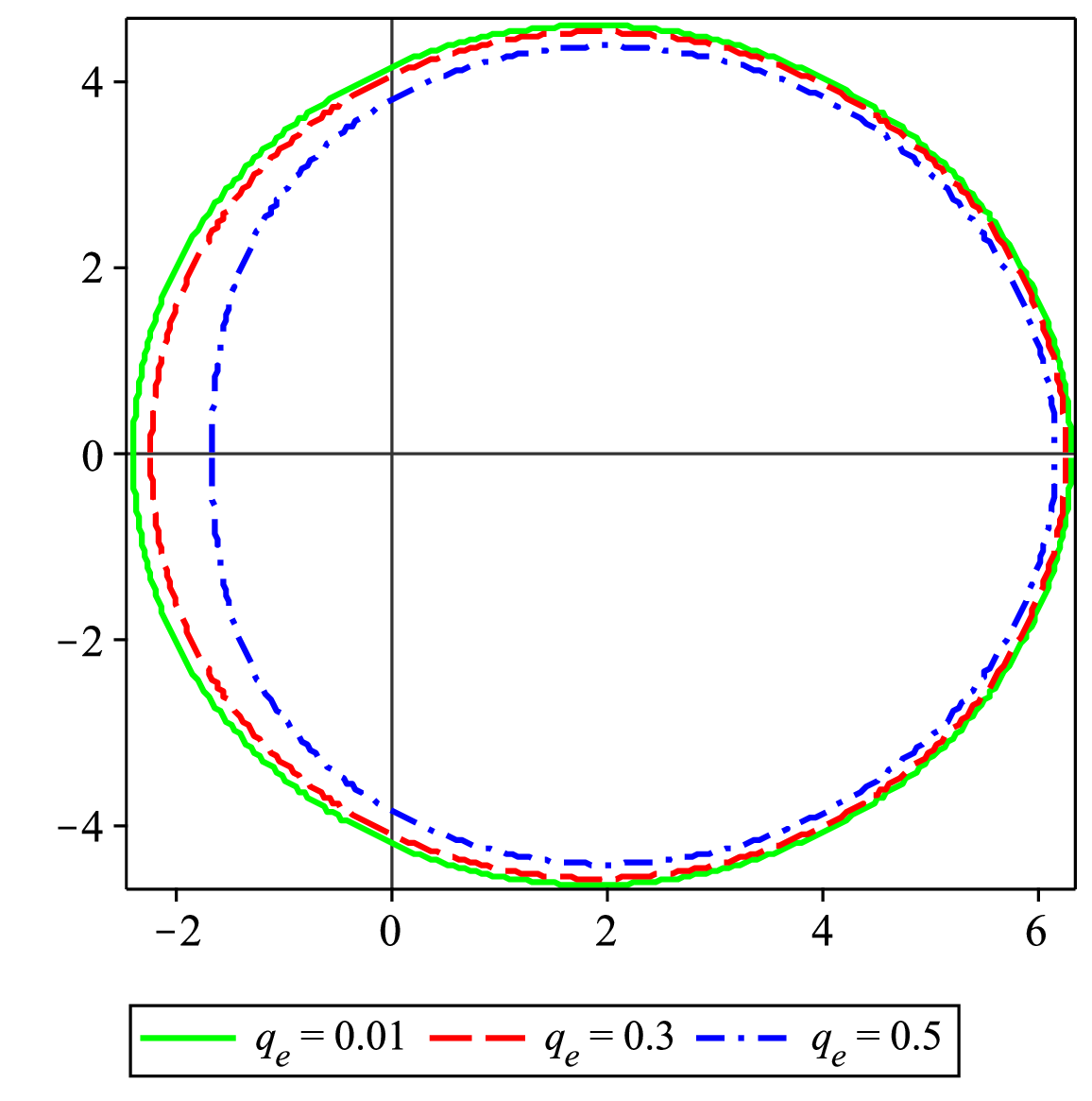}}\newline
\subfloat[ $\theta=\pi /2$, $ a=0.8$, $q_{e}=0.2 $, $ e_{1}=-0.5$ and $ \Lambda=-0.02 $]{
        \includegraphics[width=0.31\textwidth]{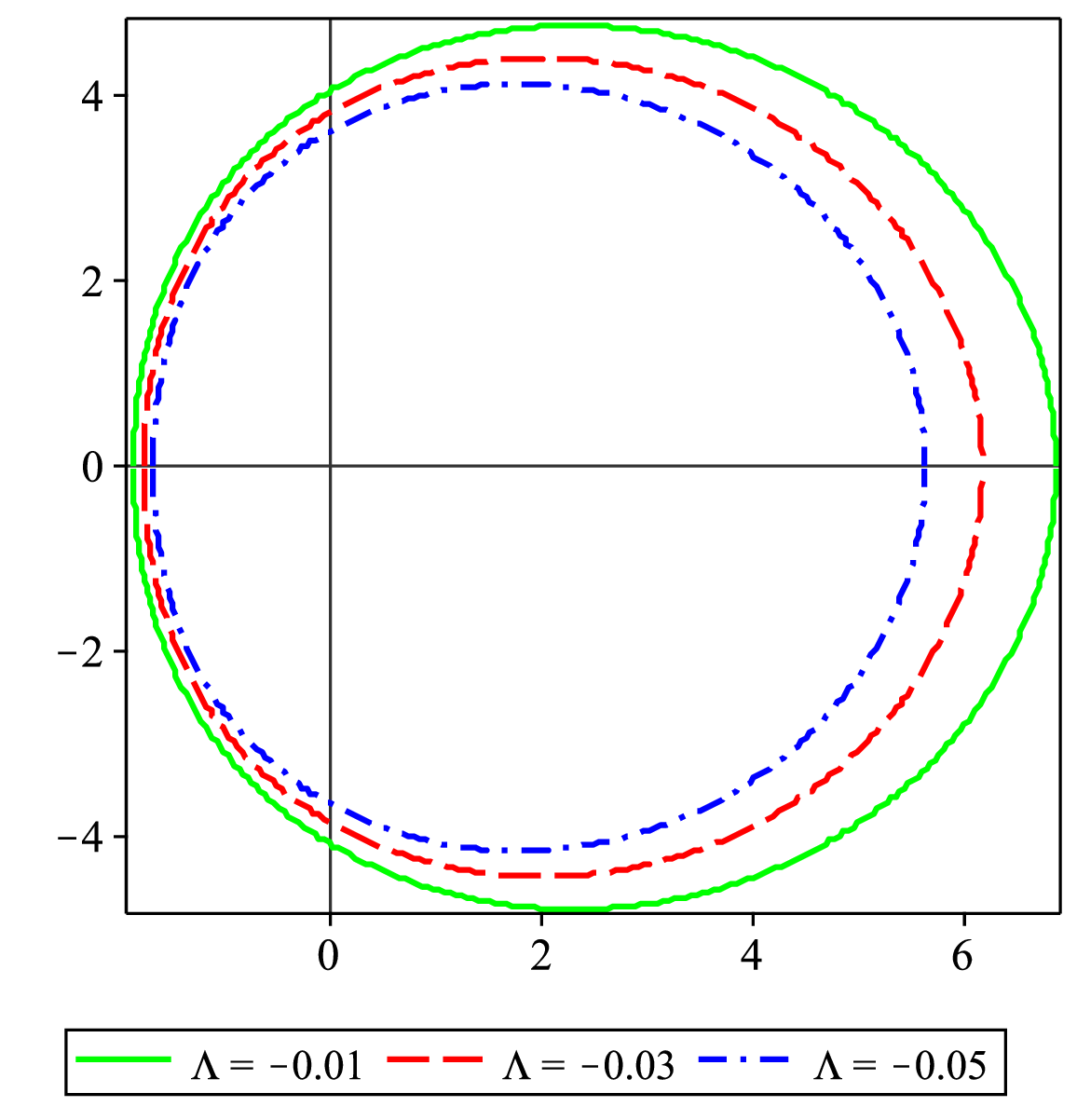}}
\subfloat[ $\theta=\pi /2$, $ a=0.8$, $ e_{1}=-0.5$ and $q_{e}=k_{e}=0.2 $]{
        \includegraphics[width=0.325\textwidth]{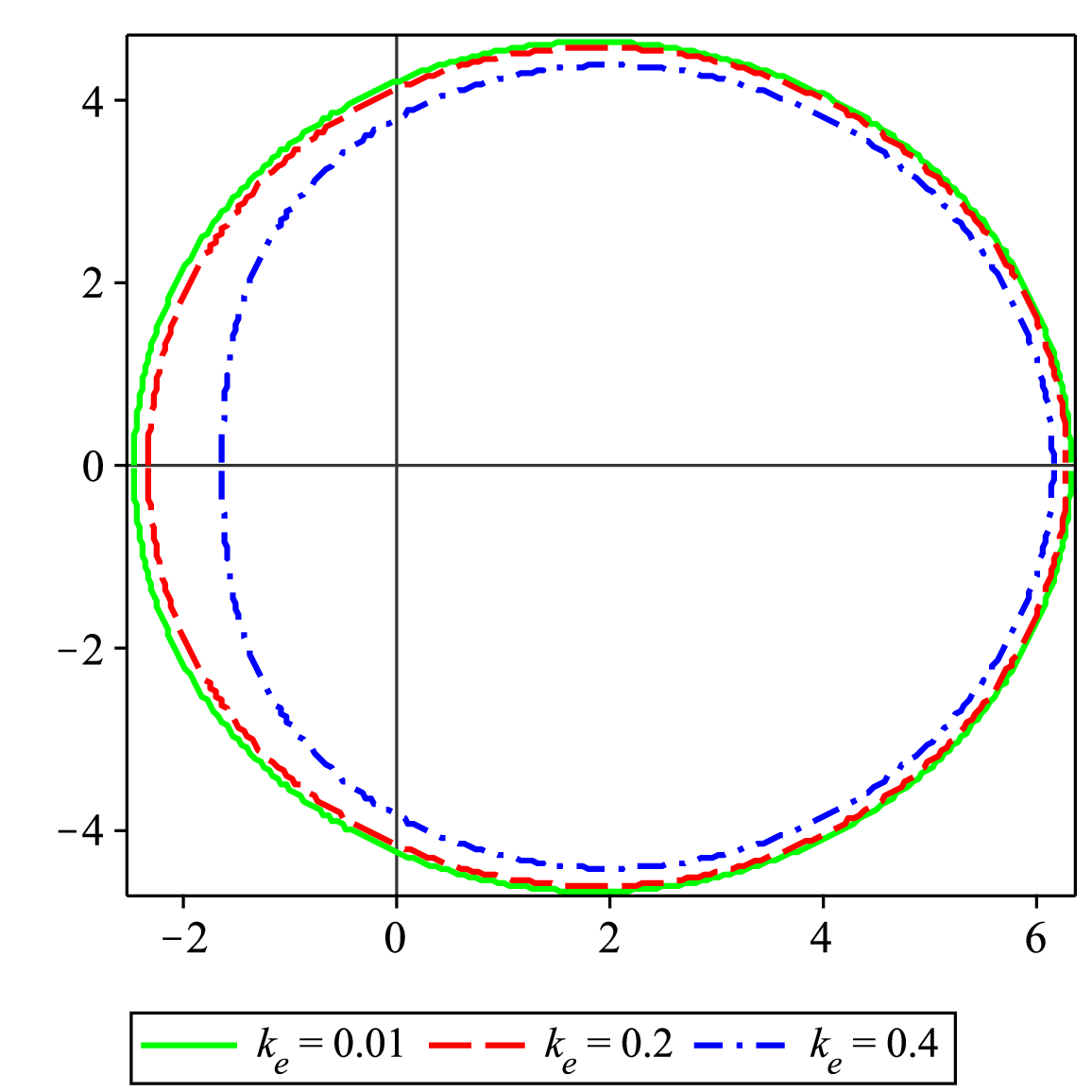}}
\subfloat[ $\theta=\pi /2$, $ a=0.8$, $q_{e}=k_{e}=0.2 $ and $ \Lambda=-0.02 $]{
        \includegraphics[width=0.31\textwidth]{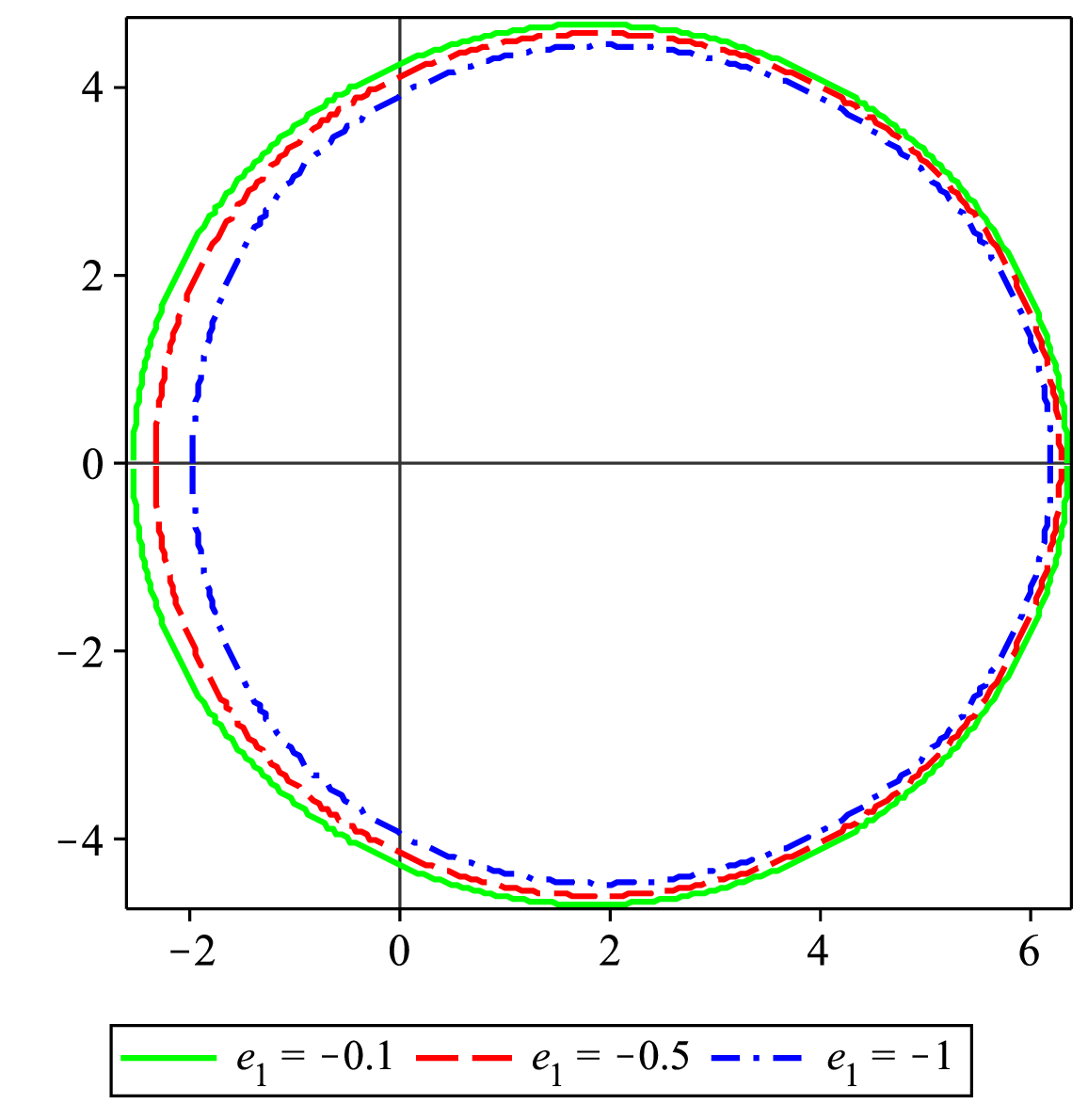}}
\newline
\caption{Shadow cast by a charged rotating BH in WCT with $ k_{s}=10^{-6} $, $ q_{m}=k_{m}=0.2 $, $d_{1}=-0.5 $, $M=1$ and different values of $\theta_{0} $, $ a $, $q_{e} $,  $k_{e}$ and $\Lambda$.}
\label{Fig2}
\end{figure}

Figure \ref{Fig2} displays the effect of BH parameters on
the shadow size for the case II. From Figs. \ref{Fig2}(a) and \ref{Fig2}(b), it
can be seen that deformation in shapes of the shadow gets more
significant with the increasing the inclination angle $\theta_0$
and rotation parameter $ a $. Regarding the influence of electric
charge,  electric dilation parameter, and parameter $e_{1}$ on the
shape of shadow, we see that although increasing $q_{e}$,
$k_{e}$ and $\vert e_{1} \vert$  makes more distortion on the shadow
size, this deformation is negligible as compared to $\theta_0$ and
$ a $. To avoid repetition, we omit drawing the figure for the case I.

\begin{figure}[!htb]
\centering
\subfloat[$ a=0.7$, $ Q=q_{e}=0.2 $]{
        \includegraphics[width=0.33\textwidth]{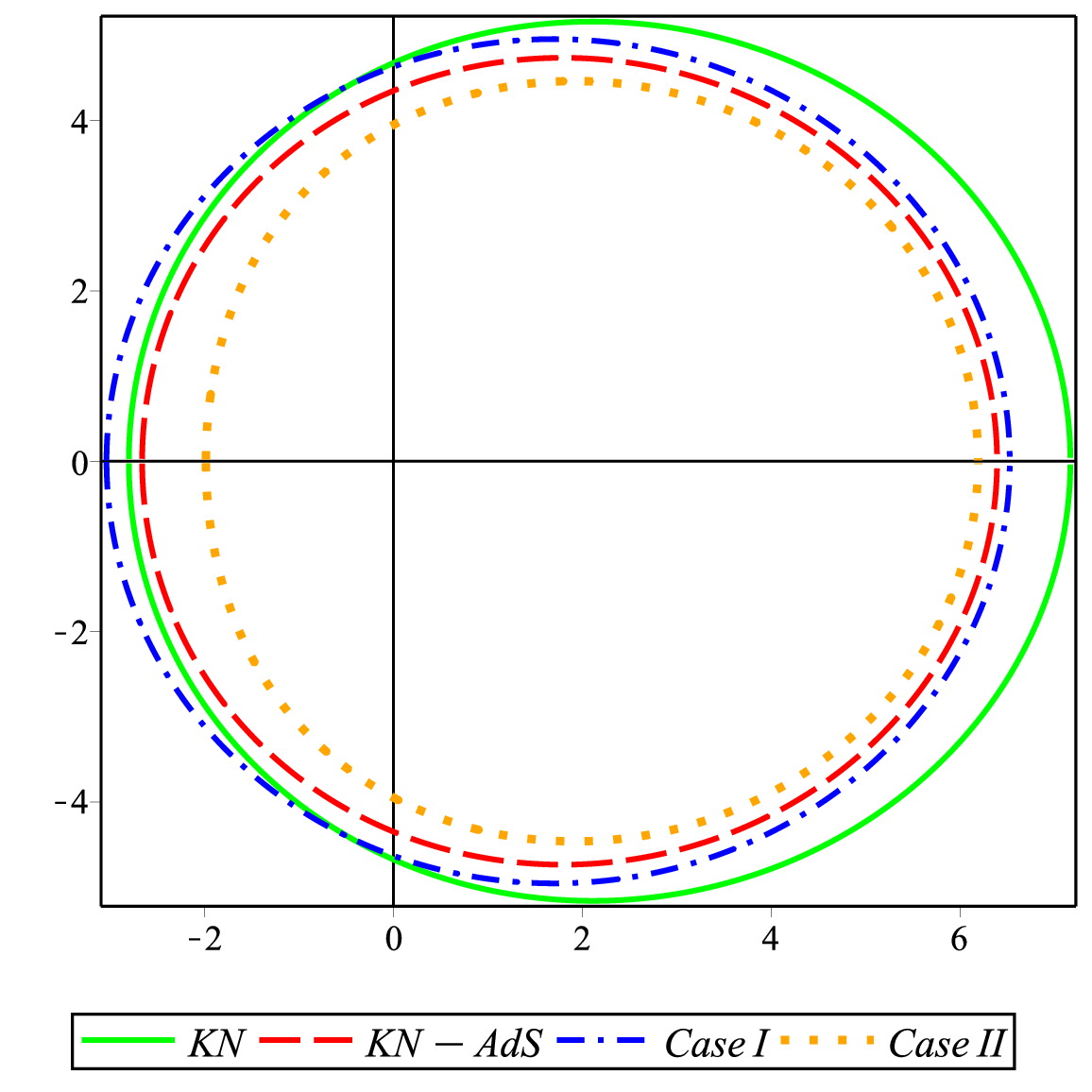}}
\subfloat[$ a=0.2$, $ Q=q_{e}=0.7 $]{
     \includegraphics[width=0.33\textwidth]{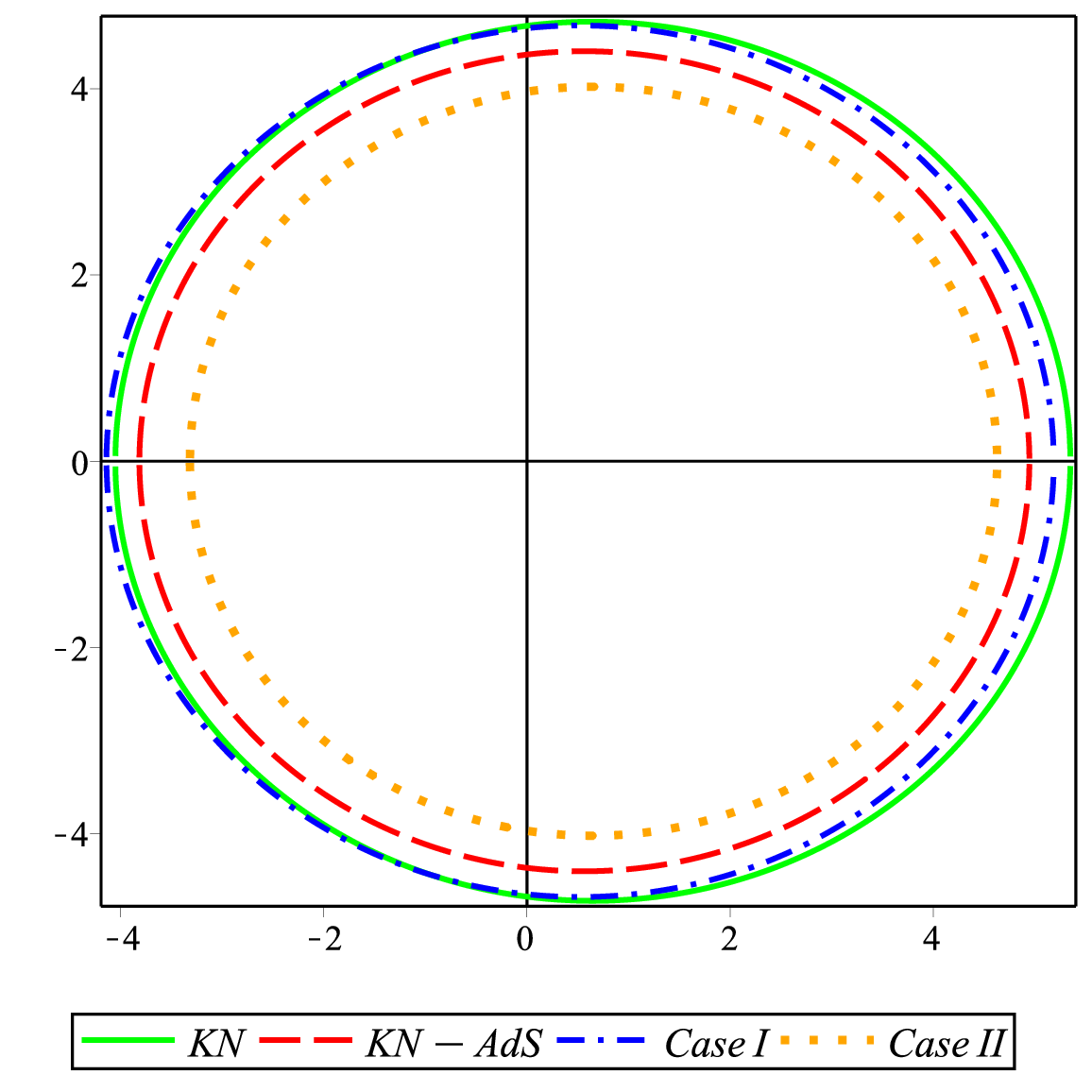}}
\newline
\caption{Shadows of KN BHs (green solid lines), KN-AdS BHs (red dashed lines), rotating BHs in WCT case I (blue dashdotted lines), and case II (orange dotted lines)
for different values of $ a $ and $ Q $. For rotating BHs in WCT, we have fixed $ k_{s}=10^{-6} $, $ q_{m}=k_{m}=0.2 $, $\Lambda=-0.02$,   $e_{1}=d_{1}=1 $ (case I), and $e_{1}=d_{1}=-1 $ (case II).}
\label{RC}
\end{figure}

Now we would like to compare rotating black holes in WC theory with their counterpart in GR. To have a better understanding we have plotted Fig. (\ref{RC}). As we expected, the cosmological constant has a decreasing contribution to the shadow size. By close looking at this figure, one can find that case I (case II) black holes in WC theory have larger (smaller) shadow radius as compared to KN-AdS black holes in GR. In Ref. \cite{Belhaj:09058}, the authors studied the optical properties of  Kerr-Newman and Kerr-Sen black holes in AdS spacetime. Their analysis represented that shadow size shrinks with the increase of both the rotation parameter and electric charge which is similar to our study. The shadow cast of Kerr-Newman-Kasuya black holes (rotating dyon BHs) has been also investigated in Ref.\cite{Saavedra:041}. According to the obtained results, both electric and magnetic charges dramatically decrease the size of the BH shadow cast which confirms the validity of our results. The optical features of rotating BHs have been investigated in different modified gravities, which have provided us with interesting results. For instance, the shadow of rotating BHs in $f(R)$ gravity was explored in Ref \cite{Dastan:1002} which was proved that by increasing the effect of the modified gravity parameter, the size of shadow image increases and the symmetry of the black hole’s shadow improves. While the study in the context of dRGT massive gravity showed that massive parameters decrease the shadow size \cite{Hendi:022}. Such a result was also observed for rotating charged black holes in 4D Einstein-Gauss-Bonnet gravity \cite{Papnoi:100916}, meaning that the modified gravity parameter reduces the shadow radii. For rotating BHs in Horndeski gravity, it has been illustrated that the shadow and its distortion get more as the effect of the Horndeski parameter gets stronger \cite{Ghosh:2023174}. In the study of spinning black holes in Symmergent gravity, it was found that increasing the Symmergent gravity parameter leads to decreasing the shadow radius \cite{Pantig:250}. Whereas in the scalar-tensor-vector gravity,  the shadow size increases with the growth of the modified gravity parameter \cite{Kuang:064012}. As a result, one can state that the shadow of the BH can increase/decrease by modified gravity parameters depending on the sort of modified gravity.

\subsection{Energy emission rate}
\label{SubsecB}

The formation and destruction of an excessive number of particles very close to the horizon are known as emission energy. In fact,  the emission of radiation from a black hole is realized through the creation of a virtual pair of particles just outside the horizon. The antiparticle falls inside the black hole,  while the particle can propagate away from the black hole whose mass has decreased due to the negative amount of energy it received. Hawking radiation process is a classical phenomenon but with a quantum origin, meaning that quantum fluctuations in the interior of BHs are the source of emission energy. Here, we are interested in exploring the energy emission rate
associated with the considered BH.

It has been shown that the emission rates of particles $  N$ and energy $  E$ per unit time
and unit frequency around a black hole in D-dimensional spacetime is given by \cite{Kanti:2009}

\begin{equation}
\frac{d^{2}}{d\omega dt}\left( 
\begin{array}{cc}
N  \\ 
\\
E     
\end{array}\right) =\frac{\sum_{\ell=0}^{\infty}\frac{(\ell+D-4)!}{\ell !}(N_{\ell}+D-4)\vert \mathcal{A}_{\ell}(\omega)\vert^{2}}{2^{D-2} \pi^{\frac{1}{2}}\Gamma\left( \frac{D-1}{2}\right)\Gamma\left( \frac{D-2}{2}\right)\left(e^{\frac{\omega }{T}}- 1 \right)}\left( 
\begin{array}{cc}
1  \\ 
\\
\omega    
\end{array}\right)
\label{dE}
\end{equation}
where $N_{\ell}=2\ell +1$ is the multiplicity of states that have the same value of the angular
momentum number $ \ell $; the quantity $ \vert \mathcal{A}_{\ell}(\omega)\vert^{2} $ is the absorption probability (a dimensionless quantity)
for a particle with energy $ \omega $ and angular momentum $ \ell $. This quantity plays the role of a greybody factor when we consider the emission of particles by the black hole. A dimensionful quantity may be constructed out of the absorption probability, namely the absorption cross-section, which is defined as \cite{GuoLiu:2011}
\begin{equation}
\sigma_{abs}(\omega)=\frac{\pi^{\frac{D-2}{2}} }{\Gamma\left( \frac{D-2}{2}\right)\omega^{D-2} }\times\sum_{\ell=0}^{\infty}\frac{(\ell+D-4)!}{\ell !}(N_{\ell}+D-4)\vert \mathcal{A}_{\ell}(\omega)\vert^{2}
\end{equation}

Therefor, for 4-dimensional BHs, Eq. (\ref{dE}) can be rewritten as \cite{Decanini:044032}
\begin{equation}
\frac{d^{2}}{d\omega dt}\left( 
\begin{array}{cc}
N  \\ 
\\
E     
\end{array}\right)=\frac{\omega^{2}}{2 \pi^{2}}\frac{\sigma_{abs}(\omega)}{e^{\frac{\omega }{T}}- 1}\left( 
\begin{array}{cc}
1  \\ 
\\
\omega    
\end{array}\right)
\label{dE1}
\end{equation}
 
 At high energy limit, the absorption cross section oscillates around
a limiting constant value $\sigma_{lim}$ which is equal  to the geometrical cross section of its photon sphere \cite{Mashhoon:1973,Wei:2013}. Since the shadow
measures the optical appearance of a black hole, it can be equal to the limiting constant
value of the high-energy absorption cross section. Hence, the limiting constant value $\sigma_{lim}$ can be approximately obtained as 
\begin{equation}
\sigma_{lim}\approx \pi r_{sh}^{2}
\label{sigmalimit}
\end{equation}

The observable $r_{\text{sh}}  $ which designates
approximately the size of the shadow, can be calculated as
\begin{equation}
r_{\text{sh}}=\frac{(x_{t}-x_{r})^{2}+y_{t}^{2}}{2|x_{t}-x_{r}|}.
\end{equation}

In Eq. (\ref{dE1}), $ T $
represents the temperature of the black hole.  Since Hawking temperature depends on the geometrical properties of the spacetime (surface gravity) such as metric tensor and its symmetrical properties, we need to obtain Killing vectors to calculate Hawking temperature. The corresponding Killing vector field is $ \chi =\partial
/\partial t +\Omega \partial /\partial \phi $ in which 
$ \Omega=-g_{\varphi t} /g_{\varphi \varphi}$ is 
the rotational velocity of the BH horizon determined as
 \begin{equation}
 \Omega=\frac{a}{a^{2}+r_{e}^{2}}.
 \end{equation}
 
The surface gravity which is related to the Hawking temperature of BHs $ (T=\kappa /2\pi)$ is defined as
 \begin{equation}
\kappa =\sqrt{-\frac{1}{2}(\partial_{\mu}\chi_{\nu})(\partial^{\mu}\chi^{\nu})}.
\end{equation}

Hence, Hawking temperature is calculated as

\begin{equation}
T=\frac{-3\Lambda r_{e}^{4}+(3-\Lambda a^{2})r_{e}^{2}-3d_{1}k_{s}^{2}+12e_{1}(k_{e}^{2}+k_{m}^{2})-3a^{2}-3(q_{e}^{2}+q_{m}^{2})}{12\pi r_{e}(a^{2}+r_{e}^{2})} .
 \label{TH}
\end{equation}

\begin{figure}[!htb]
\centering
\subfloat[$q_{e}=k_{e}=0.2 $, $ e_{1}=-0.5$ and $\Lambda=-0.02 $]{
        \includegraphics[width=0.31\textwidth]{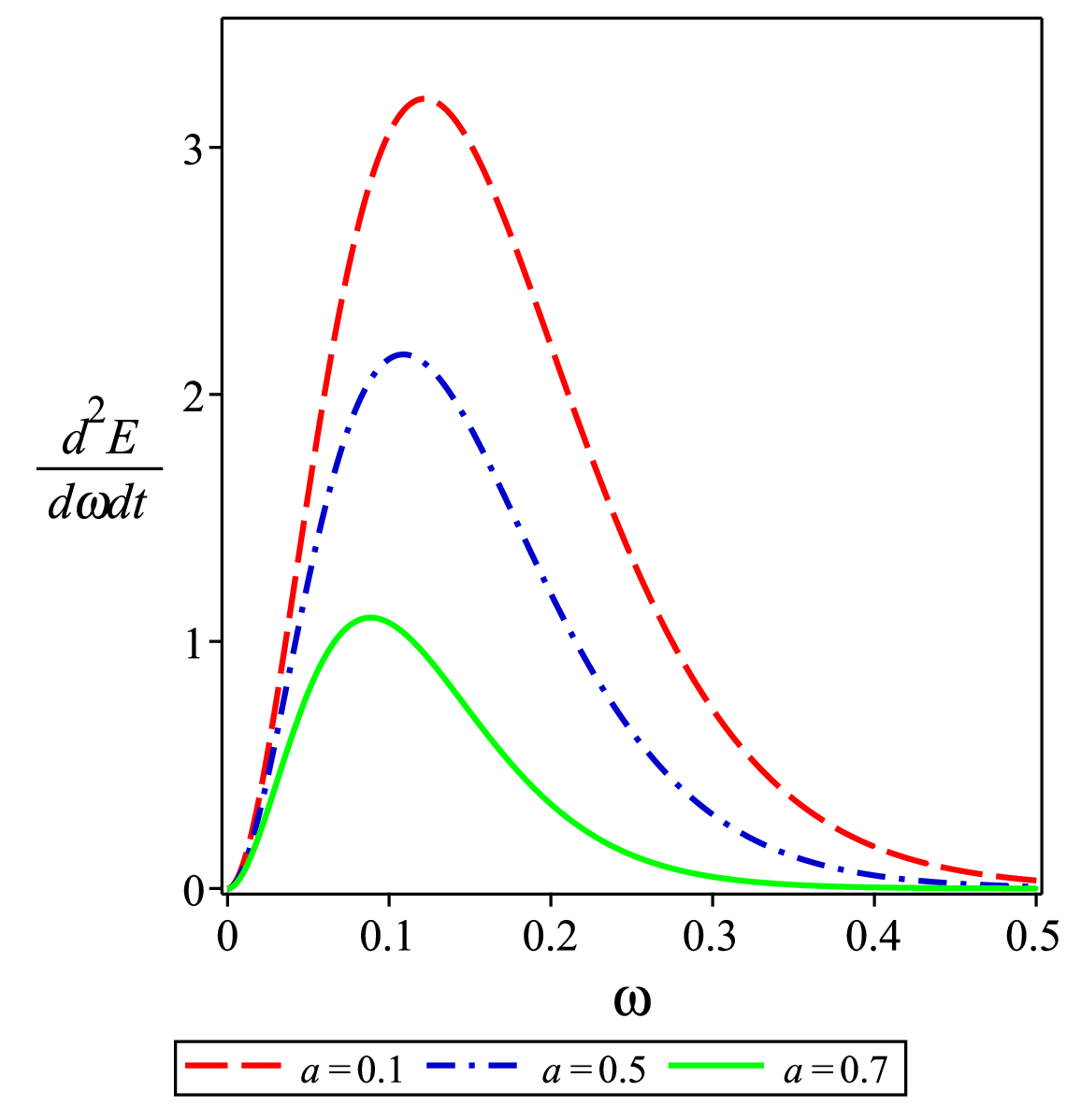}}
\subfloat[$ a=0.8$, $k_{e}=0.2 $, $ e_{1}=-0.5$ and $ \Lambda=-0.02 $]{
        \includegraphics[width=0.31\textwidth]{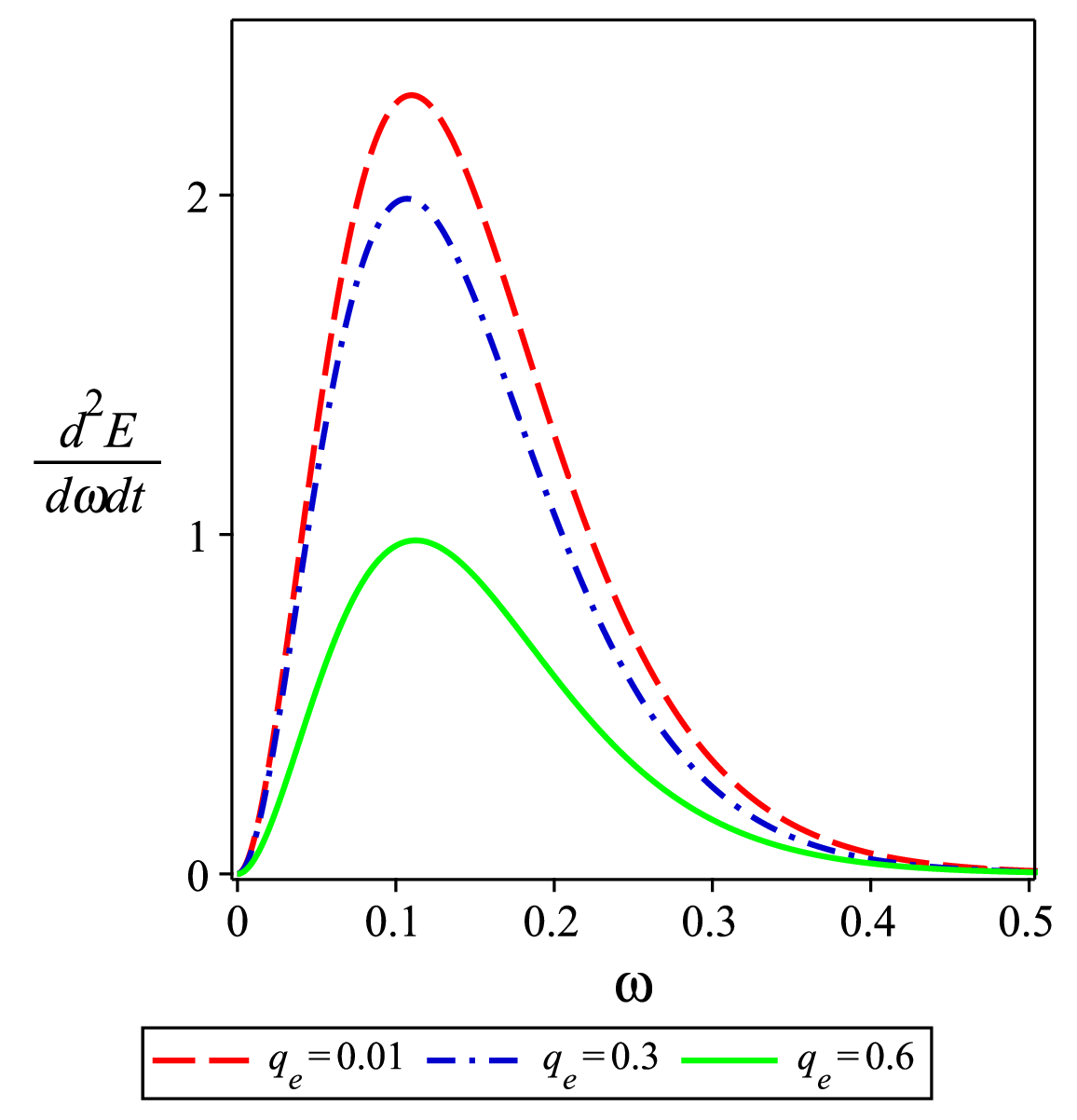}}
\subfloat[$ a=0.8$, $q_{e}=0.2 $, $ e_{1}=-0.5$ and $ \Lambda=-0.02 $]{
        \includegraphics[width=0.31\textwidth]{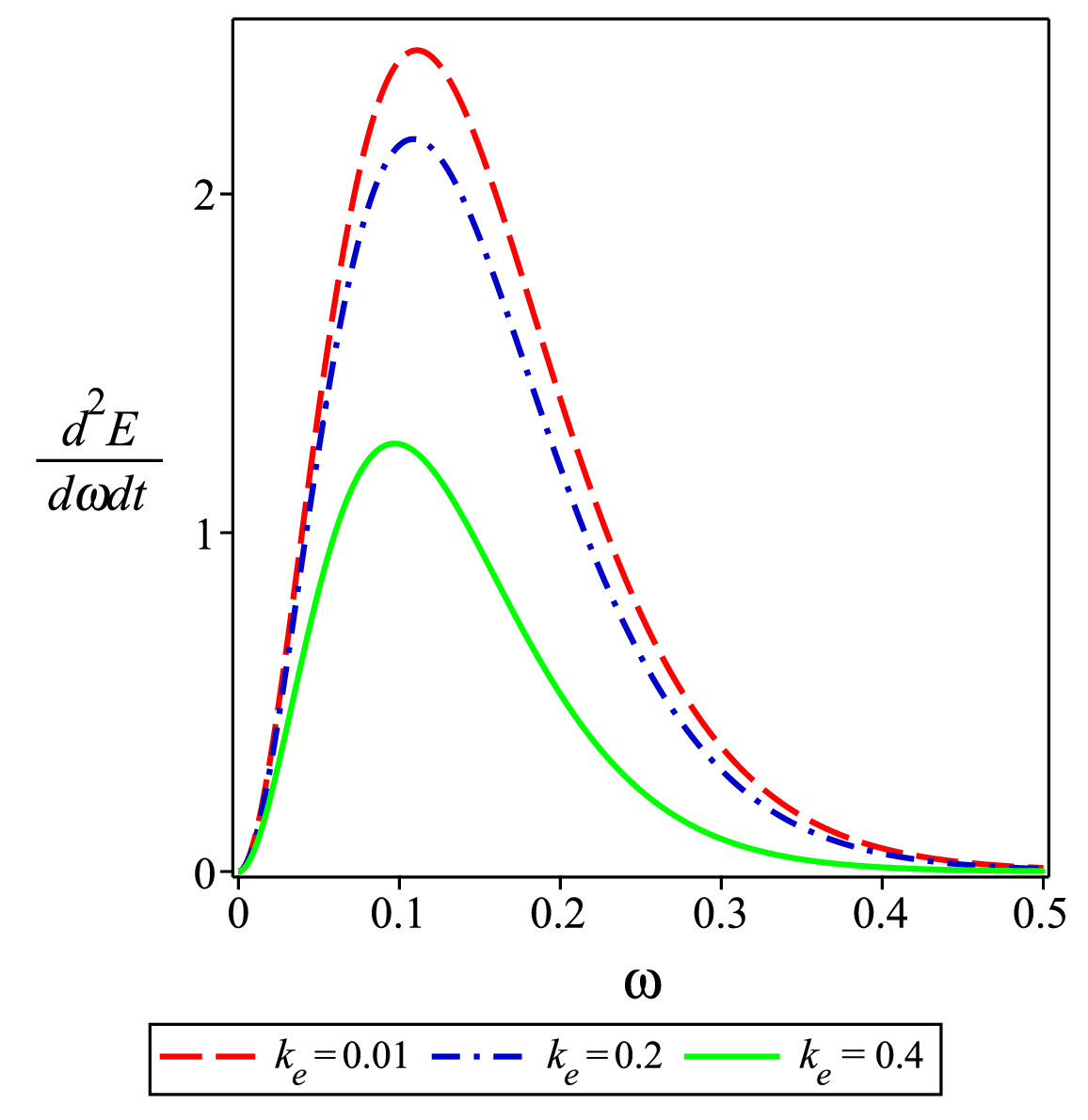}}\newline
\subfloat[ $ a=0.8$, $ e_{1}=-0.5$ and $q_{e}=k_{e}=0.2 $]{
        \includegraphics[width=0.31\textwidth]{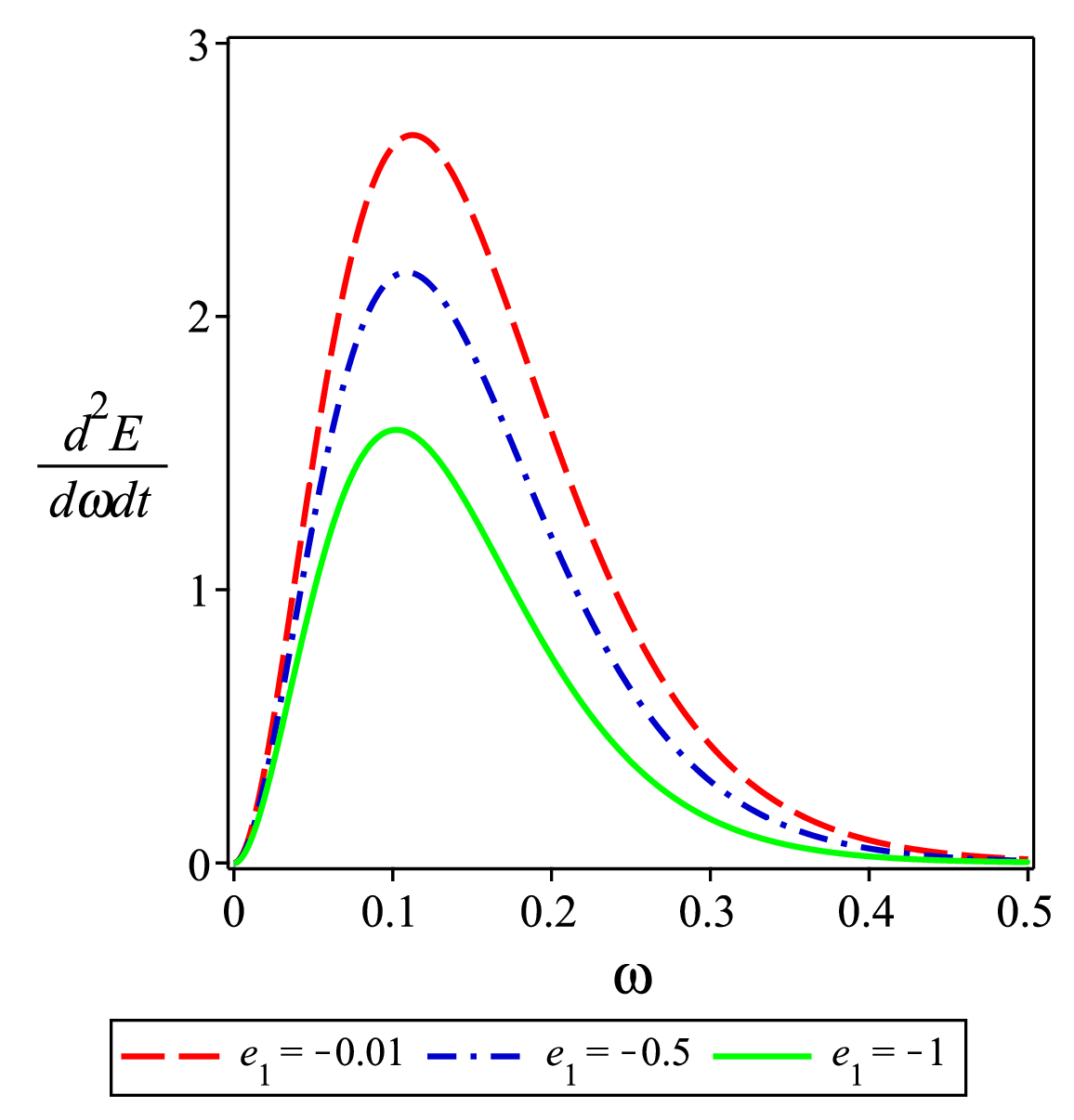}}
\subfloat[ $ a=0.8$, $q_{e}=k_{e}=0.2 $ and $ \Lambda=-0.02 $]{
        \includegraphics[width=0.31\textwidth]{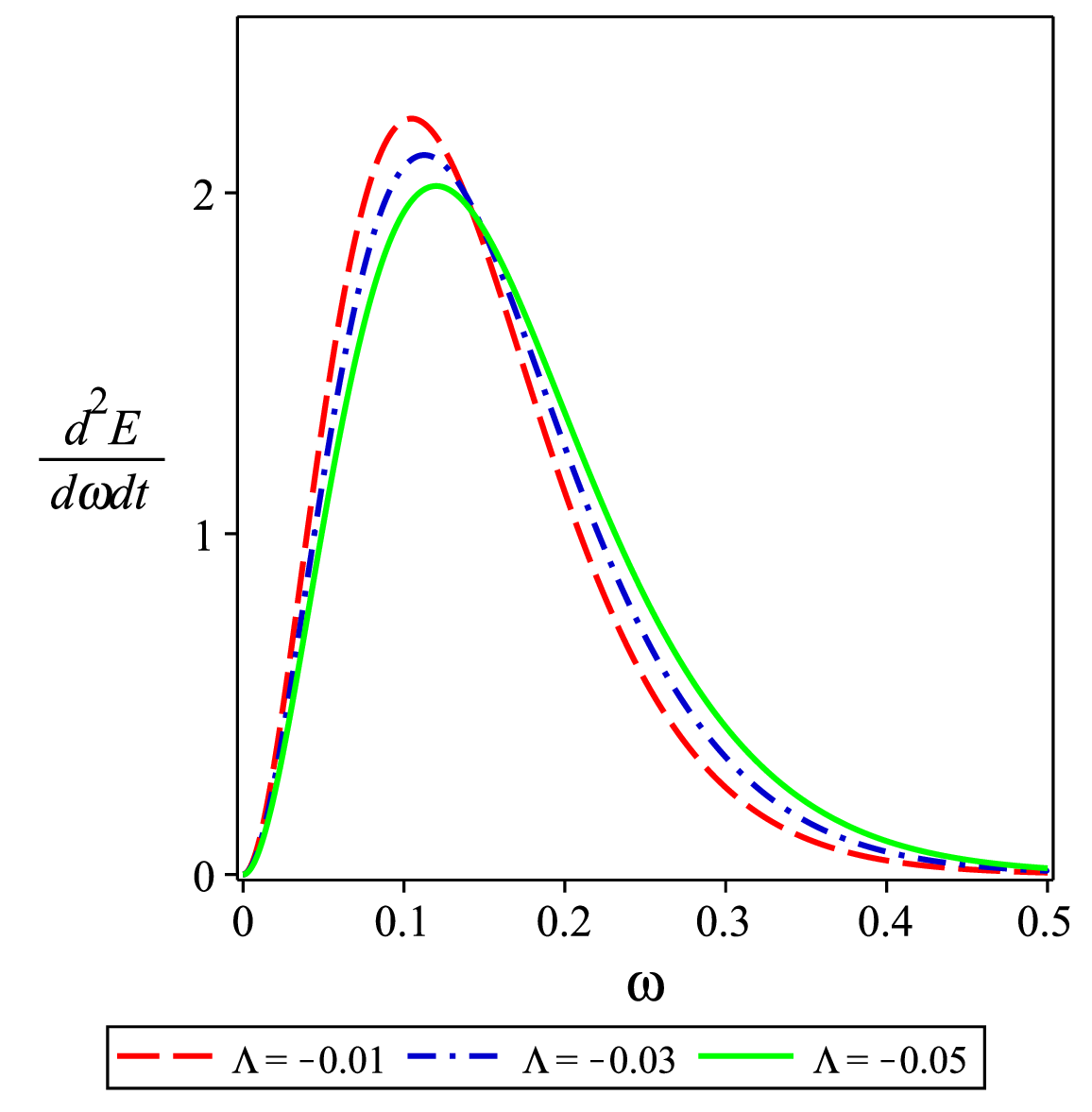}}
        \newline
\caption{The energy emission rate versus $\protect\omega $ for the
corresponding BH with $ k_{s}=10^{-6} $, $ q_{m}=k_{m}=0.2 $, $d_{1}=-0.5 $, $M=1$ and different values of $ a $, $q_{e} $,  $k_{e}$ and $\Lambda$.}
\label{Fig3}
\end{figure}

Inserting Eqs. (\ref{sigmalimit}) and (\ref{TH}) into Eq. (\ref{dE1}), we
can study the energy emission rate around the black hole solution (\ref{metric}).  In
Fig. (\ref{Fig3}), these energetic aspects are plotted as a
function of $ \omega $ for different values of the rotation
parameter, electric charge, electric dilation parameter, parameter
$ e_{1} $ and cosmological constant for the case II. From Fig. \ref{Fig3}(a),
energy emission rate decreases as the rotation parameter
increases, indicating that the evaporation process would be slow
for fast-rotating BHs. It reveals the fact that fast-rotating BHs have a longer lifetime as compared to slowly-rotating BHs.  To show the electric charge effect, we have
plotted Fig. \ref{Fig3}(b) which displays that this parameter
decreases the energy emission so that when the BH is located in a
weak electric field, the evaporation process would be faster. In other words, rotating black holes located in a powerful electric field have a longer lifetime. 
Studying the impact of the electric dilation parameter and
parameter $\vert e_{1} \vert$, we observe that their effect on the emission
rate is similar to that of the rotation parameter and electric
charge (see Figs.
\ref{Fig3}(c) and \ref{Fig3}(d)).  Therefore, one can find out that as the influence of these two parameters gets stronger the lifetime of the black hole increases. 
Regarding the influence of the cosmological
constant on the emission rate, our findings show that increasing $
\vert\Lambda \vert $ leads to decreasing this optical quantity.
Since the cosmological constant is proportional to AdS radius
which is representing the natural curvature of the spacetime, one
can find that the black hole has a longer lifetime in a high
curvature background. According to our findings, unlike case II, in case I,  increasing the electric dilation parameter and parameter $e_{1} $ leads to increasing the emission rate.

\section{CONSTRAINTS FROM THE EHT OBSERVATIONS}
\label{IV}
In this section, we try to consider Kerr-Newman black holes in
WCT as a supermassive black hole in M87* and
SgrA*, and use the EHT observations to further constrain the model
parameters. Such a study can be a possible way to estimate the BH
parameters or to distinguish modified gravity from general
relativity. In this regard, we will assume that supermassive black holes might be endowed with intrinsic properties of matter.

As already mentioned, we aim to properly understand the mentioned model to explore its validity.  To do so, we first define some astronomical observables which are helpful to test the parameters associated with the mentioned theory of gravity. To carefully study the influence of BH parameters on the distortion and size of the shadow, we need to connect them with shadow observables.  There are different observables that can be defined via the coordinate of the shadow boundary. Hioki and Maeda proposed two observables, radius $ r_{sh} $ and distortion $ \delta_{s} $ to characterize the black hole shadow silhouette \cite{Hioki:2009a}. $ r_{sh} $ is the radius of the reference circle for the distorted shadow which is defined as 

\begin{equation}
r_{sh}=\frac{(x_{t}-x_{r})^{2}+y_{t}^{2}}{2\vert x_{r}-x_{t}  \vert},
\end{equation}
and $ \delta_{s} $ is the deviation of the left edge of the shadow from the reference circle boundary given by
\begin{equation}
\delta_{s}=\frac{\vert x_{l}-x^{\prime}_{l}  \vert}{R_{s}}.
\end{equation}

Notably, for the shadow of the non-rotating case, $\delta_{s} =0$ due to the shape of a perfect circle. Another two characterized observables were proposed by Kumar and Ghosh in \cite{Kumar:2020b}. They used shadow area $ A $ and oblateness $ D $, which can describe the shadow with more general shapes. Their definitions are
\begin{equation}
A=2\int y(r_{p})dx(r_{p})=2\int_{r^{-}_{p}}^{r^{+}_{p}}\left(y(r_{p})\frac{dx(r_{p}) }{dr_{p}}\right)dr_{p}, \quad\quad\quad D=\frac{x_{r}-x_{l}}{y_{t}-y_{b}}. 
\label{AD}
\end{equation}

Here the subscripts $ l $, $  r$, $ t $ and $  b$ stand for the left and right ends of the shadow silhouette, 
where $ y(r_{p})=0 $, and the top and bottom points, 
where $ y^{\prime}(r_{p})=0 $, respectively. Note that for a spherically symmetric black hole shadow, it is straightforward to understand that $D=1$, but for a Kerr shadow  $\sqrt{3}/2 \leq D<1 $. The advantages of using two observables  $A$ and $D$ are that they do not require comparing the shadow silhouette with a perfect circle and are therefore applicable to a general shadow of any shape and size.

\subsection{Constrains on the parameters of the model from the image of M87*}
\label{IVA}
In this subsection, we compare the theoretically computed shadow outline with the observed image of M87*. As
reported by EHT collaboration \cite{Akiyama:2019a}, the
supermassive black hole M87* at the center of the galaxy M87* has
the following values
\begin{eqnarray}
\theta_{M87*} &=& (42 \pm 3) \mu as,\\
M_{M87*} &=& (6.5 \pm 0.9) \times 10^{9} M_{\odot},\\
\mathbb{D}_{M87*} &=& 16.8^{+0.8}_{-0.7} Mpc,
\label{EqdM87a}
\end{eqnarray}
where $ \theta $ and $ \mathbb{D} $ are, respectively, the angular diameter
of the shadow and the distance of the M87* from the Earth. $ M $
and $M_{\odot}$ denote the mass of the M87* and Sun mass. These
numbers imply that the diameter of the shadow in units of mass
should be \cite{Bambi1dx}
\begin{equation}
d_{M87^{*}}\equiv \frac{\mathbb{D}\theta}{M}\approx 11.0 \pm 1.5.
\label{EqdM87b}
\end{equation}
\begin{figure}[!htb]
\centering
\subfloat[$k_{e}=0.2 $, $e_{1}=0.5 $ and $\Lambda=-0.02 $]{
        \includegraphics[width=0.31\textwidth]{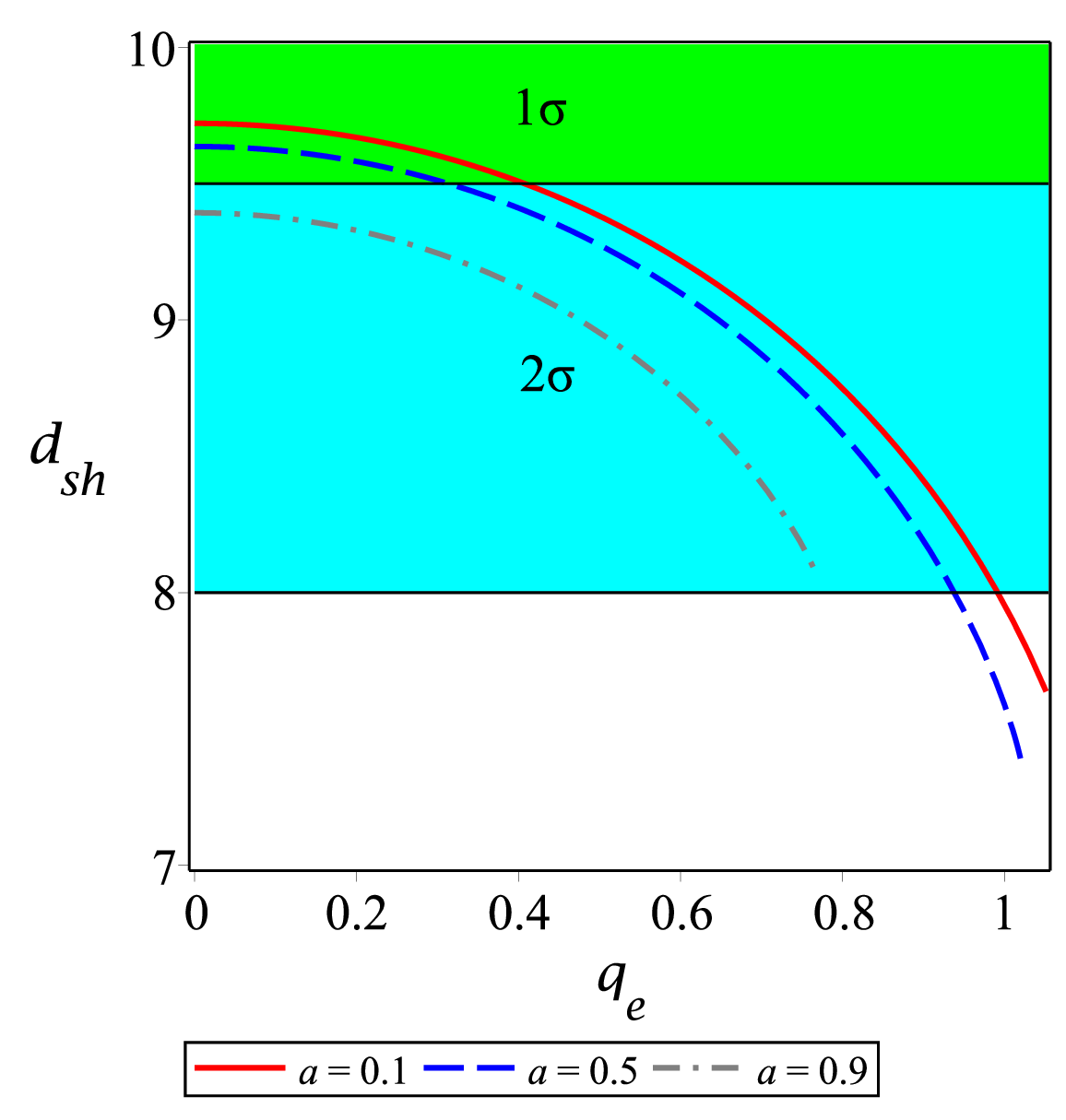}}
\subfloat[$q_{e}=0.2 $, $e_{1}=0.5 $ and $\Lambda=-0.02 $]{
        \includegraphics[width=0.322\textwidth]{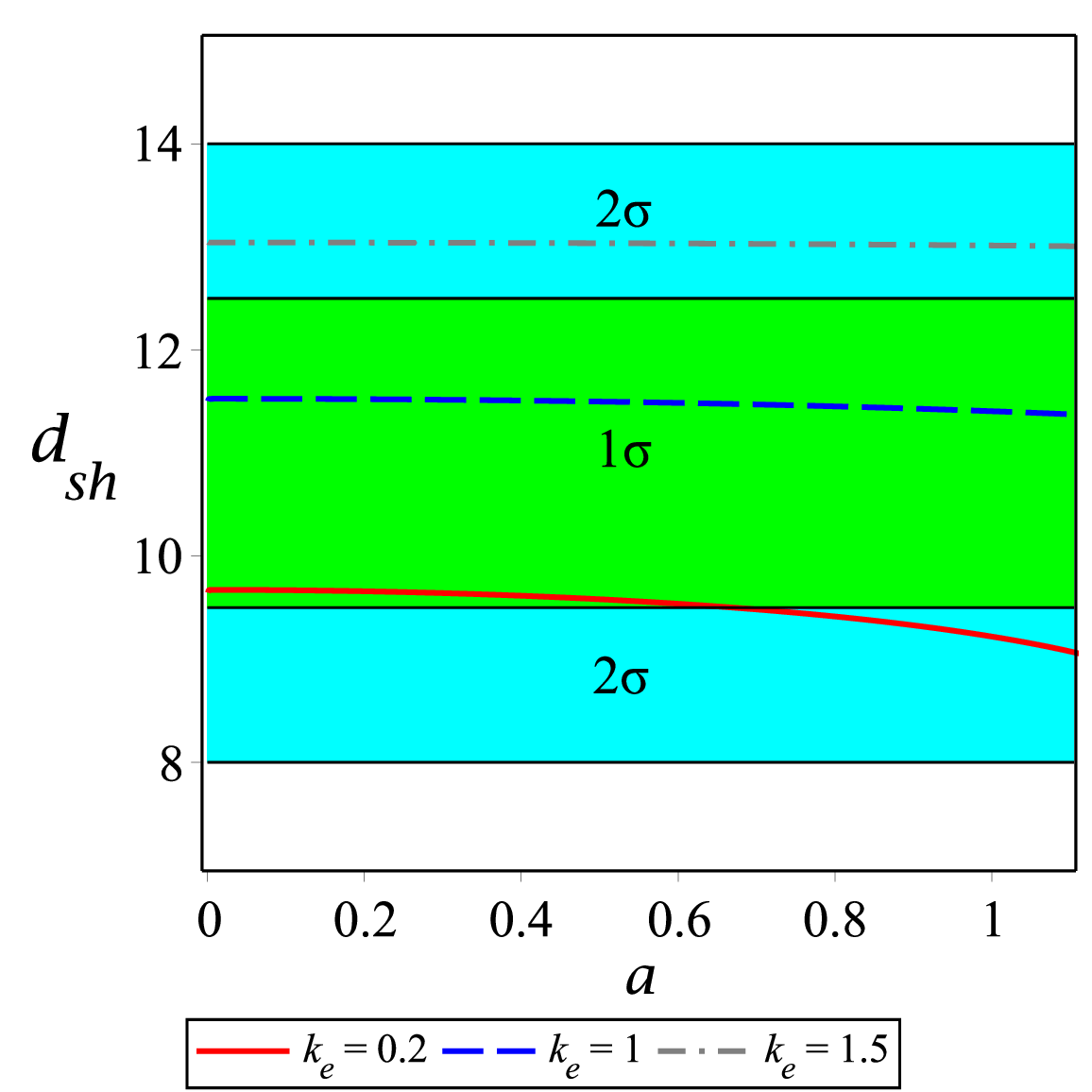}}
\newline
\subfloat[ $q_{e}=k_{e}=0.2 $ and $e_{1}=0.5 $]{
        \includegraphics[width=0.315\textwidth]{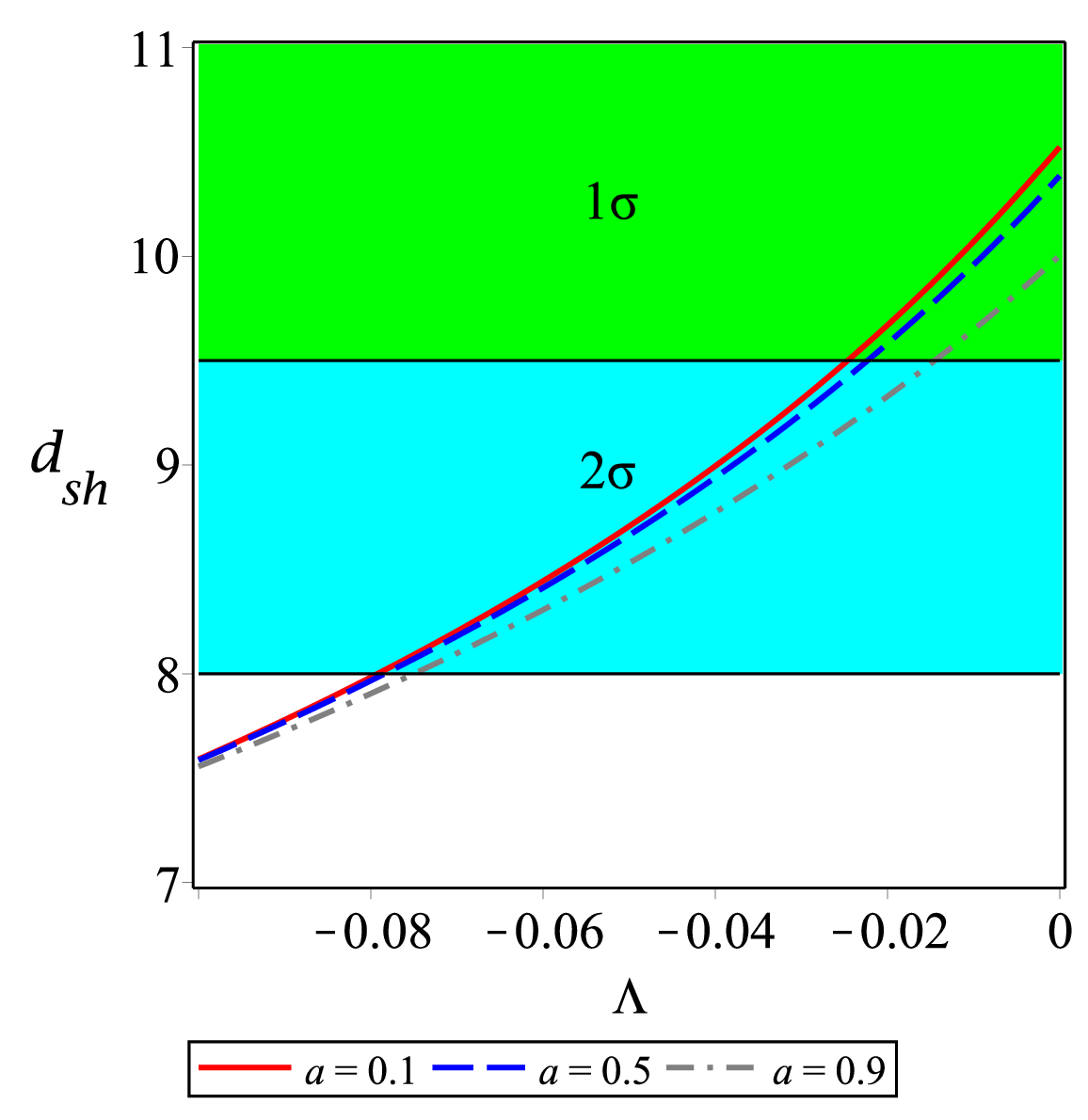}}
\subfloat[ $q_{e}=k_{e}=0.2 $ and $\Lambda=-0.02 $]{
        \includegraphics[width=0.32\textwidth]{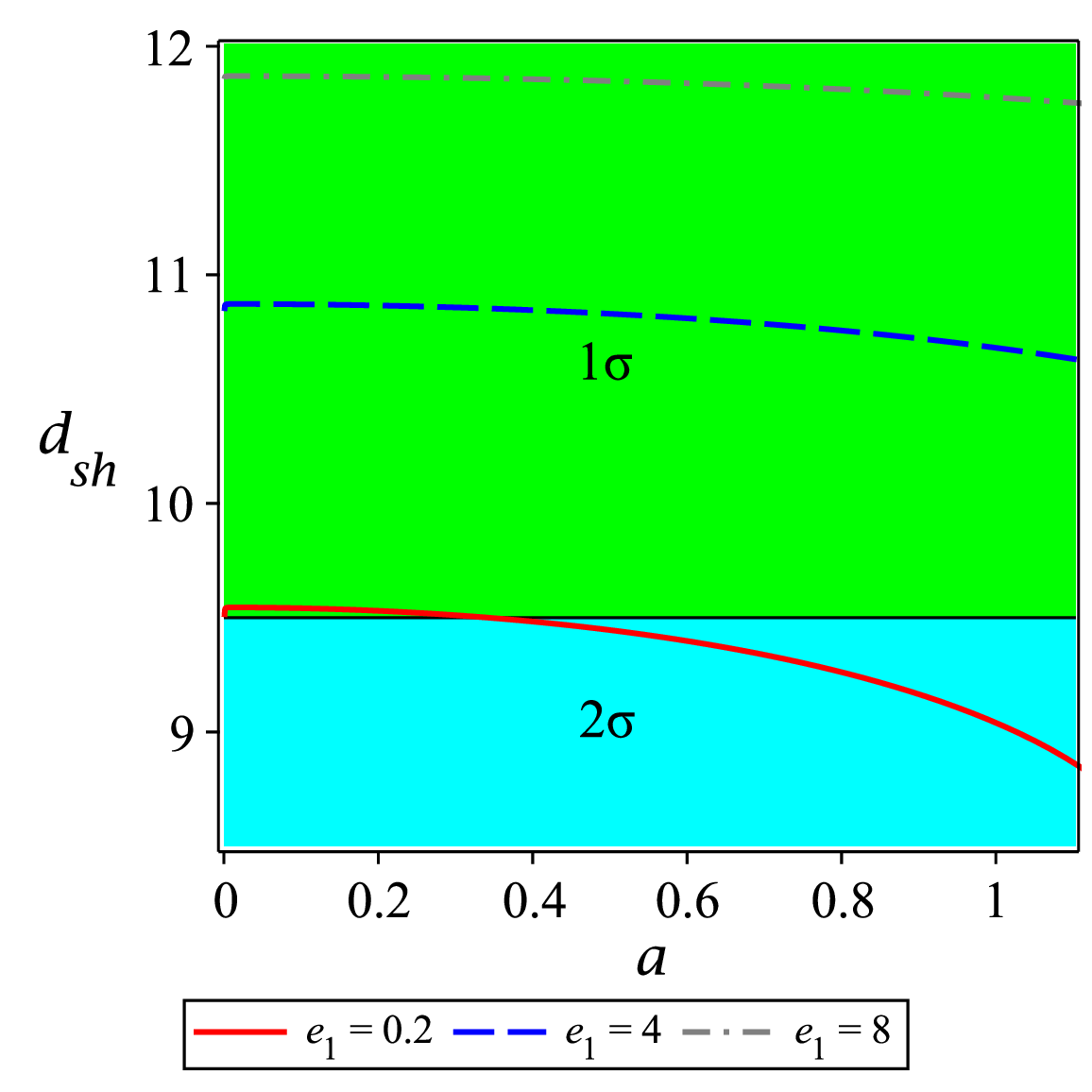}}
\newline
\caption{ The predicted diameter for a
charged rotating BH in WCT for case I with $ k_{s}=10^{-6} $, $ q_{m}=k_{m}=0.2 $, $d_{1}=0.5 $ and $M=1$ with the EHT
observations of M87*. \textbf{Left-up graph:} $ d_{sh} $ as a function of $ q_{e} $ for fixed $ k_{e}$, $ e_{1}$, $ \Lambda
$, and for several values of the rotation parameter. \textbf{Right-up graph:} $ d_{sh} $ as a function of the rotation parameter, for
fixed $ q_{e}$, $ e_{1}$,  $ \Lambda $, and  different values of
$ k_{e}$. \textbf{Left-down graph:} $ d_{sh} $ as a
function of the cosmological constant, for fixed $ q_{e}$,
$ k_{e} $, $ e_{1}$, and different values of the rotation
parameter. \textbf{Right-down graph:} $ d_{sh} $ as a function of the rotation parameter, for
fixed $ q_{e}$, $ k_{e}$, $ \Lambda $, and  different values of
$ e_{1}$. The green
shaded region gives the $ 1\sigma $ confidence region for $ d_{sh}
$, whereas the cyan shaded region gives the $ 2\sigma $ confidence
region. The inclination angle is $ \theta_{0}=0^{\circ}$.}
\label{Fig4}
\end{figure}
\begin{figure}[!htb]
\centering
\subfloat[$k_{e}=0.2 $, $e_{1}=-0.5 $ and $\Lambda=-0.02 $]{
        \includegraphics[width=0.31\textwidth]{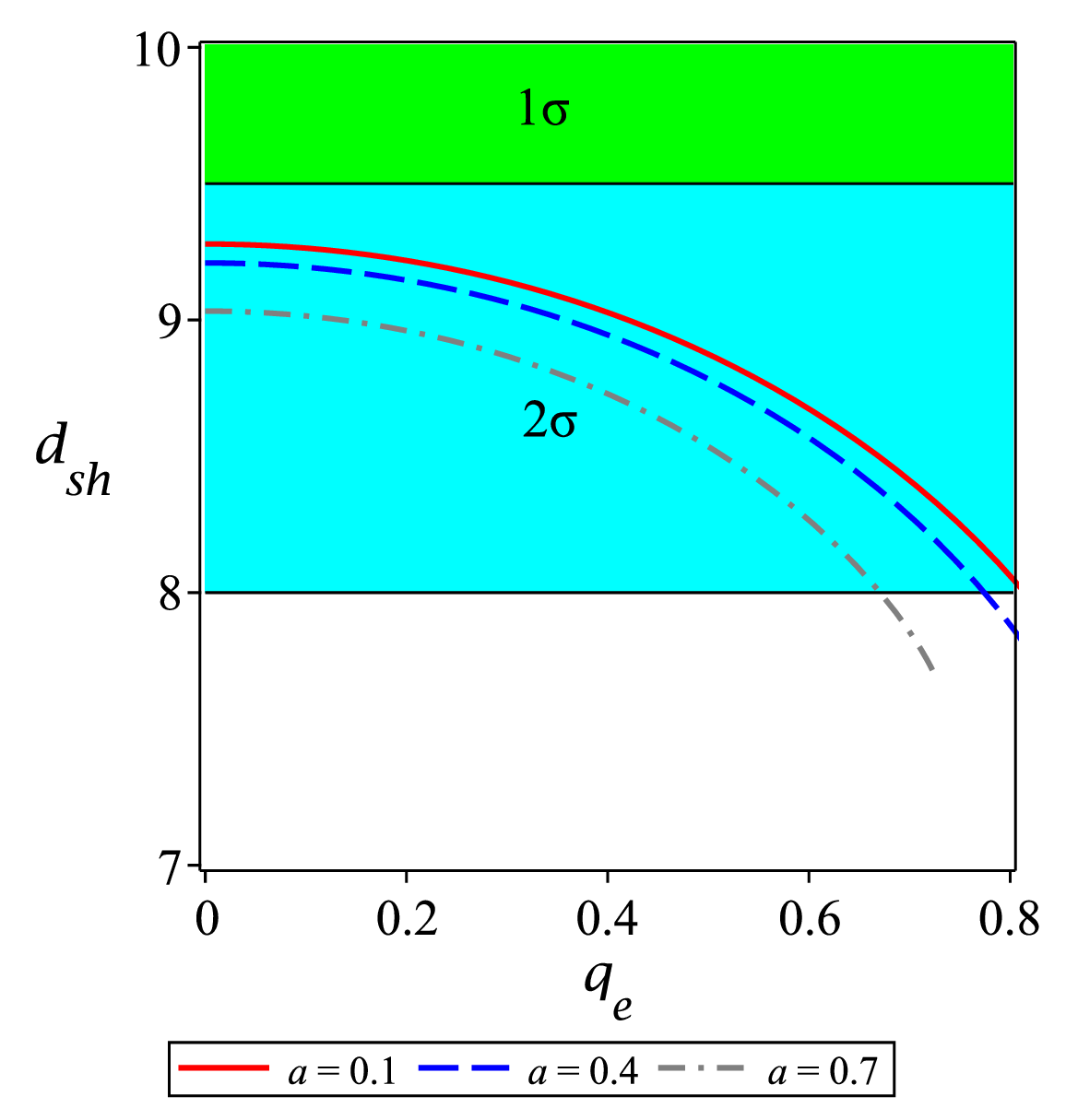}}
\subfloat[$q_{e}=0.2 $, $e_{1}=-0.5 $ and $\Lambda=-0.02 $]{
        \includegraphics[width=0.322\textwidth]{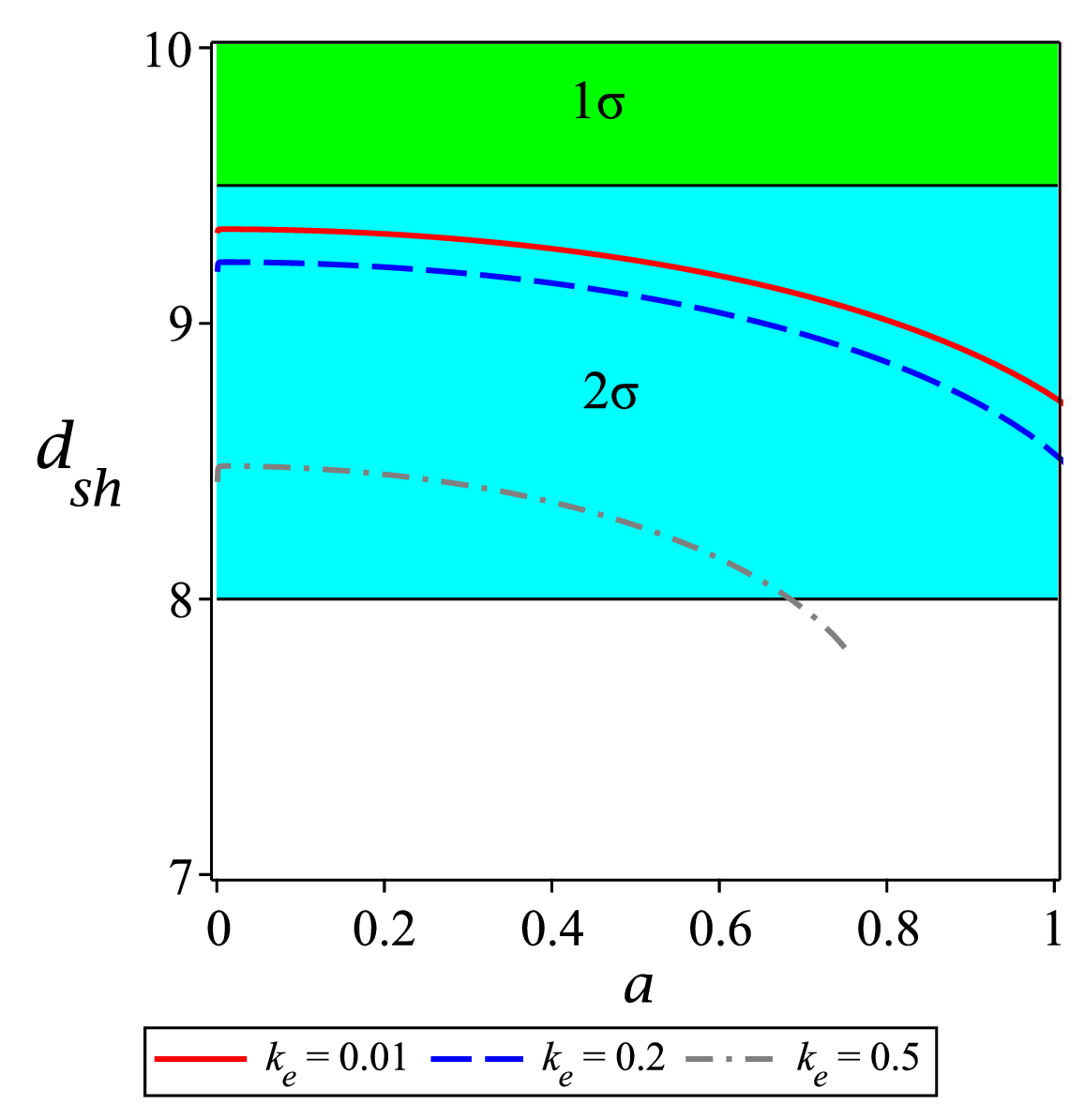}}
\newline
\subfloat[ $q_{e}=k_{e}=0.2 $ and $e_{1}=-0.5 $]{
        \includegraphics[width=0.315\textwidth]{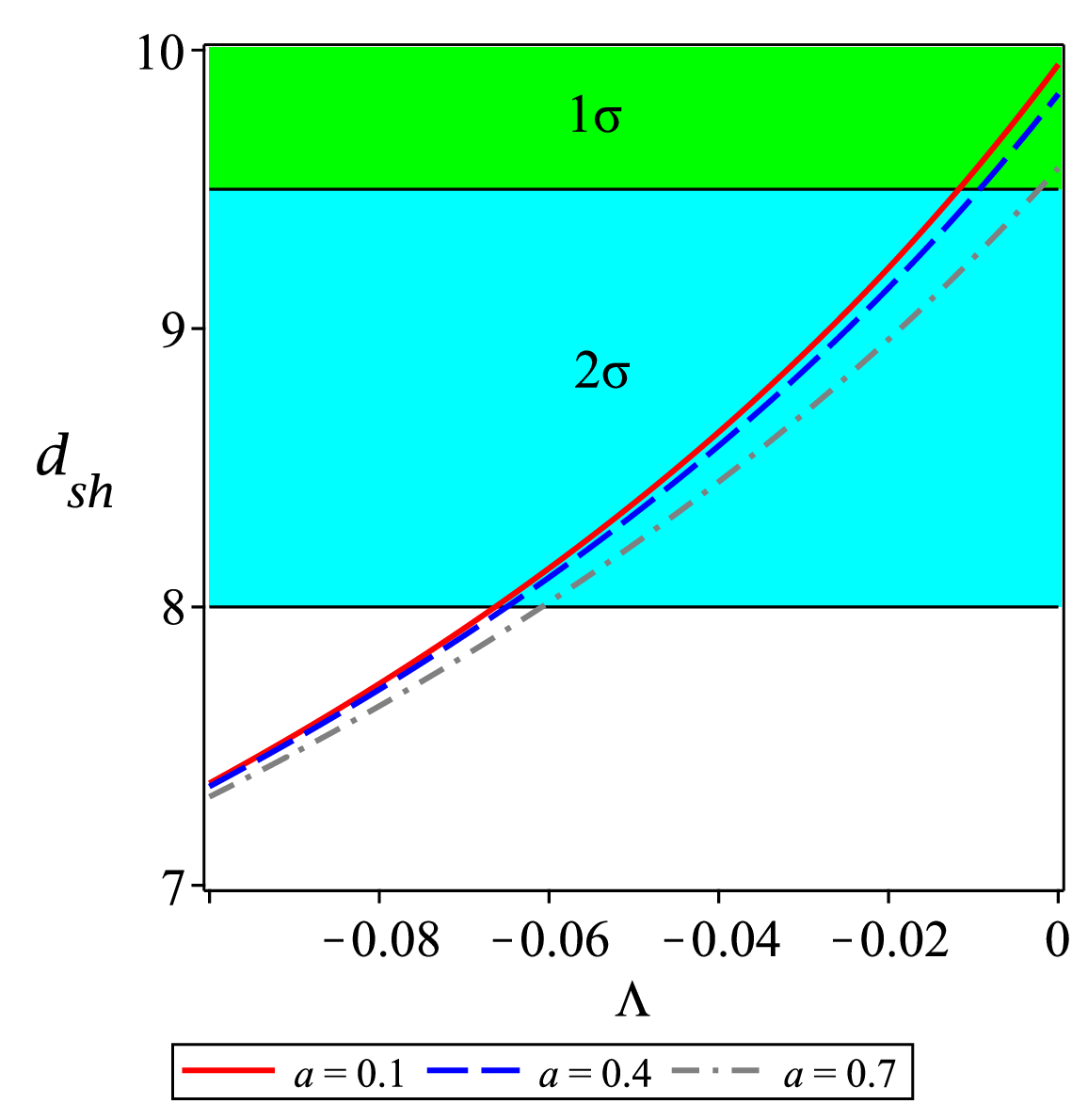}}
\subfloat[ $q_{e}=k_{e}=0.2 $ and $\Lambda=-0.02 $]{
        \includegraphics[width=0.32\textwidth]{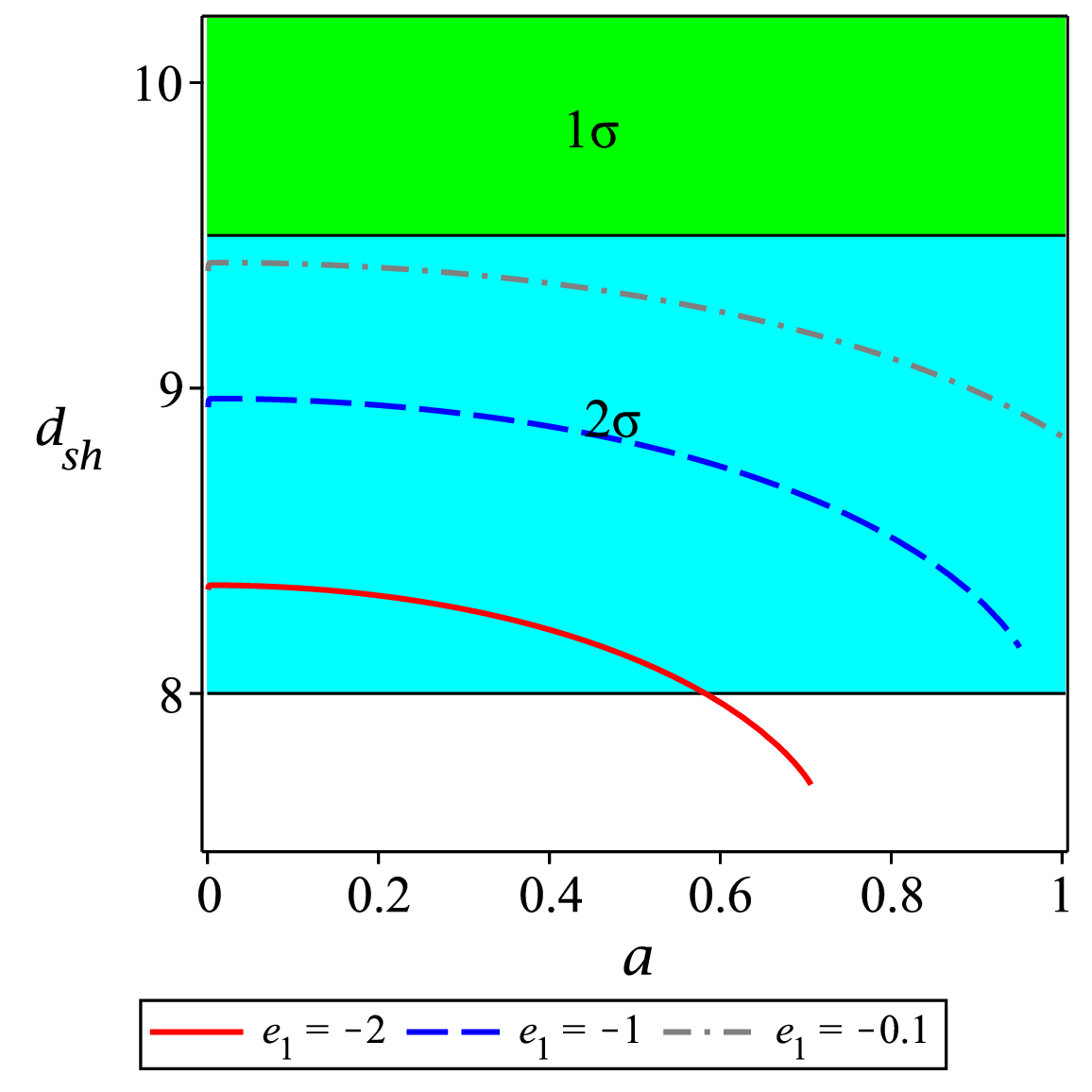}}
\newline
\caption{ The predicted diameter for a
charged rotating BH in WCT for case II with $ k_{s}=10^{-6} $, $ q_{m}=k_{m}=0.2 $, $d_{1}=-0.5 $ and $M=1$ with the EHT
observations of M87*. \textbf{Left-up graph:} $ d_{sh} $ as a function of $ q_{e} $ for fixed $ k_{e}$, $ e_{1}$, $ \Lambda
$, and for several values of the rotation parameter. \textbf{Right-up graph:} $ d_{sh} $ as a function of the rotation parameter, for
fixed $ q_{e}$, $ e_{1}$,  $ \Lambda $, and  different values of
$ k_{e}$. \textbf{Left-down graph:} $ d_{sh} $ as a
function of the cosmological constant, for fixed $ q_{e}$,
$ k_{e} $, $ e_{1}$, and different values of the rotation
parameter. \textbf{Right-down graph:} $ d_{sh} $ as a function of the rotation parameter, for
fixed $ q_{e}$, $ k_{e}$, $ \Lambda $, and  different values of
$ e_{1}$. The green
shaded region gives the $ 1\sigma $ confidence region for $ d_{sh}
$, whereas the cyan shaded region gives the $ 2\sigma $ confidence
region. The inclination angle is $ \theta_{0}=0^{\circ}$.}
\label{Fig4a}
\end{figure}

The diameter of the shadow as a function of the rotation parameter
$ a $, electric charge $ q_{e} $ and cosmological constant $
\Lambda $ is depicted in Fig. \ref{Fig4} for case I. We have plotted all graphs of this figure for the
inclination angle $ \theta_{0}=0^{\circ}$ and within $ 1\sigma $
and $ 2\sigma $ uncertainties. From Fig. \ref{Fig4}(a), data for
M87* black hole gives bound on the electric charge such that the
resulting shadow of slowly rotating black holes is within $
1\sigma $ uncertainty for $ q_{e}\leq 0.407 $ and within $ 2\sigma
$ uncertainty for $ 0.407<q_{e}\leq 0.992 $, whereas for $
q_{e}>0.992 $ the shadow is not in agreement with EHT data (see
solid curve in Fig. \ref{Fig4}(a)). For intermediate values of the
rotation parameter, $ d_{sh} $ is located in $ 1\sigma $ ($
2\sigma $) confidence region for small (large) electric charges (see dash curve in Fig.
\ref{Fig4}(a)). According to this figure,  fast-rotating BHs have
a shadow in agreement with observational data within the $ 2\sigma
$-error for all values of the electric charge (see dash dot curve
in Fig. \ref{Fig4}(a)). Fig. \ref{Fig4}(b) shows that some constraints should be imposed
on the electric dilation parameter to have  consistent results
with EHT data of M87*. According to red solid curve of this
figure, for very small values of this parameter, $ d_{sh} $ is  in
agreement with observational data within $ 1\sigma $-error ($
2\sigma $-error) for $ a \leq 0.671 $ ($ a> 0.671 $). According to our
findings, the resulting shadow becomes consistent with the
detection of EHT for M87* in the region $ 0<k_{e}< 1.835 $. Figure \ref{Fig4}(c) displays the allowed regions of the
cosmological constant for which the obtained shadow is consistent
with M87* shadow. As we see, for rotating BHs in a high curvature
background, the resulting shadow is not agreement with EHT data.
For black holes located in a low curvature background, $ d_{sh} $
is consistent with observational data within $ 1\sigma $-error.
For intermediate values of the cosmological constant, $ d_{sh} $
is located in $ 2\sigma $ confidence region for all values of the
rotation parameter. To determine
allowed regions of the parameter $e_{1}$, we have plotted Fig.
\ref{Fig4}(d), showing that for very small values of $e_{1}$, the resulting shadow of slowly (fast) rotating BHs is located in $ 1\sigma $ ($ 2\sigma $)confidence (see the solid curve of Fig.
\ref{Fig4}(d)), whereas for intermediate and large values of this parameter, $
d_{sh} $ is in agreement with observational data within $ 1\sigma$-error (see dash and dash-dotted curves of Fig.
\ref{Fig4}(d)).

\begin{figure}[!htb]
\centering
\subfloat[$k_{e}=0.2 $, $d_{1}=0.5 $]{
        \includegraphics[width=0.28\textwidth]{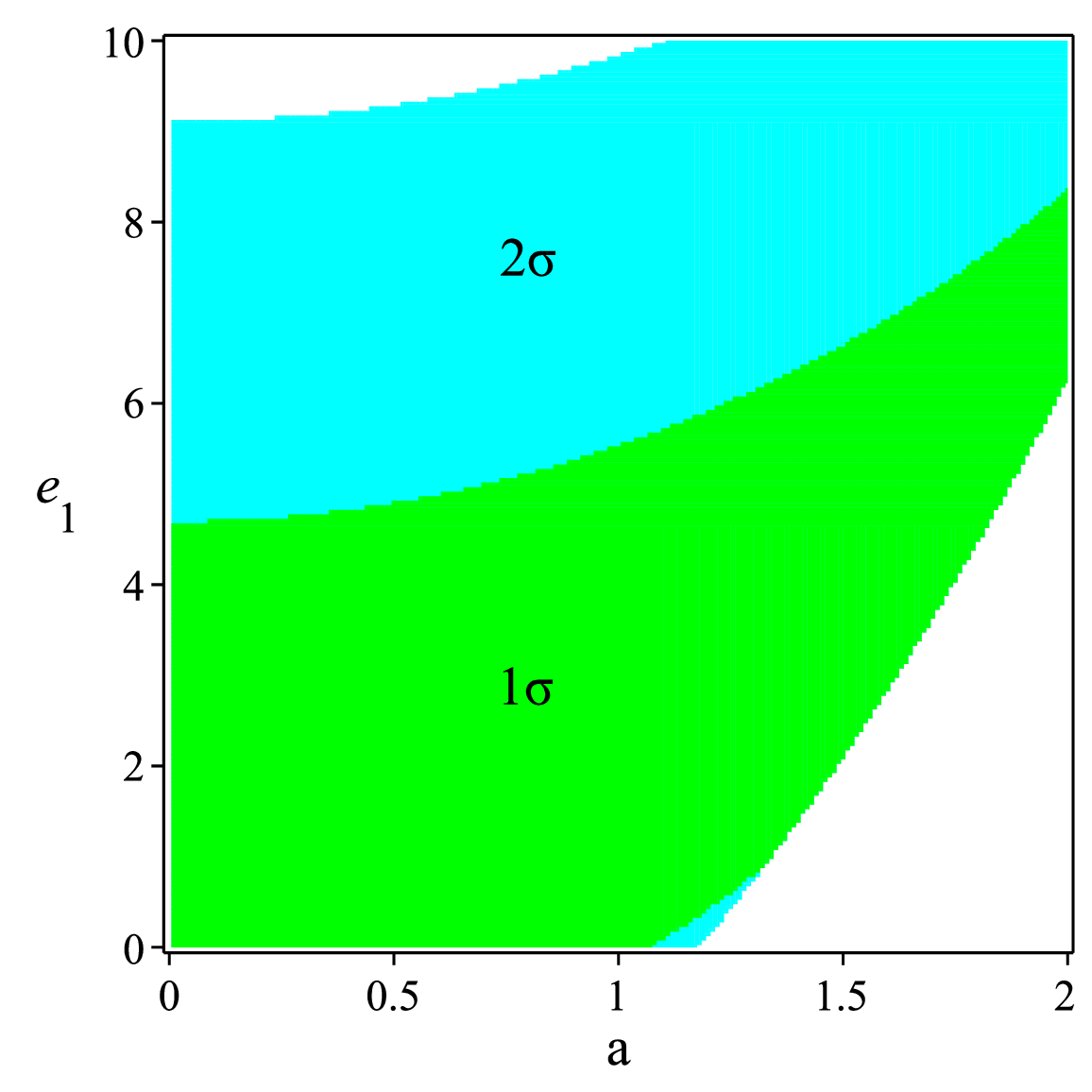}}
\subfloat[$e_{1}=d_{1}=0.5 $]{
        \includegraphics[width=0.28\textwidth]{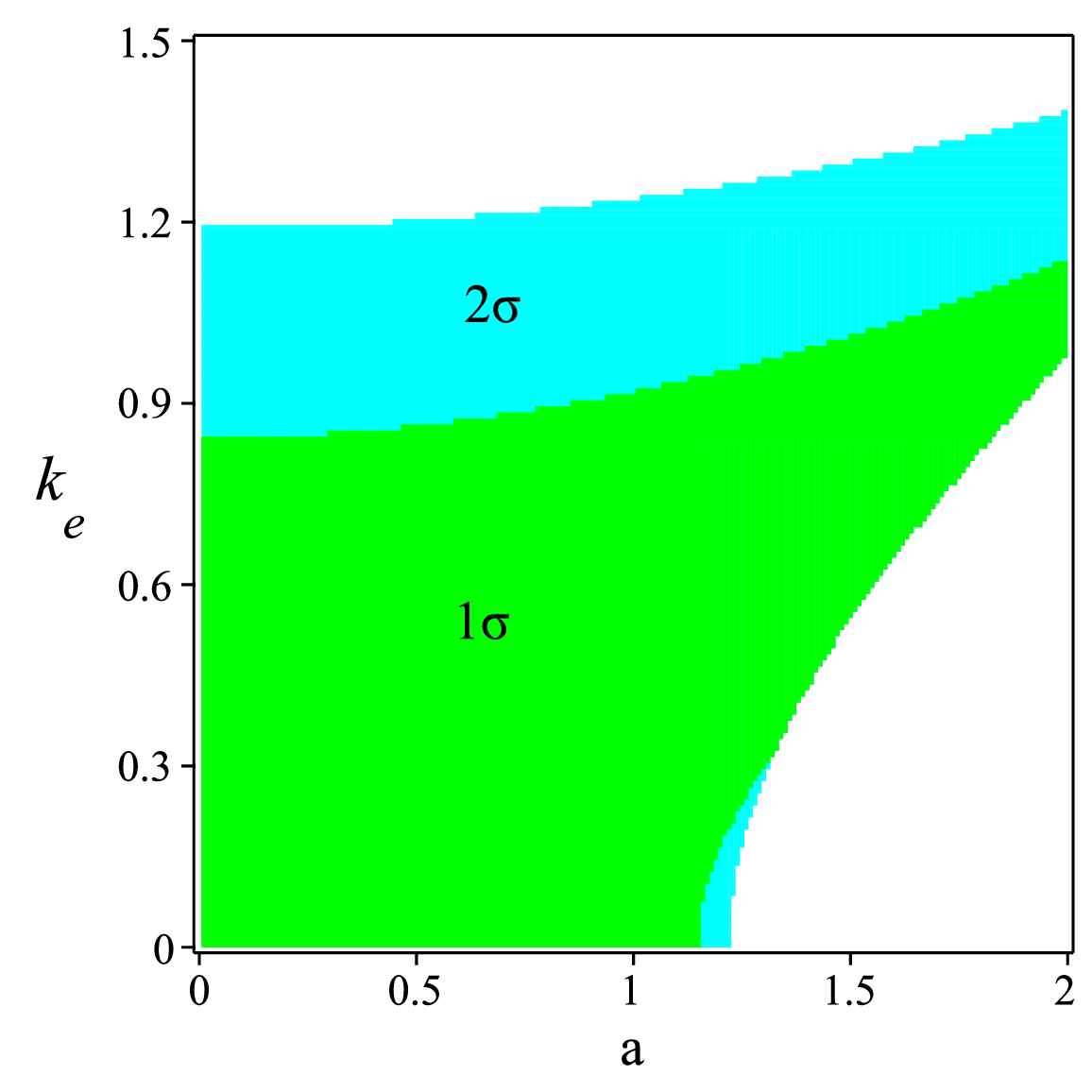}}
        \newline
\subfloat[$k_{e}=0.2 $, $d_{1}=-0.5 $]{
        \includegraphics[width=0.28\textwidth]{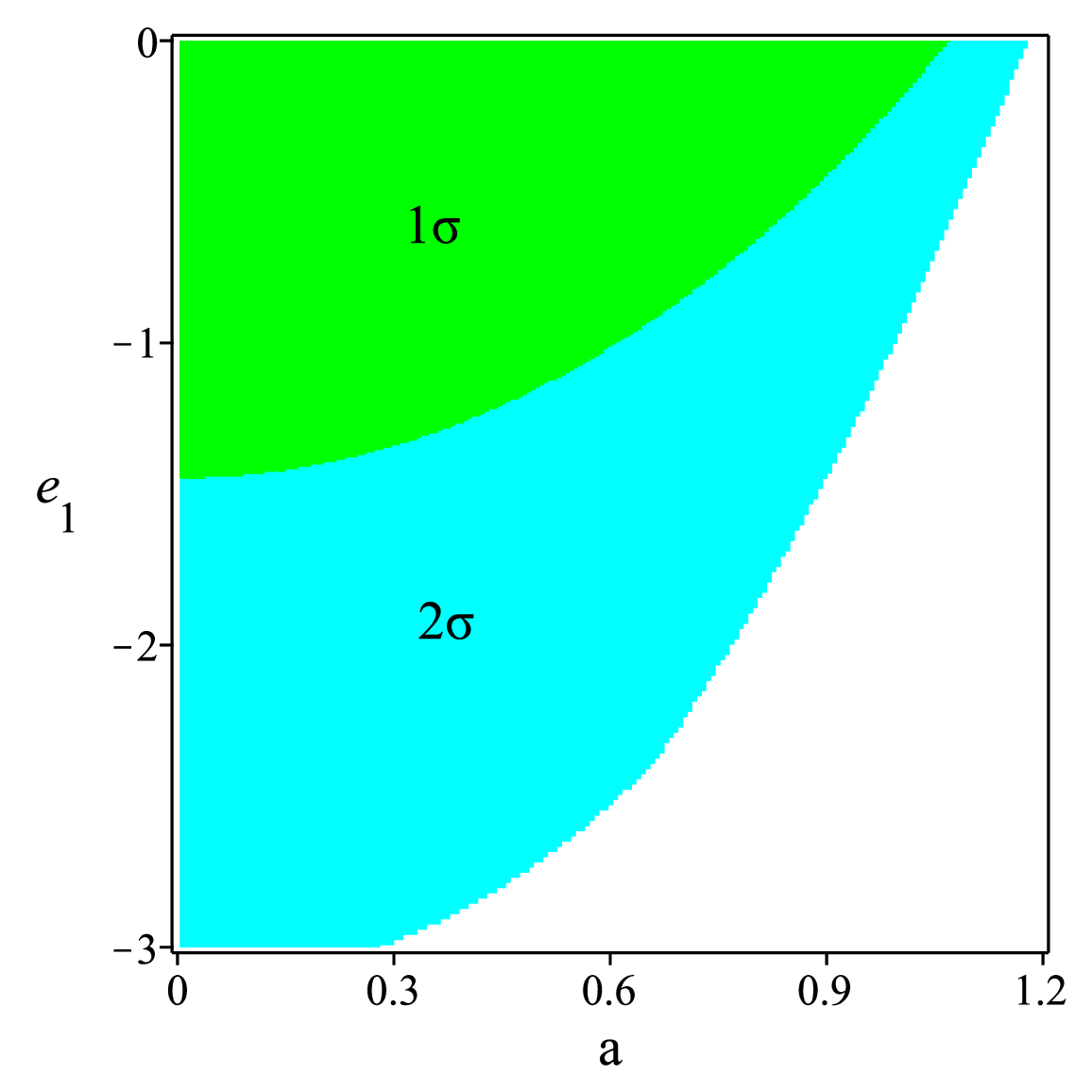}}
        \subfloat[$e_{1}=d_{1}=-0.5 $]{
        \includegraphics[width=0.28\textwidth]{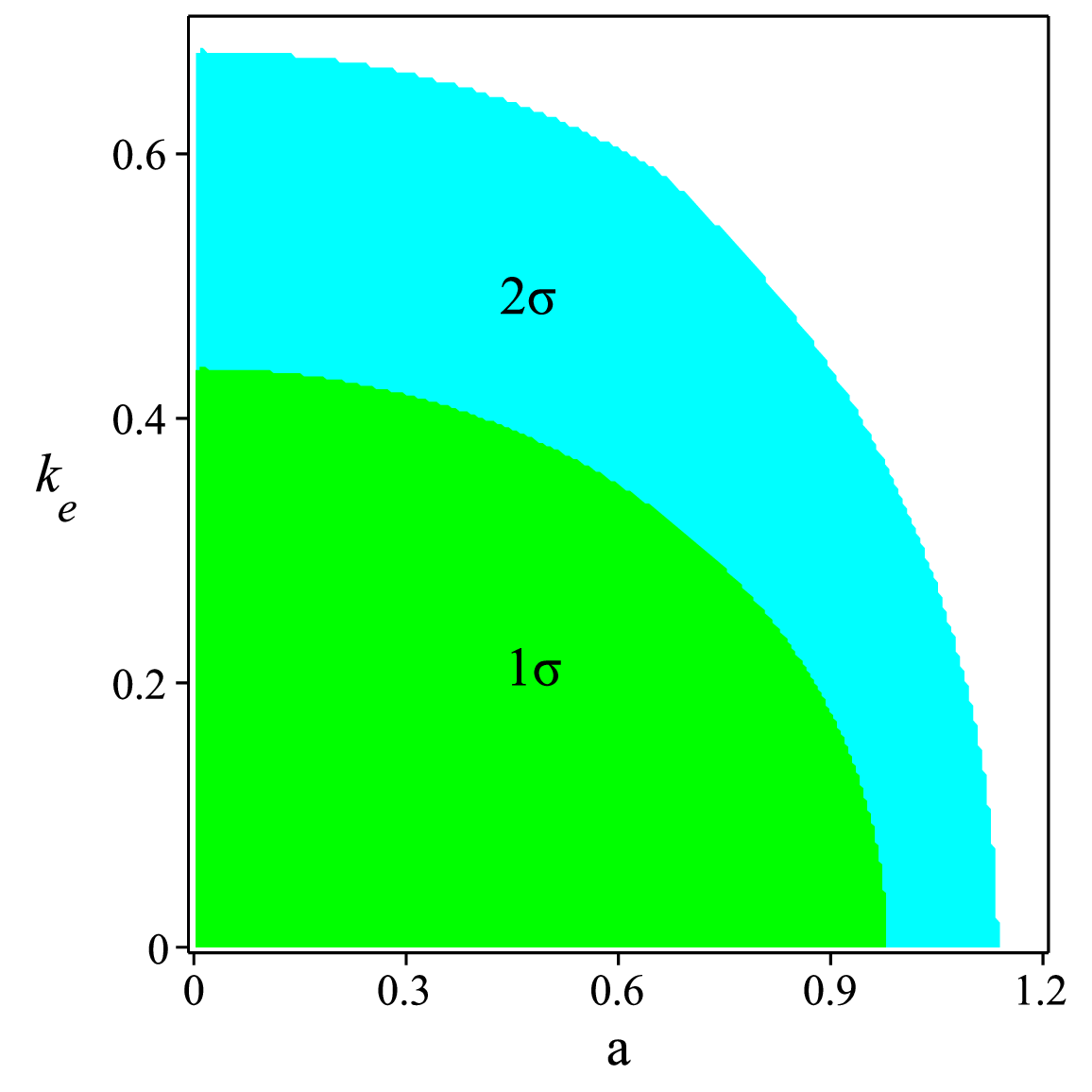}}
        \newline
\caption{ constraints on the Weyl-Cartan parameters and rotation parameter for $ k_{s}=10^{-6} $, $k_{m}=0.2 $,  $M=1$ and $ \theta_{0}=0^{\circ}$ with the EHT
observations of M87*.}
\label{FigEHT0}
\end{figure}

Fig. \ref{Fig4a} displays the variation of $ d_{sh} $ under change of BH parameters with the
inclination angle $ \theta_{0}=0^{\circ}$ for case II. According to Fig. \ref{Fig4a}(a), fast-rotating BHs located in a weak electric field have a compatible shadow within $2\sigma $ uncertainty, whereas these BHs with powerful electric charge have no compatible shadow with EHT data (see dash-dot curve in Fig. \ref{Fig4a}(a)). For small and intermediate values of the
rotation parameter, $ d_{sh} $ is located in $ 2\sigma $ confidence region for all values of the electric charge (see dash and solid curves in Fig.
\ref{Fig4a}(a)). From Fig. \ref{Fig4a}(b), one can find the allowed region of the electric dilation parameter to have  consistent results
with EHT data. According to dash-dot curve of this
figure, for intermediate values of this parameter, $ d_{sh} $ is  in
agreement with observational data within $ 2\sigma $-error for $ a \leq 0.685 $, otherwise it is not in agreement with EHT data.  This figure also shows that for rotating black holes with very small values of $ k_{e} $, the resulting shadow becomes consistent with the
detection of EHT within $ 2\sigma $-error all the time (see solid and dash curves of Fig. \ref{Fig4a}(b)). To find the allowed regions of the
cosmological constant, we have plotted Figure \ref{Fig4a}(c). According to our analysis, a compatible result occurs in the region $ -0.061<\Lambda<0 $ which shows that the allowed region of the cosmological constant is less than that of the case I. Fig. \ref{Fig4a}(d) illustrates the allowed regions of $e_{1}$ for which the obtained shadow is consistent with M87* shadow. As is seen, for small and intermediate values of $\vert e_{1} \vert$ the resulting shadow is in agreement with M87* data in $ 2\sigma $-error confidence region, whereas for large values of this parameter, a compatible $d_{sh} $ is observed for $ a \leq 0.582 $. To have a better understanding of the effects of Weyl-Cartan geometry, we set $q_{e}=q_{m}=\Lambda=0$ and determine the allowed regions of WC parameters to have consistent results. Comparing Fig. \ref{FigEHT0}(a) to Fig. \ref{Fig4}(d), one can find that the allowed region of $e_{1}$ parameter decreases in the absence of the electric and magnetic charges and cosmological constant for case I. While for case II, the allowed region of $e_{1}$ increases (compare Fig. \ref{FigEHT0}(c) to Fig. \ref{Fig4a}(d)). From Figs. \ref{FigEHT0}(b) and \ref{Fig4}(b), it is clear that the allowed region of $k_{e}$ parameter decreases for case I in the absence of $q_{e}$, $q_{m}$ and $\Lambda$ as well. While the opposite behavior is observed for case II (compare Fig. \ref{FigEHT0}(d) to Fig. \ref{Fig4a}(b)).

Now we study two other observables which are useful to determine the allowed regions of parameters. According to results obtained in \cite{Akiyama:2019a,Akiyama:2019b,Akiyama:2019c},  the image of supermassive BH M87* photographed by the EHT is crescent shaped, and the EHT observations constrain the deviation from circularity as $ \Delta C \lesssim 0.1 $ and the axis ratio as $1 < D_{x} \lesssim 4/3$. To define the circularity deviation $\Delta C$, one has to describe the  boundary of a black hole shadow in the polar coordinates
\begin{equation}
\begin{split}
\phi=\tan^{-1}\left(\frac{y-y_c}{x-x_c}\right),~~~~
\mathbb{R}(\phi)=\sqrt{(x-x_c)^2+(y-y_c)^2},
\end{split}
\end{equation}
in which the geometric center of the black hole shadow is determined by the edges of the shaped boundary via $({x_c=\frac{x_r+x_l}{2}}, y_c=0)$. The average radius of the shadow can be calculated as
\begin{equation}
\bar{\mathbb{R}}^2=\frac{1}{2\pi}\int^{2\pi}_{0} \mathbb{R}(\phi)^2d\phi.
\end{equation}
the deviation of the shadow from circularity can be defined as
\begin{equation}\label{eq-DeltaC}
{\Delta C=\frac{1}{\bar{\mathbb{R}}}\sqrt{\frac{1}{2\pi}\int^{2\pi}_{0}(\mathbb{R}(\phi)-\bar{\mathbb{R}})^2 d\phi}},
\end{equation}

and the axis ratio is given as \cite{Kumar:2020x,Banerjee:2020y}
\begin{equation}\label{eq-Dx}
D_x=D^{-1}=\frac{y_t-y_b}{x_r-x_l}.
\end{equation}

To compare with the EHT constraints on $ \Delta C $ and $ D_x $ in M87*, we have listed several values of these two quantities in Table \ref{M87} for both cases I and II. To better refer to the EHT observations of  M87*, we have considered
$ \theta_{0}=17^{\circ}$  since the jet inclination with respect to the line-of-sight for M87* is estimated to be $ 17^{\circ} $ \cite{Walker:2018h}. As can be seen from this table,  the EHT observations $ \Delta C \lesssim 0.1 $ is satisfied for the whole parameters of both cases. According to our findings, the condition  $1 < D_{x} \lesssim 4/3$ imposes a constraint on $k_{e}$ for case I such that this condition is satisfied for only $ 0\leq k_{e}\leq 0.6 $. This condition also imposes a constraint on the cosmological constant for both cases. From the fifth row of this table, one can find that the condition  $1 < D_{x} \lesssim 4/3$ is satisfied in the range $ -0.03 \leq\Lambda \leq 0 $ ($ -0.04 \leq \Lambda \leq 0 $) for case I (II).  

The recent EHT papers on M87* observation have estimated the Schwarzschild shadow deviation $(\delta)$ which measures the difference between the model shadow diameter ($ d_{metric} $) and the Schwarzschild shadow diameter and is given by \cite{sgr1}
\begin{equation}\label{eq-Dx}
\delta=\frac{d_{metric}}{6\sqrt{3}}-1,
\end{equation}
where $ d_{metric}=2R_{a} $ with $ R_{a}=\sqrt{A/\pi} $ and $ A $ is obtained by Eq.(\ref{AD}). Evidently, $ \delta $ can be positive (negative) if the black hole shadow size is greater (smaller) than the Schwarzschild black hole of the same mass. According to results reported by the EHT observations, the bound of the measured Schwarzschild deviation is as $ \delta=-0.01 \pm 0.17 $ \cite{Akiyama:2019b,Afrin:2023a}. Several values of $ \delta $ are reported in table \ref{M87}  which illustrates that for the whole parameters of both cases, the resulting shadow is smaller than the Schwarzschild black hole shadow. For case I, all values of the rotation parameter, electric dilation parameter, and parameter $e_{1}$ satisfy the $ 1\sigma $ bound. Regarding the electric charge and cosmological constant, only the range $ 0 <q_{e}<0.6 $ and $ -0.025 \leq \Lambda \leq 0 $ can satisfy the $ 1\sigma $ bound,  otherwise they satisfy the $ 2\sigma $ bound. For case II, the $ 1\sigma $ and $ 2\sigma $ bounds can be satisfied in the following ranges 
\begin{eqnarray*}
&&~~~~~1\sigma ~~~~~~~~~~~~~~~~~~~~~~~~~~~~~~~ ~~~2\sigma \\
&&0\leq a < 0.5 ~~~~~~~~~~~~~~~~~~~~~ 0.5 \leq a \leq 0.8\\
&&0\leq q_{e}< 0.2 ~~~~~~~~~~~~~~~~~ ~~~0.2 \leq q_{e}\leq 0.7  \\
&&0\leq k_{e}< 0.2 ~~~~~~~~~~~~~~~~~~~~ 0.2 \leq k_{e}\leq 0.5\\
&&-0.5< e_{1} < 0~~~~~~~~~~~ ~~~ ~~~-2 \leq e_{1} \leq -0.5 \\
&&-0.02 < \Lambda <0~~~~~~ ~~~~~~~ ~~~-0.04 \leq \Lambda \leq -0.02 
\end{eqnarray*}

\begin{table*}[t]
\caption{ The circularity deviation $ \Delta C $, axial ratio $ D_{x} $, distortion $ \delta_{s} $ and fractional deviation parameter $ \delta $ for the variation of $a$, $q_{e}$, $k_{e} $, $e_{1}$ and $\Lambda$ for $ q_{m}=k_{m}=0.2 $, $ k_{s}=10^{-6} $,  $M =1$ and and $ \theta_{0}=17^{\circ}
$.}
      \centering
    \begin{tabular}{|c|c|c|c|c| |c|c|c|c|c|}
    \hline
         \multicolumn{5}{|c|}{Case I}& \multicolumn{5}{|c|}{Case II} \\
    \hline  
    \multicolumn{5}{|c|}{ $ q_{e}=k_{e}=0.2 $, $ e_{1} =0.5$ and $ \Lambda =-0.02 $}& \multicolumn{5}{|c|}{$ q_{e}=k_{e}=0.2 $, $ e_{1} =-0.5$ and $ \Lambda =-0.02 $} \\
    \hline  
         $  a$  & $ \Delta C $  & $ D_{x} $  &$ \delta_{s} $& $ \delta $ & $  a$  & $ \Delta C $  &  $ D_{x} $  &$ \delta_{s} $& $ \delta $ \\ \hline  
          $ 0.2 $ & $ 3.34\times 10^{-5} $ & $ 1.00009 $ &$ 1.89\times 10^{-4} $& $ -0.13547 $ &$ 0.1 $ & $ 1.61\times 10^{-5} $ & $ 1.00005 $ &$ 9.04\times 10^{-5} $& $ -0.16987 $  \\
           $ 0.4 $ & $ 1.58\times 10^{-4} $ & $ 1.00045 $ &$ 8.93\times 10^{-4} $& $ -0.13920 $ & $ 0.3 $ & $ 1.63\times 10^{-4} $ & $ 1.00045 $ &$ 9.05\times 10^{-4} $& $ -0.17310 $ \\
           $ 0.6 $ & $ 4.71\times 10^{-4} $ & $ 1.0013 $ &$ 2.64\times 10^{-3} $& $ -0.14592 $ &  $ 0.5 $ & $ 5.58\times 10^{-4} $ &$ 1.0015 $ &$ 3.14\times 10^{-3} $& $ -0.18006 $ \\
           $ 0.8 $ & $ 1.25\times 10^{-3} $&$ 1.0035 $ &$ 6.97\times 10^{-3} $& $ -0.15666 $ &  $ 0.7 $ & $ 1.60\times 10^{-3} $ & $ 1.0045 $ &$ 8.99\times 10^{-3} $& $ -0.19215 $  \\
           $ 1.0 $ & $ 3.69\times 10^{-3} $ & $ 1.0102 $ &$ 2.02\times 10^{-2} $& $ -0.17423 $ & $ 0.8 $ &$ 2.83\times 10^{-3} $ & $ 1.0079 $ &$ 1.57\times 10^{-2} $& $ -0.20122 $ \\
    \hline \hline 
    \multicolumn{5}{|c|}{$ a=0.5 $, $  k_{e}=0.2 $, $ e_{1} =0.5$ and $ \Lambda =-0.02 $}& \multicolumn{5}{|c|}{$ a=0.5 $, $  k_{e}=0.2 $, $ e_{1} =-0.5$ and $ \Lambda =-0.02 $} \\
    \hline
    $  q_{e}$ & $ \Delta C $  & $ D_{x} $  &$ \delta_{s} $& $ \delta $ & $  q_{e}$  & $ \Delta C $  &  $ D_{x} $  &$ \delta_{s} $& $ \delta $ \\ \hline  
         $ 0.1 $ & $ 2.62\times 10^{-4} $ & $ 1.0007 $ &$ 1.47\times 10^{-3} $& $ -0.13895 $ & $ 0.1 $ & $ 5.22\times 10^{-4} $ & $ 1.0014 $ &$ 2.94\times 10^{-3} $& $ -0.17615 $ \\
  $ 0.3 $ & $ 3.13\times 10^{-4} $ & $ 1.0008 $ &$ 1.76\times 10^{-3} $& $ -0.14759 $ & $ 0.3 $ & $ 6.24\times 10^{-4} $ & $ 1.0017 $ &$ 3.51\times 10^{-3} $& $ -0.18677 $ \\
   $ 0.5 $ & $ 4.40\times 10^{-4} $ & $ 1.0012 $ &$ 2.47\times 10^{-3} $&$ -0.16613 $ & $ 0.5 $ & $ 9.17\times 10^{-4} $ & $ 1.0025 $ &$ 5.15\times 10^{-3} $& $ -0.21020 $ \\   
  $ 0.7 $ & $ 7.51\times 10^{-4} $ & $ 1.0021 $ &$ 4.22\times 10^{-3} $& $ -0.19807 $ & $ 0.6 $ & $ 1.25\times 10^{-3} $ & $ 1.0035 $ &$ 7.02\times 10^{-3} $& $ -0.22856 $\\   
   $ 0.9 $ & $ 1.99\times 10^{-3} $ & $ 1.0056 $ &$ 1.11\times 10^{-2} $& $ -0.25385 $ & $ 0.7 $ & $ 1.99\times 10^{-3} $ &$ 1.0056 $ &$ 1.11\times 10^{-2} $& $ -0.25385 $ \\ \hline \hline 
   \multicolumn{5}{|c|}{$ a=0.5 $, $  q_{e}=0.2 $, $ e_{1} =0.5$ and $ \Lambda =-0.02 $}& \multicolumn{5}{|c|}{$ a=0.5 $, $  q_{e}=0.2 $, $ e_{1} =-0.5$ and $ \Lambda =-0.02 $} \\
    \hline
   $  k_{e}$  & $ \Delta C $  & $ D_{x} $  &$ \delta_{s} $& $ \delta $ & $  k_{e}$  & $ \Delta C $  &  $ D_{x} $  &$ \delta_{s} $& $ \delta $ \\ \hline  
         $ 0.01 $ & $ 3.33\times 10^{-4} $ & $ 1.00094 $ &$ 1.87\times 10^{-3} $& $ -0.15091 $ & $ 0.1 $ & $ 4.89\times 10^{-4} $ & $ 1.0013 $ &$2.75\times 10^{-3} $& $ -0.17233 $ \\
  $ 0.2 $ & $ 2.80\times 10^{-4} $ & $ 1.00079 $ &$ 1.57\times 10^{-3} $& $-0.14214  $ & $ 0.2 $ & $ 5.58\times 10^{-4} $ & $ 1.0015 $ &$ 3.14\times 10^{-3} $& $ -0.18006 $ \\
   $ 0.4 $ & $ 1.54\times 10^{-4} $ & $ 1.00045 $ &$ 8.93\times 10^{-4} $& $ -0.11793 $ & $ 0.3 $ & $ 7.10\times 10^{-4} $ & $ 1.0019 $ &$ 3.93\times 10^{-3} $& $ -0.19374 $ \\   
  $ 0.6 $ & $ 3.15\times 10^{-5} $ & $ 1.00008 $ &$ 1.59\times 10^{-4} $& $ -0.08305 $ & $ 0.4 $ & $ 9.93\times 10^{-4} $ & $ 1.0028 $ &$ 5.58\times 10^{-3} $&$ -0.21499 $ \\   
   $ 0.7 $ & $ 3.16\times 10^{-5} $ & $ 0.99996 $ &$ -1.67\times 10^{-4} $& $ -0.06313 $ & $ 0.5 $ & $ 1.76\times 10^{-3} $ &$ 1.0049 $ &$ 9.89\times 10^{-3} $& $ -0.24757 $\\ \hline \hline 
   \multicolumn{5}{|c|}{$ a=0.5 $, $  q_{e}=k_{e}=0.2 $ and $ \Lambda =-0.02 $}& \multicolumn{5}{|c|}{$ a=0.5 $, $  q_{e}=k_{e}=0.2 $ and $ \Lambda =-0.02 $} \\
    \hline
   $  e_{1}$ & $ \Delta C $  & $ D_{x} $  &$ \delta_{s} $& $ \delta $ & $  e_{1}$  & $ \Delta C $  &  $ D_{x} $  &$ \delta_{s} $& $ \delta $ \\ \hline  
         $ 0.01 $ & $ 3.93\times 10^{-4} $ & $ 1.00111 $ &$ 2.21\times 10^{-3} $& $ -0.15976 $ & $ -0.01 $ & $ 3.98\times 10^{-4} $ & $ 1.0011 $ &$ 2.24\times 10^{-3} $& $ -0.16052 $ \\
  $ 0.4 $ & $ 3.01\times 10^{-4} $ & $ 1.00085 $ &$ 1.69\times 10^{-3} $& $ -0.14561 $ & $ -0.4 $ & $ 5.20\times 10^{-4} $ & $ 1.0014 $ &$ 2.93\times 10^{-3} $& $ -0.17590 $\\
   $ 0.8 $ & $ 2.26\times 10^{-4} $ & $ 1.00064 $ &$ 1.27\times 10^{-3} $& $ -0.13210 $ & $ -0.8 $ & $ 6.93\times 10^{-4} $ & $ 1.0019 $ &$ 3.90\times 10^{-3} $& $ -0.19317 $ \\   
  $ 1.2 $ & $ 1.66\times 10^{-5} $ & $ 1.00074 $ &$ 9.32\times 10^{-4} $& $ -0.11945 $ & $ -1.2 $ & $ 9.51\times 10^{-4} $ & $ 1.0026 $ &$ 5.35\times 10^{-3} $& $ -0.21242 $ \\   
   $ 1.6 $ & $ 1.16\times 10^{-4} $ & $ 1.00032 $ &$ 6.49\times 10^{-4} $& $ -0.10755 $ & $ -1.6$ & $ 1.38\times 10^{-3} $ &$ 1.0038 $ &$ 7.76\times 10^{-3} $& $ -0.23436 $ \\ \hline \hline 
   \multicolumn{5}{|c|}{$ a=0.5 $, $  q_{e}=k_{e}=0.2 $ and $ e_{1} =0.5$}& \multicolumn{5}{|c|}{$ a=0.5 $, $  q_{e}=k_{e}=0.2 $ and $ e_{1} =-0.5$} \\
    \hline
   $  \Lambda$  & $ \Delta C $  & $ D_{x} $  &$ \delta_{s} $& $ \delta $ & $  \Lambda$  & $ \Delta C $  &  $ D_{x} $  &$ \delta_{s} $& $ \delta $ \\ \hline  
         $ -0.01 $ & $ 4.20\times 10^{-4} $ & $ 1.00119 $ &$ 2.37\times 10^{-3} $& $ -0.07930 $ & $ -0.01 $ & $ 6.78\times 10^{-4} $ & $ 1.00191 $ &$ 3.82\times 10^{-3} $& $ -0.12572 $ \\
  $ -0.02 $ & $ 2.80\times 10^{-4} $ & $ 1.00079 $ &$ 1.57\times 10^{-3} $& $ -0.14214 $ & $ -0.02 $ & $ 5.58\times 10^{-4} $ & $ 1.00157 $ &$ 3.14\times 10^{-3} $& $-0.180062  $ \\
   $ -0.03 $ & $ 9.18\times 10^{-5} $ & $ 1.00024 $ &$ 4.88\times 10^{-4} $& $ -0.19361 $ & $ -0.03 $ & $ 3.82\times 10^{-4} $ & $ 1.00107 $ &$ 2.13\times 10^{-3} $& $ -0.22533 $\\   
  $ -0.04 $ & $ 1.40\times 10^{-5} $ & $ 0.99961 $ &$ -7.74\times 10^{-4} $&$ -0.23676 $ & $ -0.04 $ & $ 1.75\times 10^{-4} $ & $ 1.00047 $ &$ 9.37\times 10^{-4} $& $ -0.26381 $ \\   
   $ -0.05$ & $ 3.79\times 10^{-4} $ & $ 0.99893 $ &$ -2.13\times 10^{-3} $& $ -0.27364 $ & $ -0.05$ & $ 8.98\times 10^{-5} $ &$ 0.99980 $ &$- 3.82\times 10^{-4} $& $ -0.29706 $ \\ \hline   
    \end{tabular} 
\vspace{1ex}
\label{M87}
\end{table*}

According to our analysis, increasing the rotation parameter and electric charge leads to decreasing the diameter of the shadow and distortion parameter ($ \delta $) which is consistent with the results obtained of \cite{Kocherlakota:104047} in which authors considered Kerr-Newman BH in GR and 
constrained the electric charge and rotation
parameter with the aid of the EHT shadow observational results of M87*. Also in Ref. \cite{Meng:106}, it was shown that the circularity deviation $ \Delta C $ and axis ratio $D_{x}$ of Kerr-Newman BH in GR increase with the growth of the rotation parameter and electric charge which confirms the validity of our results.

\subsection{Constrains on the parameters of the model from the image of Sgr A*}
\label{IVB}
In this subsection, we are going to model Sgr A* as the rotating BHs in WCT and impose the EHT-inferred bounds on  $d_{sh}$,  $D$, and $\delta$ to find the constraints on the parameters. The results obtained from EHT observations showed that the Sgr A* BH shadow images have advantages for testing the nature of astrophysical black holes, as the mass of the Sgr A* black hole bridges the gap between the stellar black holes observed by the LIGO and the M87* black hole. Therefore, it can probe a significant curvature scale ($ 10^{6} $ orders of curvature higher than M87*). According to the EHT collaboration, the image of Sgr A* is consistent with an expected appearance of the Kerr black hole  with an inferred BH mass $ M=(4.3 \pm 0.013) \times 10^{6} M_{\odot} $ and distance $\mathbb{D}= 8277\pm 33 pc $ from earth \cite{sgr1}. In contrast to the M87* black hole, for which the EHT only measured the emission ring diameter, for Sgr A*, the EHT not only estimated the emission ring angular diameter $\theta_{d} = (51.8 \pm 2.3)\mu as $ but also could provide an estimate of the shadow diameter $\theta_{sh} = (48.7 \pm 7)\mu as$ \cite{Walia:2022s}. Using these reported expressions, the diameter of the shadow image for Sgr. A* is obtained as
\begin{equation}
d_{Sgr. A*}\equiv \frac{\mathbb{D}\theta_{sh}}{M}\approx 9.5 \pm 1.4.
\label{EqdSgrb}
\end{equation}

\begin{figure}[!htb]
\centering
\subfloat[$k_{e}=0.2 $, $e_{1}=0.5 $ and $\Lambda=-0.02 $]{
        \includegraphics[width=0.31\textwidth]{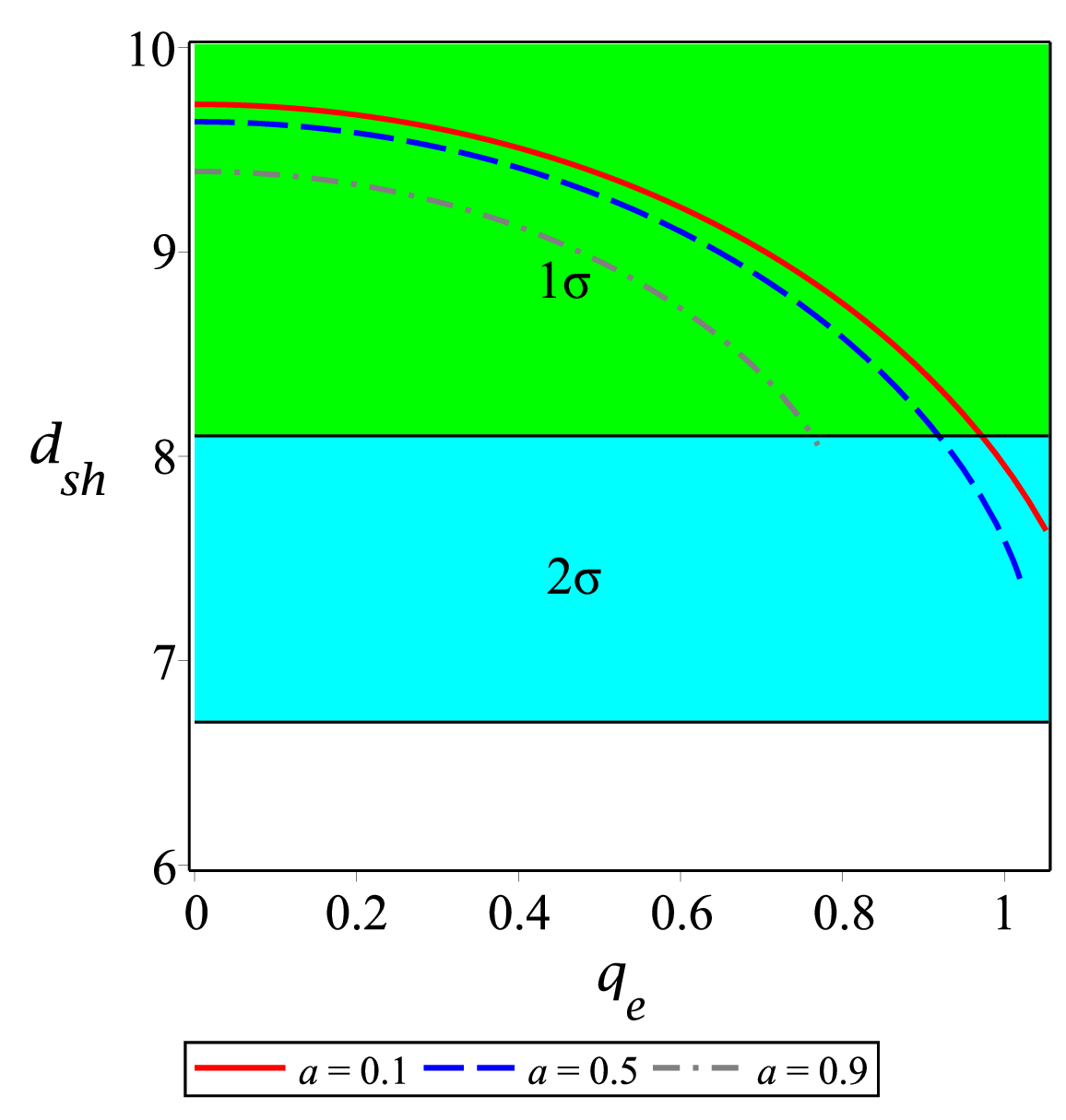}}
\subfloat[$q_{e}=0.2 $, $e_{1}=0.5 $ and $\Lambda=-0.02 $]{
        \includegraphics[width=0.322\textwidth]{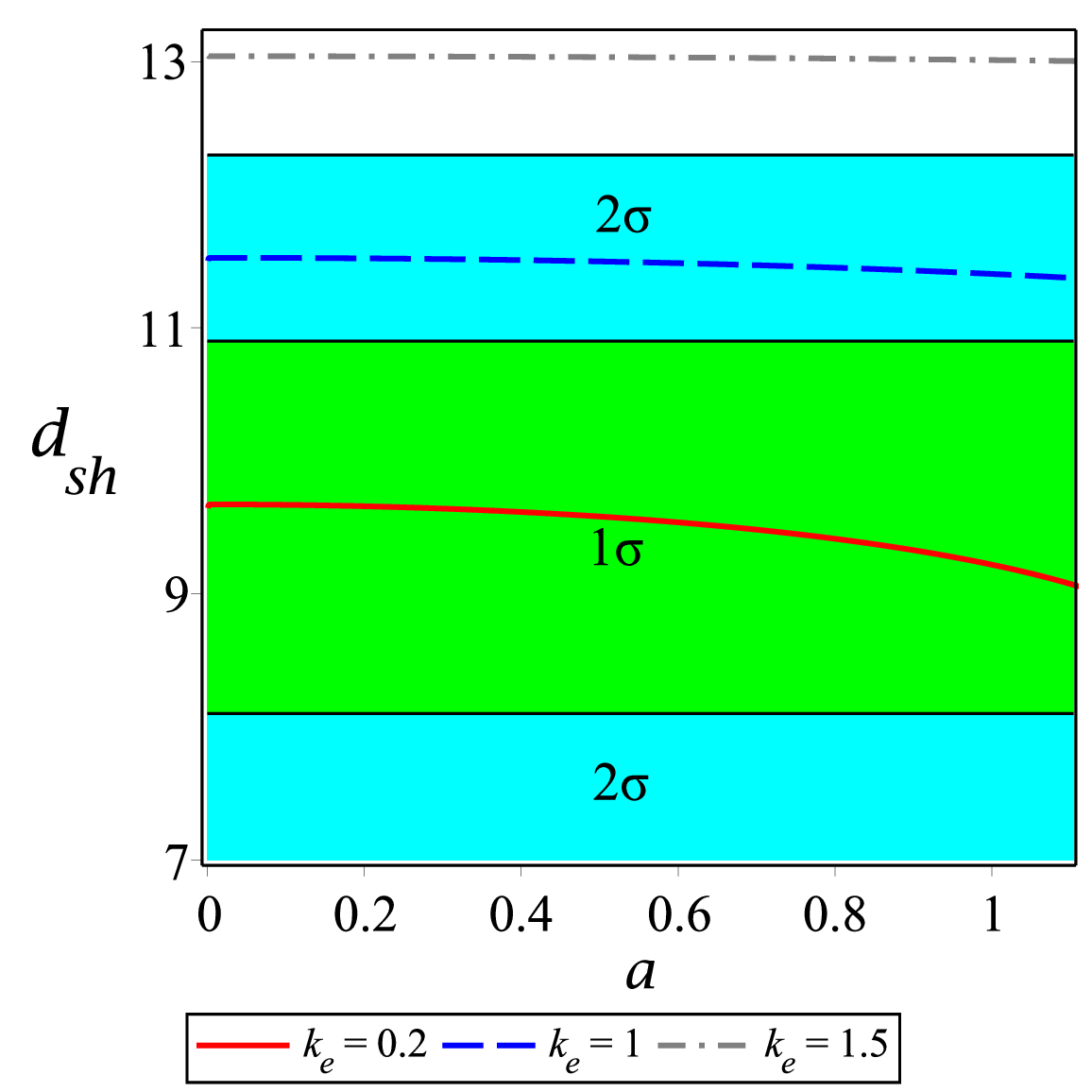}}
\newline
\subfloat[ $q_{e}=k_{e}=0.2 $ and $e_{1}=0.5 $]{
        \includegraphics[width=0.315\textwidth]{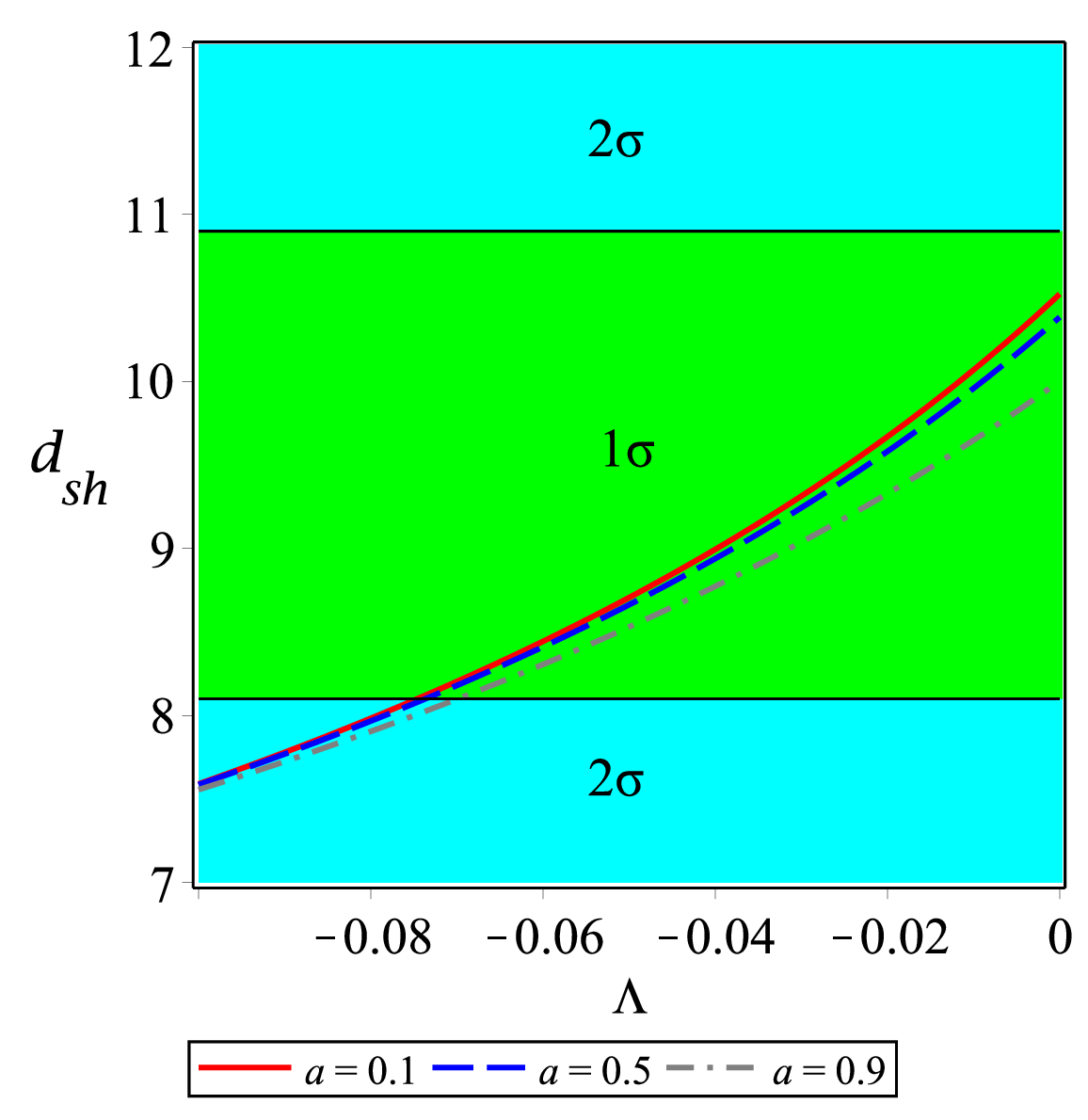}}
\subfloat[ $q_{e}=k_{e}=0.2 $ and $\Lambda=-0.02 $]{
        \includegraphics[width=0.32\textwidth]{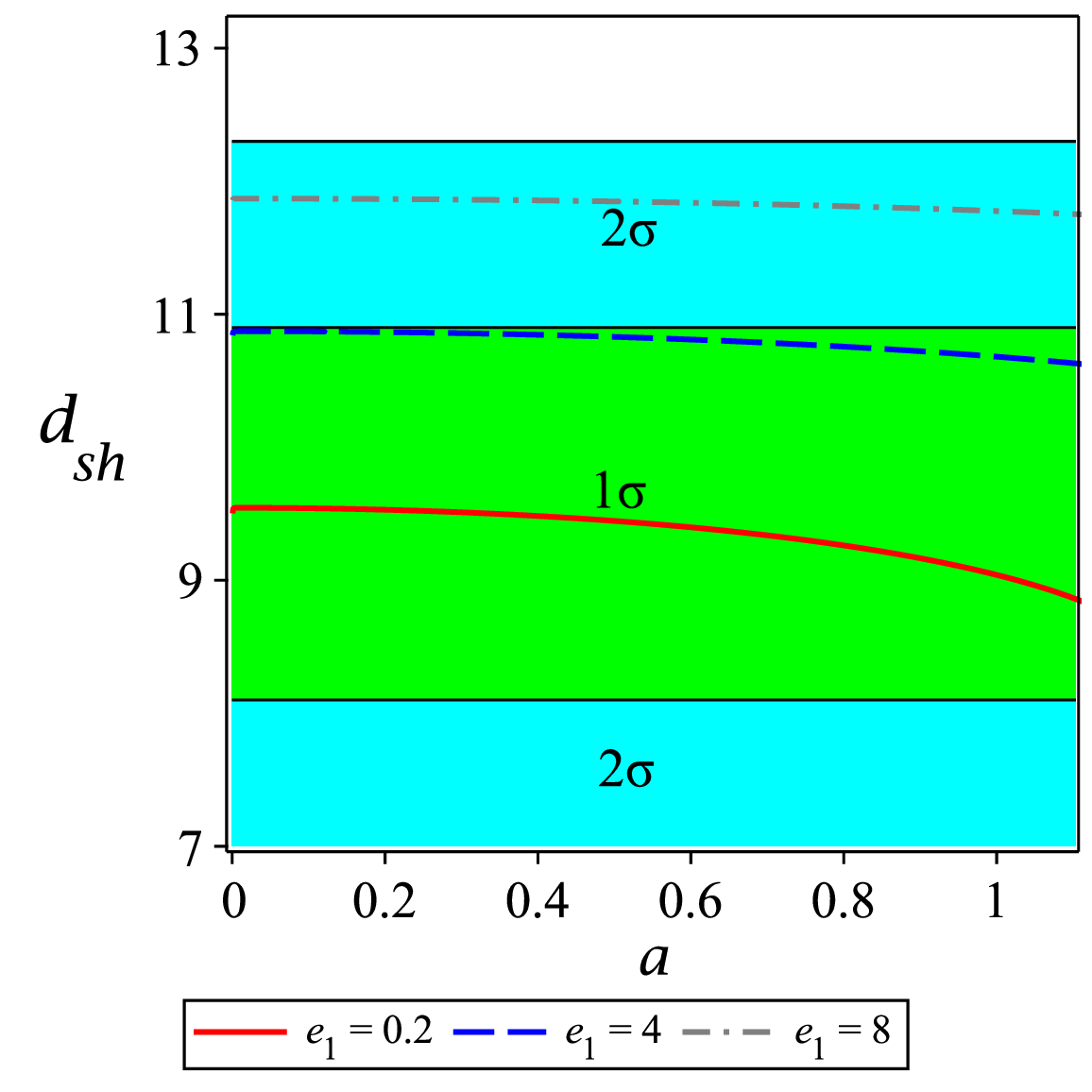}}
\newline
\caption{ The predicted diameter for a
charged rotating BH in WCT for case I with $ k_{s}=10^{-6} $, $ q_{m}=k_{m}=0.2 $, $d_{1}=0.5 $ and $M=1$ with the EHT
observations of Sgr A*. \textbf{Left-up graph:} $ d_{sh} $ as a function of $ q_{e} $ for fixed $ k_{e}$, $ e_{1}$, $ \Lambda
$, and for several values of the rotation parameter. \textbf{Right-up graph:} $ d_{sh} $ as a function of the rotation parameter, for
fixed $ q_{e}$, $ e_{1}$,  $ \Lambda $, and  different values of
$ k_{e}$. \textbf{Left-down graph:} $ d_{sh} $ as a
function of the cosmological constant, for fixed $ q_{e}$,
$ k_{e} $, $ e_{1}$, and different values of the rotation
parameter. \textbf{Right-down graph:} $ d_{sh} $ as a function of the rotation parameter, for
fixed $ q_{e}$, $ k_{e}$, $ \Lambda $, and  different values of
$ e_{1}$. The green
shaded region gives the $ 1\sigma $ confidence region for $ d_{sh}
$, whereas the cyan shaded region gives the $ 2\sigma $ confidence
region. The inclination angle is $ \theta_{0}=0^{\circ}$.}
\label{Fig5}
\end{figure}

\begin{figure}[!htb]
\centering
\subfloat[$k_{e}=0.2 $, $e_{1}=-0.5 $ and $\Lambda=-0.02 $]{
        \includegraphics[width=0.31\textwidth]{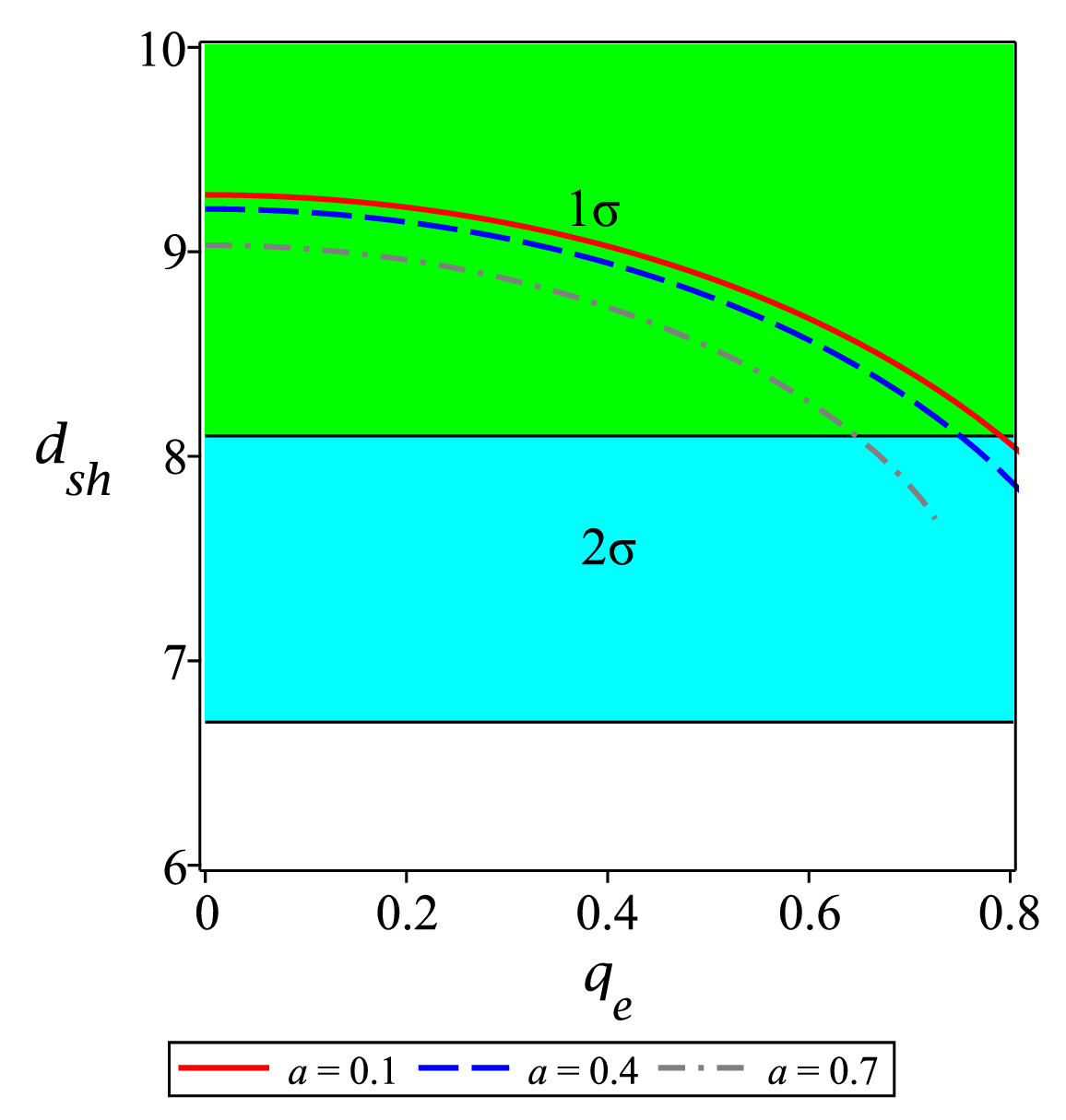}}
\subfloat[$q_{e}=0.2 $, $e_{1}=-0.5 $ and $\Lambda=-0.02 $]{
        \includegraphics[width=0.322\textwidth]{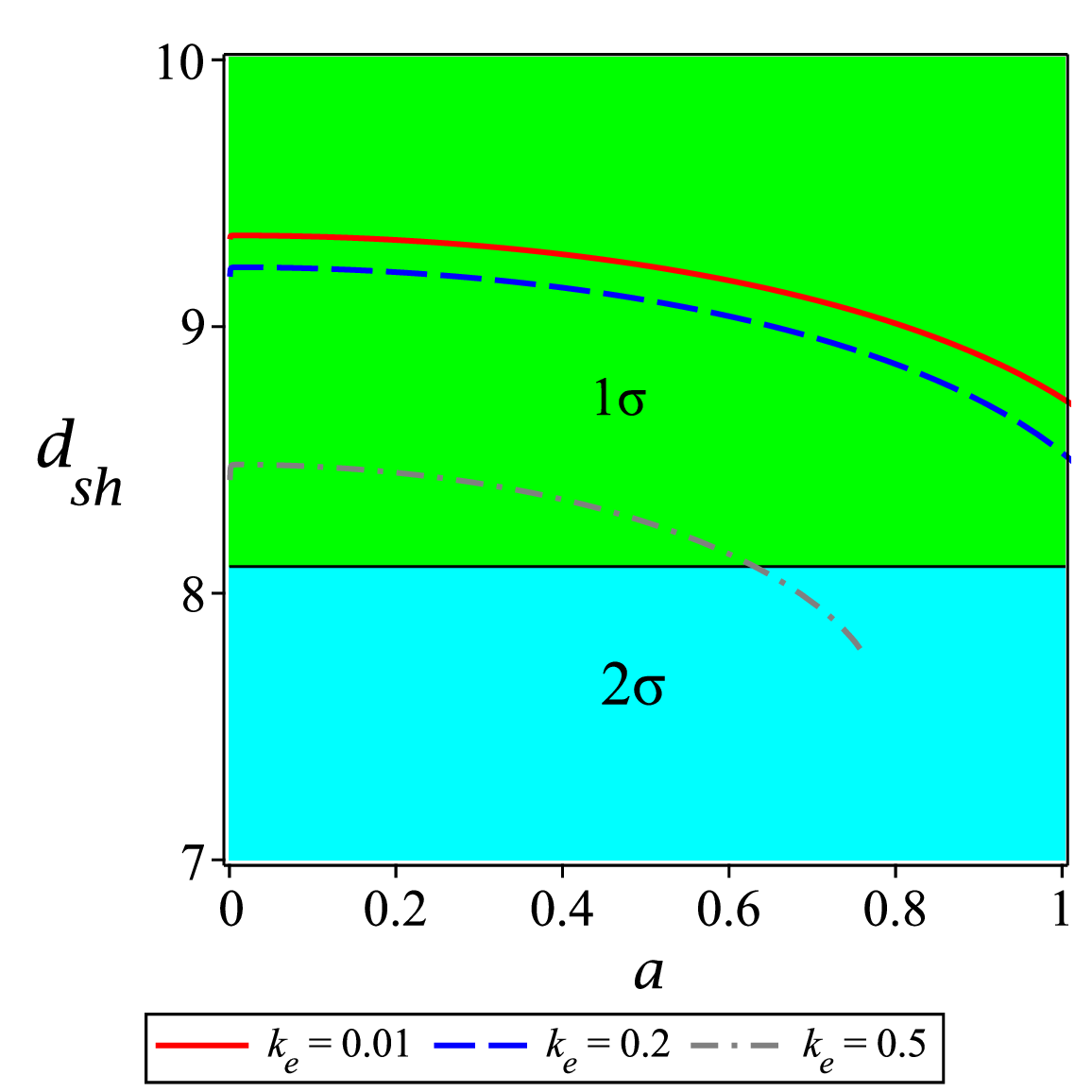}}
\newline
\subfloat[ $q_{e}=k_{e}=0.2 $ and $e_{1}=-0.5 $]{
        \includegraphics[width=0.315\textwidth]{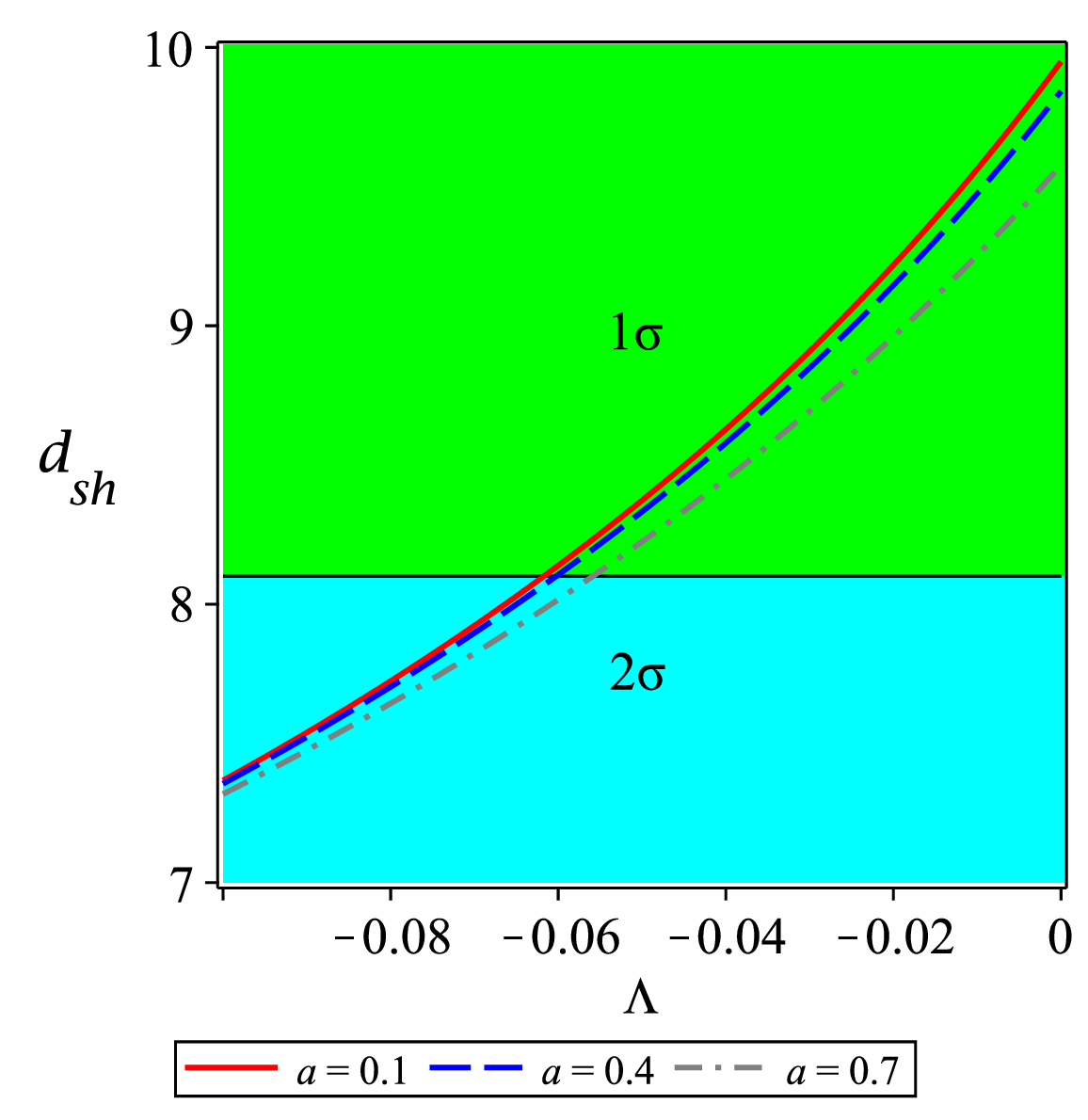}}
\subfloat[ $q_{e}=k_{e}=0.2 $ and $\Lambda=-0.02 $]{
        \includegraphics[width=0.32\textwidth]{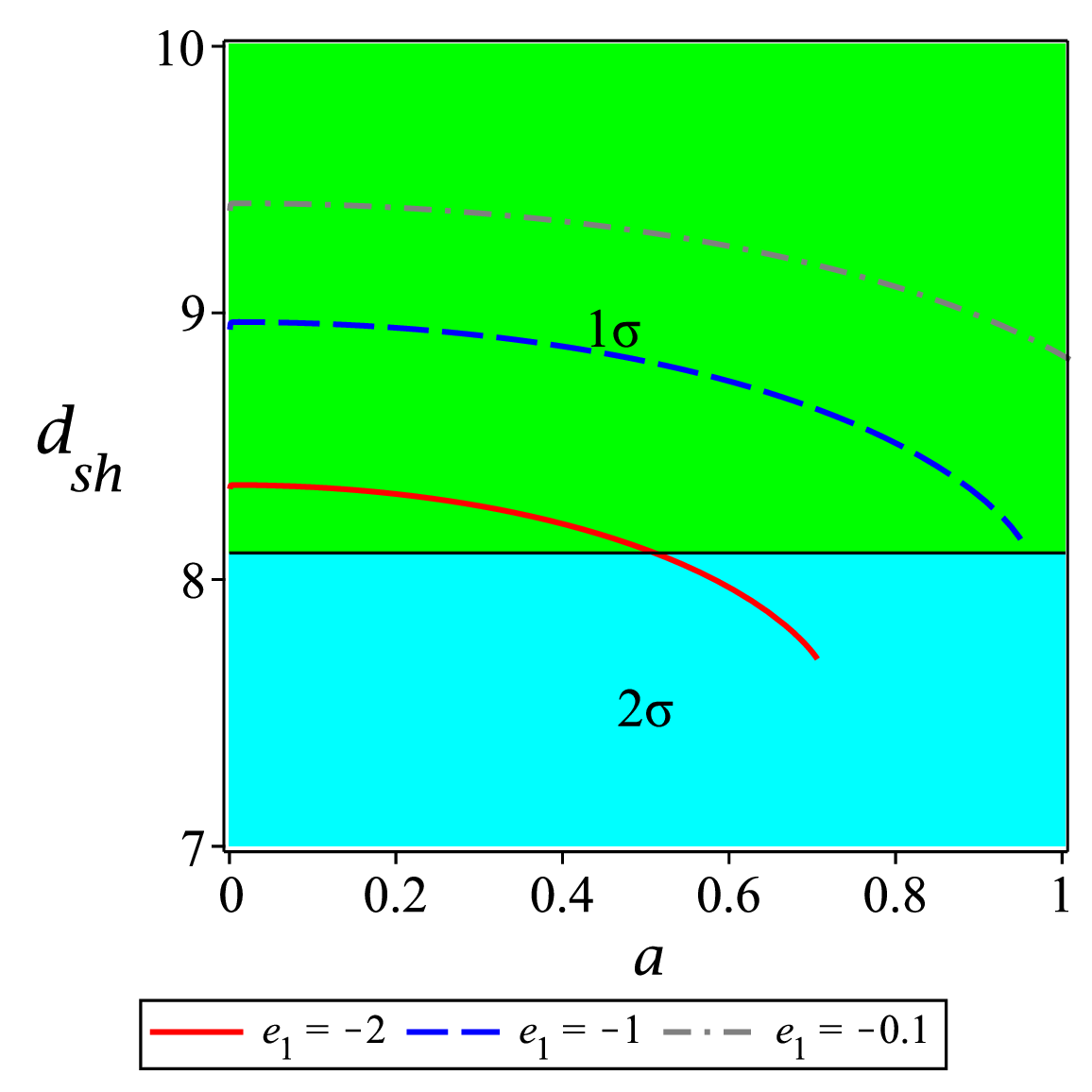}}
\newline
\caption{ The predicted diameter for a
charged rotating BH in WCT for case II with $ k_{s}=10^{-6} $, $ q_{m}=k_{m}=0.2 $, $d_{1}=-0.5 $ and $M=1$ with the EHT
observations of Sgr A*. \textbf{Left-up graph:} $ d_{sh} $ as a function of $ q_{e} $ for fixed $ k_{e}$, $ e_{1}$, $ \Lambda
$, and for several values of the rotation parameter. \textbf{Right-up graph:} $ d_{sh} $ as a function of the rotation parameter, for
fixed $ q_{e}$, $ e_{1}$,  $ \Lambda $, and  different values of
$ k_{e}$. \textbf{Left-down graph:} $ d_{sh} $ as a
function of the cosmological constant, for fixed $ q_{e}$,
$ k_{e} $, $ e_{1}$, and different values of the rotation
parameter. \textbf{Right-down graph:} $ d_{sh} $ as a function of the rotation parameter, for
fixed $ q_{e}$, $ k_{e}$, $ \Lambda $, and  different values of
$ e_{1}$. The green
shaded region gives the $ 1\sigma $ confidence region for $ d_{sh}
$, whereas the cyan shaded region gives the $ 2\sigma $ confidence
region. The inclination angle is $ \theta_{0}=0^{\circ}$.}
\label{Fig5a}
\end{figure}

\begin{figure}[!htb]
\centering
\subfloat[$k_{e}=0.2 $, $d_{1}=0.5 $]{
        \includegraphics[width=0.28\textwidth]{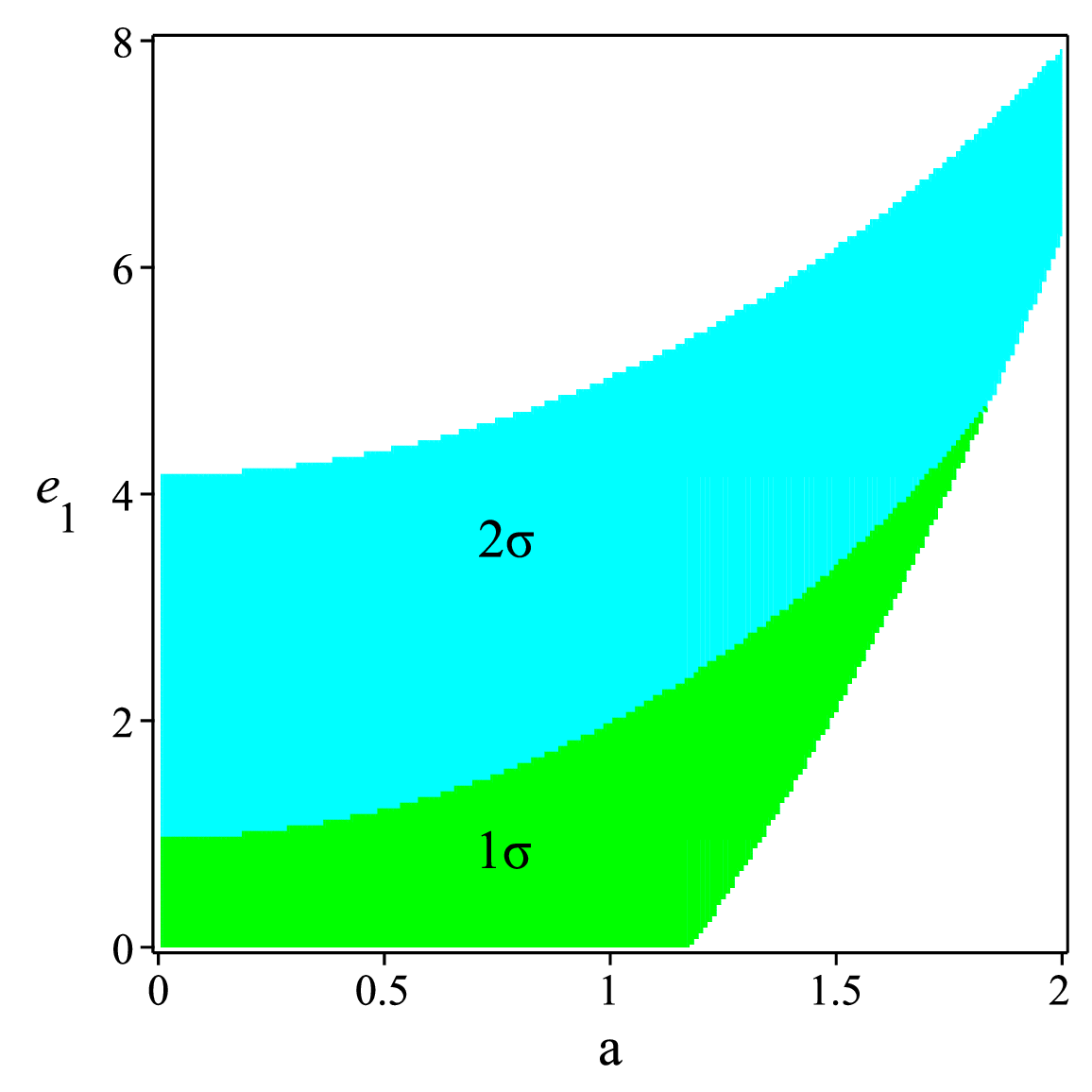}}
\subfloat[$e_{1}=d_{1}=0.5 $]{
        \includegraphics[width=0.28\textwidth]{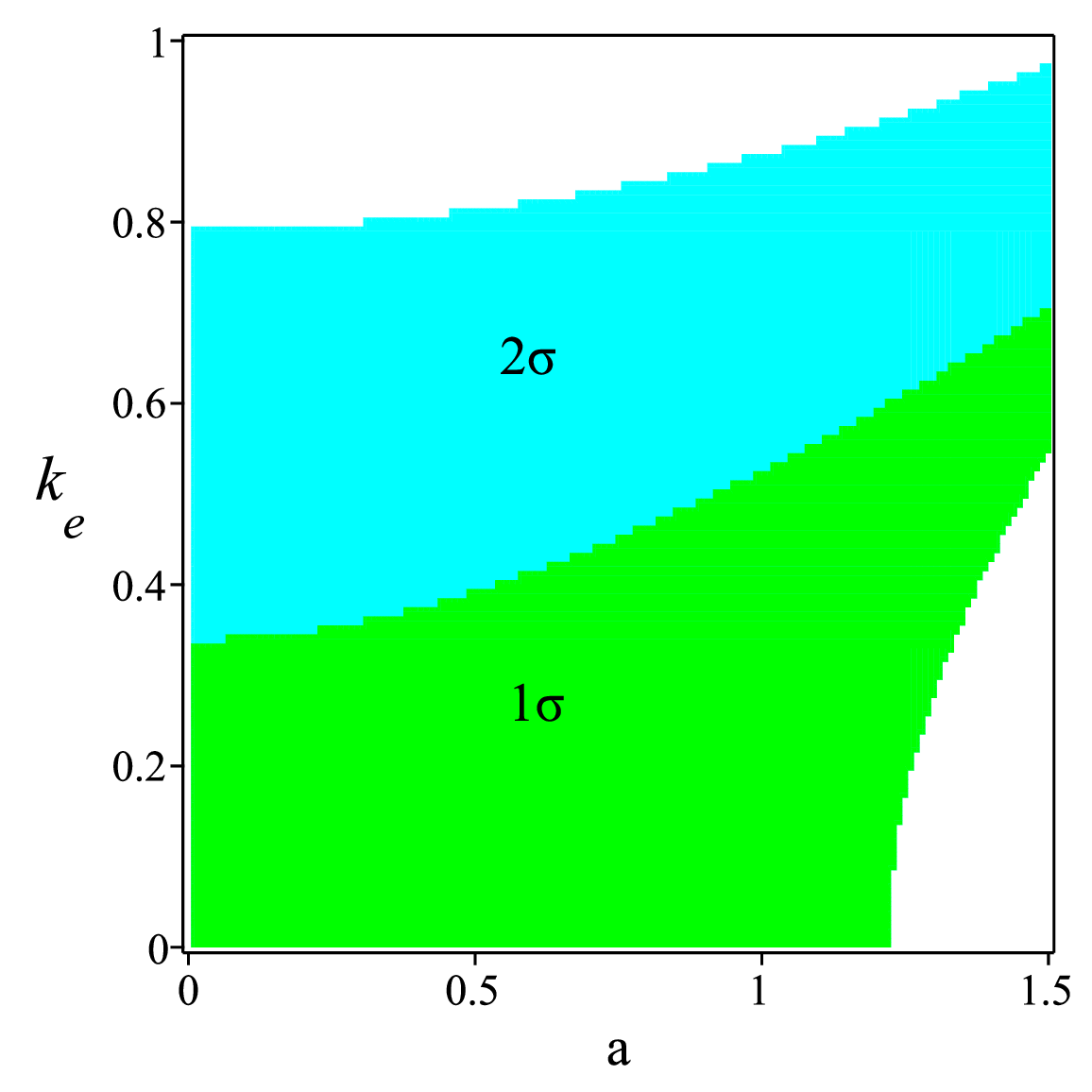}}
        \newline
\subfloat[$k_{e}=0.2 $, $d_{1}=-0.5 $]{
        \includegraphics[width=0.28\textwidth]{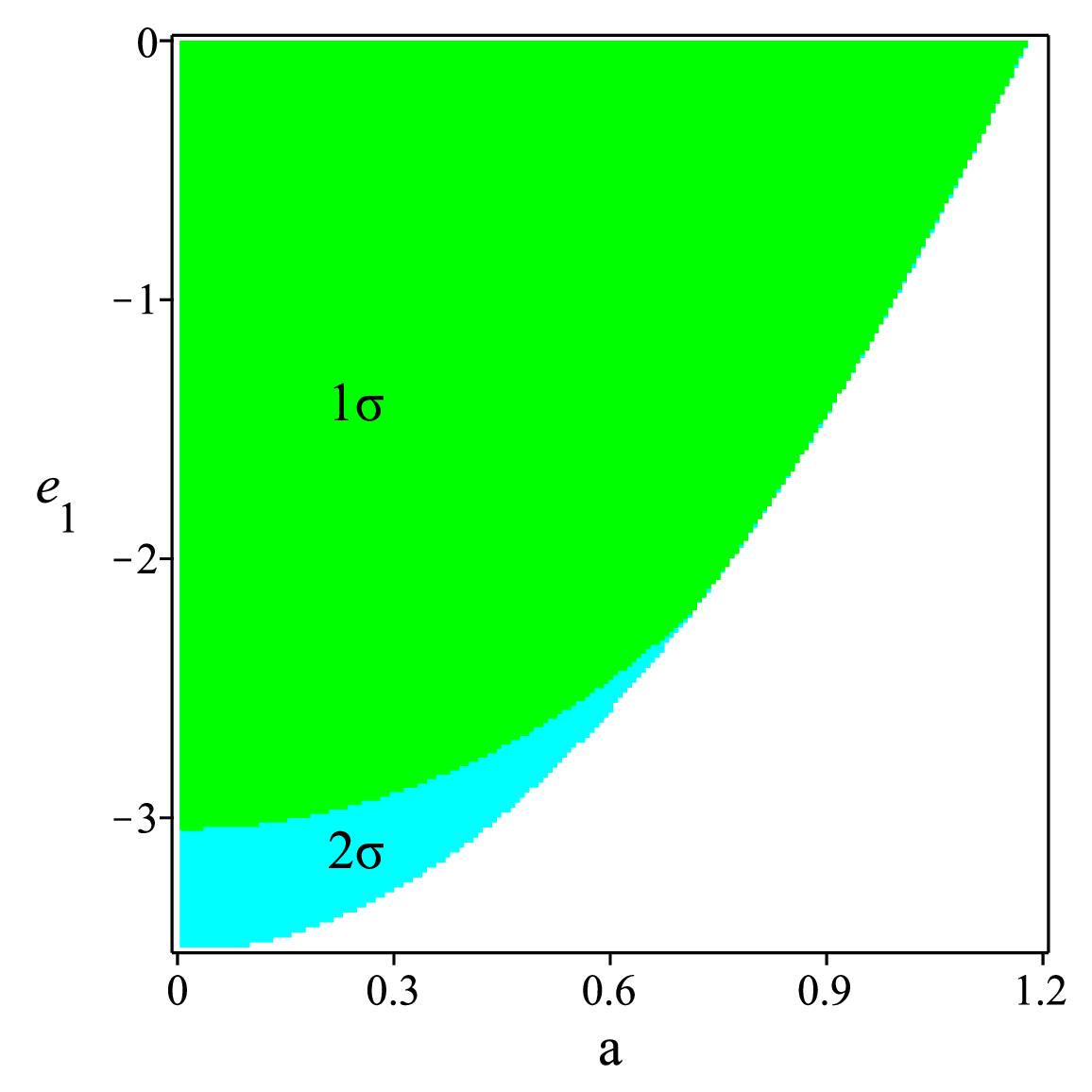}}
        \subfloat[$e_{1}=d_{1}=-0.5 $]{
        \includegraphics[width=0.28\textwidth]{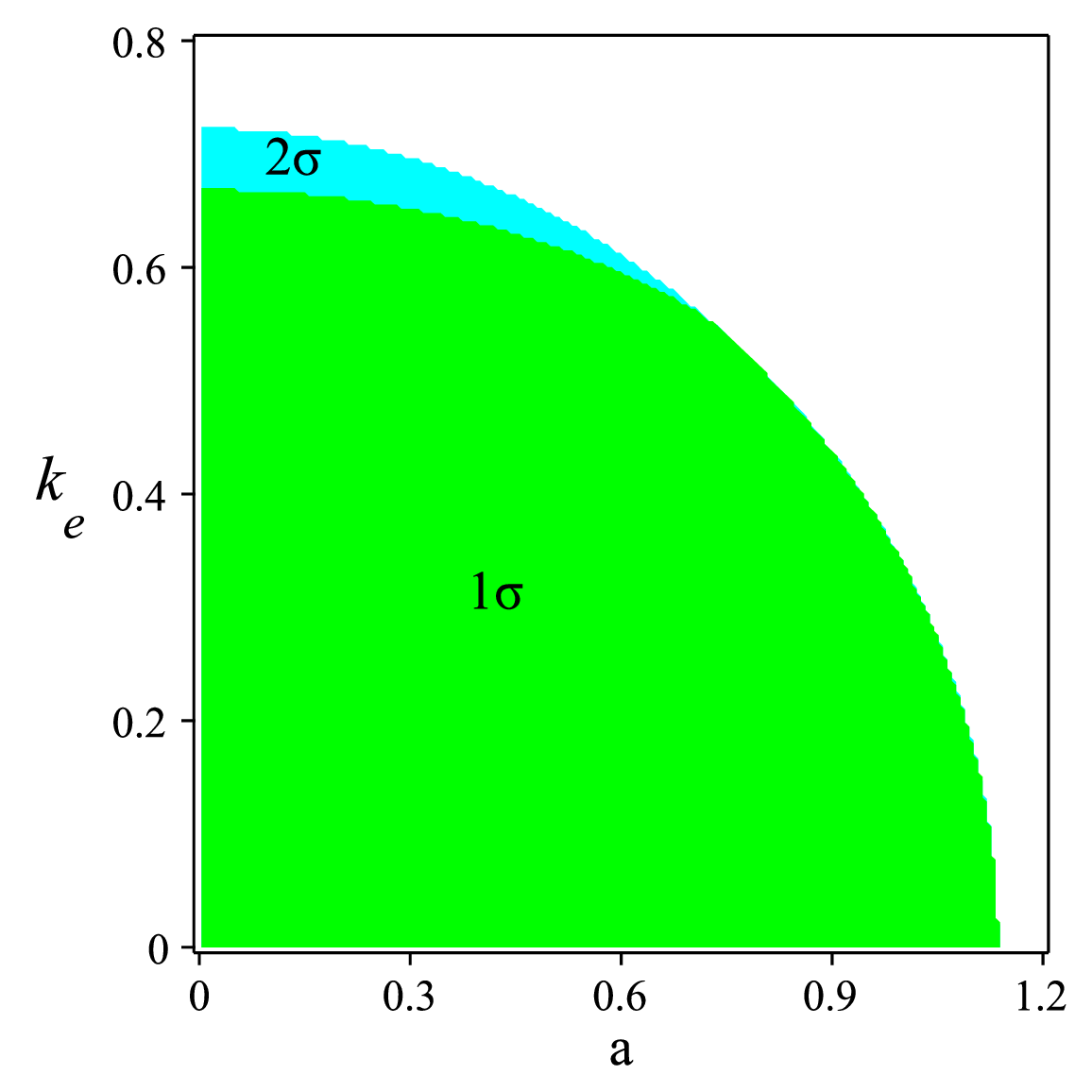}}
        \newline
\caption{ constraints on the Weyl-Cartan parameters and rotation parameter for $ k_{s}=10^{-6} $, $k_{m}=0.2 $,  $M=1$ and $ \theta_{0}=0^{\circ}$ with the EHT
observations of Sgr A*.}
\label{FigEHT01}
\end{figure}
To explore the effect of BH parameters on $d_{sh}$,  Figs. \ref{Fig5} (case I) and \ref{Fig5a} (case II) have been drawn for the inclination angle $ \theta_{0}=0^{\circ}$.
 Taking a look at Fig. \ref{Fig5}, one can find that the resulting shadow of fast-rotating black holes is
within $ 1\sigma $ uncertainty according to data for Sgr A* (see
dash dot curve in Fig. \ref{Fig5}(a)). For slowly rotating black
holes, $ d_{sh} $ becomes consistent with EHT data within $
1\sigma $-error  for  $ q_{e}\leq 0.972 $, otherwise it is in
agreement with observations within $ 2\sigma $
uncertainty (see solid curve in Fig. \ref{Fig5}(a)). It is clear from this figure that the resulting shadow of Kerr-Newman black holes in WCT for case I is consistent with Sgr A* shadow for all values of the electric charge. To find some constraints on the electric dilation parameter to have consistent results with EHT data of Sgr A*, we have plotted Fig. \ref{Fig5}(b) which shows that rotating black holes with
very small values of $ k_{e} $ have a compatible result with EHT
data within $ 1\sigma $ uncertainty. According to our analysis,  no compatible result can be observed with Sgr A* data for $ k_{e}> 1.27 $. Figure \ref{Fig5}(c) displays the allowed regions of the
cosmological constant for which the obtained shadow is consistent with Sgr A* shadow. As we see, for rotating BHs in a high curvature back ground, $d_{sh} $ becomes consistent with EHT data within $2\sigma $ uncertainty, otherwise it is in agreement with observations in $ 1\sigma $-error confidence region. From Fig  \ref{Fig5}(d), one finds allowed regions of the parameter $e_{1}$, showing that for small and intermediate values of  $e_{1}$, the resulting shadow is in agreement with observational data within $ 1\sigma
$-error, whereas for large values  $ d_{sh} $ is located within $ 2\sigma $-error.

\begin{table*}[t]
\caption{shadow observables $ D $, $ \delta_{s} $, $ \delta $ for the variation of $a$, $q_{e}$, $k_{e} $, $e_{1}$ and $\Lambda$ for $ q_{m}=k_{m}=0.2 $, $ k_{s}=10^{-6} $, $M =1$ and and $ \theta_{0}=50^{\circ}
$.}
      \centering
    \begin{tabular}{|c|c|c|c||c|c|c|c|}
    \hline
         \multicolumn{4}{|c|}{Case I}& \multicolumn{4}{|c|}{Case II} \\
    \hline 
     \multicolumn{4}{|c|}{ $ q_{e}=k_{e}=0.2 $, $ e_{1} =0.5$ and $ \Lambda =-0.02 $}& \multicolumn{4}{|c|}{$ q_{e}=k_{e}=0.2 $, $ e_{1} =-0.5$ and $ \Lambda =-0.02 $} \\
    \hline 
     $ a $ &\quad \quad \quad$ D $ \quad\quad \quad & \quad\quad$ \delta_{s} $\quad\quad &$\delta$ & $  a$  & \quad \quad\quad$ D $\quad\quad\quad    &\quad\quad $ \delta_{s} $\quad\quad & $ \delta $ \\ \hline
$ 0.2 $  & $ 0.99779 $&$ 0.00441 $ & $ -0.13618 $ &$ 0.1 $ & $ 0.99939 $ &$ 0.00141 $& $ -0.17005 $ \\
$ 0.4 $  & $ 0.99083 $ &$ 0.01839 $ & $ -0.14208 $ &$ 0.3 $ & $ 0.99345$ &$ 0.01311 $& $ -0.17477 $ \\
$ 0.6 $  & $ 0.97780 $ &$ 0.04473 $ & $ -0.15256 $ &$ 0.5 $ & $0.97905$ &$ 0.03981 $& $ -0.19168 $ \\
$ 0.8 $  & $ 0.95490 $ & $ 0.09128 $& $ -0.16918 $ &$ 0.7 $ & $ 0.95393 $ &$ 0.09317 $& $ -0.20259 $ \\
$ 1.0 $  & $ 0.90239 $ & $ 0.19799 $& $ -0.19840 $&$ 0.8 $ & $ 0.92846 $ &$ 0.14489 $& $ -0.21671$ \\  \hline  \hline
\multicolumn{4}{|c|}{$ a=0.5 $, $  k_{e}=0.2 $, $ e_{1} =0.5$ and $ \Lambda =-0.02 $}& \multicolumn{4}{|c|}{$ a=0.5 $, $  k_{e}=0.2 $, $ e_{1} =-0.5$ and $ \Lambda =-0.02 $} \\
    \hline 
$  q_{e}$  & $ D $& $ \delta_{s} $ & $ \delta $ & $  q_{e}$  &  $ D $  &$ \delta_{s} $ & $ \delta $ \\ \hline
$ 0.1 $  & $ 0.98554$ &$ 0.02905 $& $ -0.14347 $ &$ 0.1 $ & $ 0.98085 $ &$ 0.03853 $& $ -0.18094 $ \\
$ 0.3 $  & $ 0.98463 $ &$ 0.03089 $& $ -0.15216 $ &$ 0.3 $ & $ 0.97905$ &$ 0.04218 $&  $-0.19168 $ \\
$ 0.5 $  & $ 0.98234 $ &$ 0.03553 $& $ -0.17084 $ &$ 0.5 $ & $ 0.97385 $ &$ 0.05269 $&  $ -0.21553 $ \\
$ 0.7 $  & $ 0.97678$ &$ 0.04676 $& $ -0.20315 $ &$ 0.6 $ & $ 0.96794 $ &$ 0.06465 $&  $ -0.23448 $ \\
$ 0.9 $  & $ 0.95385 $ &$ 0.09315 $& $ -0.26160 $ &$ 0.7 $ & $0.95385 $ &$ 0.09311 $&  $ -0.26160 $ \\  \hline  \hline 
\multicolumn{4}{|c|}{$ a=0.5 $, $  q_{e}=0.2 $, $ e_{1} =0.5$ and $ \Lambda =-0.02 $}& \multicolumn{4}{|c|}{$ a=0.5 $, $  q_{e}=0.2 $, $ e_{1} =-0.5$ and $ \Lambda =-0.02 $} \\
    \hline
$  k_{e}$  & $ D $& $ \delta_{s} $ & $ \delta $ & $  k_{e}$  &  $ D $  &$ \delta_{s} $ & $ \delta $ \\ \hline
$ 0.2 $  & $ 0.98522 $ &$ 0.02971 $& $ -0.14668 $ &$ 0.1 $ & $ 0.98144 $ &$ 0.03734 $& $ -0.17709 $  \\
$ 0.4 $  & $0.98744 $ &$ 0.02523 $& $-0.12235 $ &$ 0.2 $ & $ 0.97905 $ &$ 0.03981 $& $ -0.19168 $ \\
$ 0.6 $  & $ 0.98984$ &$ 0.02039 $& $ -0.08735 $ &$ 0.3 $ & $ 0.97772 $ &$ 0.04497$& $ -0.19875 $  \\
$ 0.8 $  & $ 0.99189 $ &$ 0.01627 $& $ -0.04641 $ &$ 0.4 $ & $0.97251 $ &$ 0.05541 $& $ -0.22044 $  \\
$ 1.0 $  & $ 0.99349 $ &$ 0.01305 $& $ -0.00304 $ &$ 0.5 $ & $ 0.95846$ &$ 0.08381 $& $ -0.25466 $  \\  \hline  \hline 
\multicolumn{4}{|c|}{$ a=0.5 $, $  q_{e}=k_{e}=0.2 $ and $ \Lambda =-0.02 $}& \multicolumn{4}{|c|}{$ a=0.5 $, $  q_{e}=k_{e}=0.2 $ and $ \Lambda =-0.02 $} \\
    \hline
$  e_{1}$  & $ D $& $ \delta_{s} $ & $ \delta $ & $  e_{1}$  &  $ D $  &$ \delta_{s} $ & $ \delta $ \\ \hline
$ 0.01 $  & $ 0.98318 $ &$ 0.03382 $& $ -0.16441 $ &$ -0.01 $ & $ 0.98309 $&$ 0.03402 $& $ -0.16517 $  \\
$ 0.4 $  & $ 0.98485 $ &$ 0.03046 $& $ -0.15017 $ &$ -0.4 $ & $ 0.98089$ &$ 0.03845 $& $ -0.18068 $  \\
$ 0.8 $  & $ 0.98621 $ &$ 0.02772 $& $-0.13659 $ &$-0.8 $ & $ 0.97781 $ &$ 0.04468$& $ -0.19818 $  \\
$ 1.2 $  & $ 0.98731 $ &$0.02548 $& $ -0.12389 $ &$ -1.2$ & $ 0.97324 $ &$ 0.05392 $& $ -0.21780 $ \\
$ 1.6 $  & $ 0.98824 $ &$ 0.02362 $& $ -0.11194 $ &$ -1.6 $ & $ 0.96555 $ &$ 0.06951 $& $ -0.24055 $ \\  \hline  \hline 
\multicolumn{4}{|c|}{$ a=0.5 $, $  q_{e}=k_{e}=0.2 $ and $ e_{1} =0.5$}& \multicolumn{4}{|c|}{$ a=0.5 $, $  q_{e}=k_{e}=0.2 $ and $ e_{1} =-0.5$} \\
    \hline
$  \Lambda$  & $ D $& $ \delta_{s} $ & $ \delta $ & $  \Lambda$  &  $ D $  &$ \delta_{s} $ & $ \delta $ \\ \hline
$ -0.01 $  & $ 0.98740 $ &$ 0.02530 $& $ -0.08111 $ &$ -0.01$ & $ 0.98281 $ &$0.03456$& $ -0.12744 $ \\
$ -0.03$  & $0.98522 $ &$ 0.03212 $& $ -0.19965 $ &$ -0.02 $ & $ 0.97905 $ &$ 0.03981 $& $ -0.19168 $  \\
$ -0.05 $  & $ 0.98403 $ &$ 0.03396 $& $ -0.28099 $ &$ -0.03 $ & $ 0.97870 $ &$ 0.04289$& $ -0.23198 $ \\
$ -0.07 $  & $ 0.98312$ &$ 0.03395 $& $ -0.34143 $ &$-0.04 $ & $ 0.97782 $ &$ 0.04467 $& $ -0.27155 $ \\
$ -0.09 $  & $ 0.98352 $ &$0.03314$& $ -0.38868 $ &$ -0.05$ & $ 0.97735 $ &$ 0.04564 $& $ -0.30544 $  \\  \hline  
\end{tabular} 
\vspace{1ex}
\label{SgrA}
\end{table*}

To find the allowed regions of parameters for case II, we have plotted Fig. \ref{Fig5a}. From Fig. \ref{Fig5a}(a), data for
Sgr A* black hole gives bound on the electric charge such that  the resulting shadow of fast-rotating black holes is
within $ 1\sigma $ uncertainty for  $ q_{e}\leq 0.63 $, otherwise it is within $ 2\sigma $ uncertainty (see
dash dot curve in Fig. \ref{Fig5a}(a)). For slowly rotating black
holes, $ d_{sh} $ becomes consistent with EHT data within $
1\sigma $-error  for  all electric charges (see solid curve in Fig. \ref{Fig5a}(a)). Fig. \ref{Fig5a}(b) shows the influence of the electric dilation parameter on $ d_{sh} $. According to
dash-dot curve of this figure, rotating black holes with
intermediate values of $ k_{e} $ have a compatible result with EHT
data within $ 1\sigma $ ($ 2\sigma $) uncertainty for  $ a \leq 0.631 $ ($ a>0.631 $), whereas for very small values of this parameter, $ d_{sh} $ is located in $ 1\sigma $ confidence region according to Sgr A* data. From Fig. \ref{Fig5a}(c), we find the compatible range of the cosmological constant with EHT
data. According to our analysis, for the range of $ -0.056\leq \Lambda <0 $, the diameter of shadow is located in $ 1\sigma $  confidence region, otherwise it is located within $ 2\sigma $ uncertainty. To determine
allowed regions of the parameter $e_{1}$, we have plotted Fig. \ref{Fig5a}(d). According to this figure, fast-rotating BHs with large values of $\vert e_{1} \vert$ have a consistent shadow with data of Sgr A* within $2\sigma $ uncertainty, whereas for  small and intermediate values of $\vert e_{1} \vert$, $ d_{sh} $ is located within $ 1\sigma $-error. In Fig. \ref{FigEHT01},  we set $q_{e}=q_{m}=\Lambda=0$  and constrain Weyl-Cartan parameters to have consistent results. Comparing this figure to figures \ref{Fig5} and \ref{Fig5a}, we notice that the allowed regions of WC parameters decrease (increase) for case I (case II) in the absence of the electric and magnetic charges and the cosmological constant.

\begin{table*}[t]
\caption{Constraints on BH parameters for the two cases, Set by the EHT results.}
      \centering
      \begin{tabular}{|c| |c|c| |c|c||}
\hline
  &\multicolumn{2}{c||}{M87*} & \multicolumn{2}{c||}{Sgr A*} \\
\cline{2-5}
Case & \multicolumn{2}{c||}{ $  q_{e}=k_{e}=0.2 $, $ e_{1} =0.5$ and $ \Lambda=-0.02 $} & \multicolumn{2}{c||}{$  q_{e}=k_{e}=0.2 $, $ e_{1} =-0.5$ and $ \Lambda=-0.02 $} \\
\cline{2-5}
 & \multicolumn{1}{l|}{\quad\quad 1$\sigma$ \quad\quad}            & \quad\quad 2$\sigma$ \quad\quad           & \multicolumn{1}{l|}{\quad\quad 1$\sigma$ \quad\quad}            & \quad\quad 2$\sigma$ \quad\quad           \\ \hline
I                & \multicolumn{1}{l|}{~~~~~~~~$--$} & $a\in [0, 1.0)$ & \multicolumn{1}{l|}{$a\in [0, 1.0)$} & $--$ \\ \hline
II                & \multicolumn{1}{l|}{~~~~~~~~$--$} & $a\in [0, 0.8)$ & \multicolumn{1}{l|}{$a\in [0, 0.8)$} & $--$ \\ \hline \hline
 Case& \multicolumn{2}{c||}{$ a=0.5 $, $k_{e}=0.2 $, $ e_{1} =0.5$ and $ \Lambda=-0.02 $} & \multicolumn{2}{c||}{$ a=0.5 $, $k_{e}=0.2 $, $ e_{1} =-0.5$ and $ \Lambda=-0.02 $} \\
\cline{2-5}
 & \multicolumn{1}{l|}{\quad\quad 1$\sigma$ \quad\quad}            & \quad\quad 2$\sigma$ \quad\quad           & \multicolumn{1}{l|}{\quad\quad 1$\sigma$ \quad\quad}            & \quad\quad 2$\sigma$ \quad\quad           \\ \hline
 I                & \multicolumn{1}{l|}{~~~~~~~~$--$} & $q_{e}\in [0, 0.82]$ & \multicolumn{1}{l|}{$q_{e}\in [0, 0.85]$} & $q_{e}\in (0.85, 0.9]$ \\ \hline
II                & \multicolumn{1}{l|}{~~~~~~~~$--$} & $q_{e}\in [0, 0.6]$ & \multicolumn{1}{l|}{$q_{e}\in [0, 0.64]$} & $q_{e}\in (0.64, 0.7]$ \\ \hline \hline
Case& \multicolumn{2}{c||}{$ a=0.5 $, $q_{e}=0.2 $, $ e_{1} =0.5$ and $ \Lambda=-0.02 $} & \multicolumn{2}{c||}{$ a=0.5 $, $q_{e}=0.2 $, $ e_{1} =-0.5$ and $ \Lambda=-0.02 $} \\
\cline{2-5}
 & \multicolumn{1}{l|}{\quad\quad 1$\sigma$ \quad\quad}            & \quad\quad 2$\sigma$ \quad\quad           & \multicolumn{1}{l|}{\quad\quad 1$\sigma$ \quad\quad}            & \quad\quad 2$\sigma$ \quad\quad           \\ \hline
I                & \multicolumn{1}{l|}{~~~~~~~~$--$} & $k_{e}\in [0, 0.65]$ & \multicolumn{1}{l|}{$k_{e}\in [0, 1.26]$} & $k_{e}\in (1.26, 1.92]$ \\ \hline
II                & \multicolumn{1}{l|}{~~~~~~~~$--$} & $k_{e}\in [0, 0.45]$ & \multicolumn{1}{l|}{$k_{e}\in [0, 0.47]$} & $k_{e}\in (0.47, 0.5]$ \\ \hline \hline
Case& \multicolumn{2}{c||}{$ a=0.5 $, $q_{e}=k_{e}=0.2 $ and $ \Lambda=-0.02 $} & \multicolumn{2}{c||}{$ a=0.5 $, $q_{e}=k_{e}=0.2$ and $ \Lambda=-0.02 $} \\
\cline{2-5}
 & \multicolumn{1}{l|}{\quad\quad 1$\sigma$ \quad\quad}            & \quad\quad 2$\sigma$ \quad\quad           & \multicolumn{1}{l|}{\quad\quad 1$\sigma$ \quad\quad}            & \quad\quad 2$\sigma$ \quad\quad           \\ \hline
I                & \multicolumn{1}{l|}{$e_{1}\in [2.5, 24)$} & $e_{1}\in [0, 2.5)$ & \multicolumn{1}{l|}{$e_{1}\in (0, 10)$} & $e_{1}\in [10, 23]$ \\ \hline
II                & \multicolumn{1}{l|}{~~~~~~~~$--$} & $e_{1}\in [-1.48, 0)$ & \multicolumn{1}{l|}{$e_{1}\in (-1.28, 0)$} & $e_{1}\in [-2, -1.28]$ \\ \hline \hline
Case& \multicolumn{2}{c||}{$ a=0.5 $, $q_{e}=k_{e}=0.2 $ and $ e_{1} =0.5$} & \multicolumn{2}{c||}{$ a=0.5 $, $q_{e}=k_{e}=0.2 $ and $ e_{1} =-0.5$} \\
\cline{2-5}
 & \multicolumn{1}{l|}{\quad\quad 1$\sigma$ \quad\quad}            & \quad\quad 2$\sigma$ \quad\quad           & \multicolumn{1}{l|}{\quad\quad 1$\sigma$ \quad\quad}            & \quad\quad 2$\sigma$ \quad\quad           \\ \hline
I                & \multicolumn{1}{l|}{$\Lambda\in [-0.011, 0)$} & $\Lambda\in (-0.034, -0.011)$ & \multicolumn{1}{l|}{$\Lambda\in [-0.034, 0)$} & $\Lambda\in [-0.075, -0.034]$ \\ \hline
II                & \multicolumn{1}{l|}{~~~~~~~~$--$} & $\Lambda\in [-0.031, 0)$ & \multicolumn{1}{l|}{$\Lambda\in [-0.029, 0)$} & $\Lambda\in (-0.07, -0.029)$ \\ \hline
\end{tabular}
\vspace{1ex}
\label{dsh}
\end{table*}

Now, we calculate Schwarzschild shadow deviation $ \delta $ based on observations of SgrA*. EHT used the two separate priors for the Sgr A* shadow size
from the Keck and Very Large Telescope
Interferometer (VLTI) observations to estimate the bounds on
the fraction deviation observable $ \delta $ as follows \cite{Akiyama:2022f}
\begin{equation}
\delta=\left\{
  \begin{array}{ll}
  -0.08^{+0.09}_{-0.09}  & \;\;(\mbox{VLTI}) \\
  -0.04^{+0.09}_{-0.10}  & \;\;(\mbox{Keck})
  \end{array}
\right..
\end{equation}

It is worthwhile to mention that the Event Horizon Telescope collaboration does not impose any constrains on $ \Delta C $ and $ D_{x} $ for the image of Sgr A*.

Table \ref{SgrA} lists several values of $ D $, $ \delta_{s} $ and $ \delta $ under variation of BH parameters for both cases I and II. Here we have considered the case with
$ \theta_{0}=50^{\circ}
$ because as addressed in \cite{Islam:2023k} the inclination angle (with respect to the line of sight) is estimated to be
$ \theta_{0}=50^{\circ}
$ in the Sgr A*. As is seen from this table, the condition $\sqrt{3}/2 \leq D<1 $ is satisfied for the whole parameters for both cases I and II. The distortion $ \delta_{s} $ is non-zero and becomes larger for larger rotation parameter, electric charge and cosmological constant in both cases. Regarding to the electric dilation charge and $ e_{1} $ parameter, $ \delta_{s} $ increases as these two parameters decrease (increase) in case I (case II). The fractional deviation $ \delta $ increases by increasing the rotation parameter. For case I, the range $ 0< a< 0.4$ satisfies $ 1\sigma $ Keck bound, while the range $ 0\leq a \leq 0.8$ satisfies $ 1\sigma $ VLTI bound. For case II, no values of the rotation parameter can satisfy  $ 1\sigma $ Keck bound. Only the range $ 0< a< 0.1$ satisfies $ 1\sigma $ VLTI bound. But all values of $a$ satisfy both VLTI and Keck bounds  within $ 2\sigma $ uncertainty. Regarding the electric charge, as is seen from the second row of table \ref{SgrA}, the $ 1\sigma $ Keck bound is not satisfied for any values of the electric charge for case I. While VLTI measurements
within the $ 1\sigma $ confidence interval, constraint $0< q_{e}<0.5 $. For case II,  $ 1\sigma $  bound is never satisfied, but the range $ 0< q_{e}\leq 0.6  $ ($ 0< q_{e}< 0.7  $) can satisfy $ 2\sigma $ confidence interval of Keck (VLTI) measurements. From third row of table \ref{SgrA}, we see that in case I, all values of the electric dilation parameter $ k_{e} $ satisfy $ 1\sigma $ VLTI bound, while according to Keck measurements, for $ k_{e}>0.2 $ this constraint can be satisfied. In case II, only $ 2\sigma $ Keck (VLTI) bound is satisfied in the range $ 0<k_{e}\leq 0.45 $ ($ 0<k_{e}\leq 0.5 $). According to forth row of table \ref{SgrA}, in case I,  $ 1\sigma $ VLTI bound is satisfied for all $ e_{1} $, while $ e_{1}\geq 0.8 $ is the allowed range of this parameter to satisfy $ 1\sigma $ keck bound. For case II, only  $ 1\sigma $ VLTI bound is satisfied in the range $ -0.15<e_{1}<0 $, while  $ 2\sigma $ Keck (VLTI) bound can be satisfied in the range $ -1.6< e_{1}<0 $ ($ -1.8\leq e_{1}<0 $). Regarding the cosmological constant, the range $ -0.017< \Lambda <0 $ ($ -0.024< \Lambda <0 $) satisfies $ 1\sigma $ Keck (VLTI) bound in case I. According to our analysis, for $ \Lambda \leq -0.043 $, the fractional deviation $ \delta $ is beyond Keck/VLTI bound. For case II, the allowed range of $ \Lambda$ is $(-0.012,0)$ for $ 1\sigma $ Keck bound and $(-0.017,0)$ for $ 1\sigma $ VLTI bound. Our findings show that for $ \Lambda \leq -0.04 $, the fractional deviation $ \delta $ is located outside Keck/VLTI measurements. 

Table \ref{dsh} shows the allowed regions of BH parameters for both cases for which the resulting shadow diameter is consistent with EHT data of M87*/SgrA* within $ 1\sigma $/$ 2\sigma $ uncertainty. As we see, for no values of the rotation parameter, electric charge and electric dilation parameter, $ d_{sh} $ is not in agreement with observations for M87* within $ 1\sigma $ uncertainty. Only for case I and in the range $ 2.5 < e_{1}<24 $ or $ -0.01\leq\Lambda<0 $, a rotating BH located in a weak electric field has a compatible shadow with EHT data of M87* within $ 1\sigma $ uncertainty. But for given regions of parameters, the resulting shadow is located in $ 1\sigma $ confidence region of Sgr A* data for both cases I and II.
So SgrA* black hole can be a suitable model for Kerr-Newman BHs in WCT.

As a final point, it is worthwhile to point out that in Ref. \cite{Ghosh:2023174}, authors employed the EHT observational results of Sgr A* to investigate the constraints on the distortion parameter and diameter of the shadow of Kerr-Newman BHs in GR. They found that these two observables $d_{sh}$ and $\delta $ reduce with the increase of the rotation parameter and electric charge which is in agreement with our results. 


\section{Conclusion}
The field of research in BH imaging and the optical appearance of black holes has experienced significant growth recently. Numerous studies have been conducted to explore how data from the Event Horizon Telescope (EHT) can be used to constrain various parameters of black holes. In this study, we specifically investigated Kerr-Newman black holes within the framework of WCT by examining their geometric and optical properties. Studying
geometrical property, we found that there is an essential
singularity at $ r=0 $ and $ \theta =\pi/2 $ which is covered by
an event horizon, confirming the existence of black holes in this
theory of gravity. Then, we studied the optical features of black
holes such as  photon sphere radius and shadow size, and noticed
that some constraints should be imposed on parameters of the model
to have acceptable optical behavior. Studying the effect of
parameters on the size and shape of the BH shadow, we found that
the rotation parameter causes deformations to both of them. We
also saw that the shape of the shadow depends on the inclination
angle $\theta_{0}$ of the far distant observer and is skewed as
$\theta_{0}$ increases. For the case of $\theta_{0}=0$,  the
shadow is a round disk whereas at the equatorial plane
($\theta_{0}=\pi/2$) it becomes most distorted. We examined the
influence of electric charge on the BH shadow and observed that
increasing this parameter leads to decreasing the shadow size.
Regarding the electric dilation parameter, its effect on the
radius of shadow was opposite (similar) to the electric charge in case I (case II). Investigating
the cosmological constant effect, we found that $\vert
\Lambda\vert$ decreases the shadow radii. We continued by
calculating the energy emission rate and observed that the
rotation parameter, electric charge,  and $\vert
\Lambda\vert$ decrease the emission rate, indicating that the evaporation process would be slow with an increase of these three parameters. Regarding the effect of the electric dilation parameter and $e_{1}$ on the energy emission rate, our findings showed that increasing $ k_{e} $ and $\vert e_{1}\vert$ makes increasing (decreasing) the emission rate in case I (case II).

Finally, we considered M87* BH and Sgr A* BH as models for Kerr-Newman BHs in WCT and imposed some constraints on the parameters of the model to have consistent results with observational data. Calculating the shadow diameter, we noticed that the resulting shadow of these BHs is not located within $ 1\sigma $ uncertainty of M87* data for no values of the rotation parameter, electric charge, and electric dilation parameter. While for given ranges of parameters, the resulting shadow was in agreement with observational data of Sgr A*  in $ 1\sigma $ confidence region. This reveals the fact Sgr A* BH can be a suitable model for corresponding BHs.We also obtained the circularity deviation $ \Delta C $ and axial ratio $ D_{x} $ and found that the constraint $ \Delta C \lesssim 0.1 $ is satisfied for the whole parameters of the model. While the condition  $1 < D_{x} \lesssim 4/3$ imposes some constraints on $k_{e}$ for case I and cosmological constants for both cases. Moreover, we computed the Schwarzschild shadow deviation $(\delta)$ by using measurements of M87* and SgrA*.  According to our analysis, all values of the rotation parameter, electric dilation parameter, and parameter $e_{1}$ satisfy the $ 1\sigma $ bound of M87* measurements in case I, whereas in case II, $ 1\sigma $ bound can be satisfied only for special ranges of these parameters. Comparing the fraction deviation observable $ \delta $ with values reported by SgrA* measurements, we found that resulting $ \delta $ lies within both the VLTI
and Keck bounds. For the case I, its $ \delta $ is located in $ 1\sigma $ uncertainty, whereas for case II, $ \delta $ lies in $ 2\sigma $ uncertainty. 

It is also important to note that the solution studied in this paper arises from a theory with dynamical torsion and the Weyl part of nonmetricity, making the theory quite complex. Indeed, to fully consider this solution as a potentially realistic description of black holes, one would need to verify its stability under linear perturbations. However, conducting such an analysis is beyond the scope of our paper, as it would require a comprehensive decomposition of torsion and nonmetricity into odd and even parity modes around black hole configurations. Such a study has not yet been undertaken in the literature, which is why there are no papers exploring effects beyond black hole backgrounds in non-Riemannian theories of gravity. To address this, one would need to adopt a similar approach as presented in \cite{Aoki:2023sum}, where the theory of cosmological perturbations was initially developed. We leave these investigations for future research.

\begin{acknowledgements}
SHH and KhJ thank Shiraz University Research Council. KhJ is
grateful to the Iran Science Elites Federation for the financial
support. SB was supported by JSPS Postdoctoral Fellowships for Research in Japan and KAKENHI Grant-in-Aid for Scientific Research No. JP21F21789 (final part of the fellowship). SB is now supported by “Agencia Nacional de Investigación y Desarrollo” (ANID), Grant “Becas Chile postdoctorado
al extranjero” No. 74220006.
\end{acknowledgements}

\end{document}